\documentclass{article}
\usepackage[utf8]{inputenc}
\usepackage{amsmath,amsthm,amsfonts}
\usepackage[margin=1in]{geometry}
\usepackage{mathrsfs}
\usepackage{graphicx}
\usepackage[linktocpage,linkcolor=red]{hyperref}
\hypersetup{
    colorlinks,
    allcolors=red
}
\usepackage{makecell}
\usepackage{newtxmath}
\usepackage{thm-restate}
\usepackage{csquotes}
\usepackage[capitalise]{cleveref}
\usepackage{algpseudocode}

\allowdisplaybreaks

\usepackage{fancybox, pdfsync}

\newtheorem{definition}{Definition}

\usepackage[noend,linesnumbered]{algorithm2e}

\newenvironment{fminipage}%
  {\begin{Sbox}\begin{minipage}}%
  {\end{minipage}\end{Sbox}\fbox{\TheSbox}}

\usepackage{bbm}

\newtheorem{theorem}{Theorem}[section]
\newtheorem{corollary}[theorem]{Corollary}
\newtheorem{lemma}[theorem]{Lemma}
\newtheorem{note}[theorem]{Note}
\newtheorem{fact}[theorem]{Fact}
\newtheorem{remark}[theorem]{Remark}
\newtheorem*{remark*}{Remark}

\DeclareMathOperator*{\argmax}{arg\,max}
\DeclareMathOperator*{\argmin}{arg\,min}

\DeclareMathOperator*{\st}{s.t.}

\usepackage{float}
\usepackage{color}

\def\floor#1{\left\lfloor #1 \right\rfloor}
\def\ceil#1{\left\lceil #1 \right\rceil}

\def\abs#1{\left|#1  \right|}

\def\norm#1{\left\| #1 \right\|}

\newcommand\R{\mathbb{R}}
\newcommand\Z{\mathbb{Z}}
\newcommand\Q{\mathbb{Q}}
\newcommand\N{\mathbb{N}}

\newcommand\mtime{\textsc{MM}}
\newcommand\Otil{\tilde{O}}

\newcommand\omegadual{\alpha}
\newcommand\omegadualcap{\widehat{\alpha}}
\newcommand\nnz{\textsc{nnz}}
\newcommand\ds{\textsc{ds}}
\newcommand\pds{\textsc{pds}}
\newcommand\dds{\textsc{dds}}
\newcommand\sds{\textsc{sds}}
\newcommand\initialize{\textsc{Initialize}}
\newcommand\update{\textsc{Update}}
\newcommand\query{\textsc{Query}}
\newcommand\ifltwoipm{\textsc{InverseFreeL2IPM}}
\newcommand\genericipm{\textsc{GenericIPM}}
\newcommand\old{\text{old}}
\newcommand\new{\text{new}}
\newcommand\final{\text{end}}
\newcommand\ptope{\mathcal{P}}
\newcommand\dualptope{\mathcal{D}}
\newcommand\phil{\Phi_{\lambda}}
\newcommand\qround{\textsc{QRound}}
\newcommand\round{\textsc{Round}}

\newcommand{\vertiii}[1]{{\left\vert\kern-0.25ex\left\vert\kern-0.25ex\left\vert #1 
    \right\vert\kern-0.25ex\right\vert\kern-0.25ex\right\vert}}

\let\emptyset\varnothing

\newcommand{\fro}{\textnormal{F}}
\newcommand{\veps}{\varepsilon}
\DeclareMathOperator*{\poly}{poly}

\newcommand{\opt}{\mathtt{opt}}

\newcommand{\vb}{\mathbf{b}}
\newcommand{\vbtil}{\widetilde{\mathbf{b}}}
\newcommand{\vbbar}{\overline{\mathbf{b}}}
\newcommand{\vc}{\mathbf{c}}
\newcommand{\vcbar}{\overline{\mathbf{c}}}
\newcommand{\vctil}{\widetilde{\mathbf{c}}}
\newcommand{\vg}{\mathbf{g}}
\newcommand{\vp}{\mathbf{p}}
\newcommand{\vq}{\mathbf{q}}
\newcommand{\vqtil}{\widetilde{\mathbf{q}}}
\newcommand{\vr}{\mathbf{r}}
\newcommand{\vrhat}{\widehat{\mathbf{r}}}
\newcommand{\vrbar}{\overline{\mathbf{r}}}
\newcommand{\vrtil}{\widetilde{\mathbf{r}}}
\newcommand{\vs}{\mathbf{s}}
\newcommand{\vstil}{\widetilde{\mathbf{s}}}
\newcommand{\vsbar}{\overline{\mathbf{s}}}
\newcommand{\vt}{\mathbf{t}}
\newcommand{\vthat}{\widehat{\mathbf{t}}}
\newcommand{\vu}{\mathbf{u}}
\newcommand{\vutil}{\widetilde{\mathbf{u}}}
\newcommand{\vecv}{\mathbf{v}}
\newcommand{\vecvbar}{\overline{\mathbf{v}}}
\newcommand{\vw}{\mathbf{w}}
\newcommand{\vwtil}{\widetilde{\mathbf{w}}}
\newcommand{\vwhat}{\widehat{\mathbf{w}}}
\newcommand{\vx}{\mathbf{x}}
\newcommand{\vh}{\mathbf{h}}
\newcommand{\vxhat}{\widehat{\mathbf{x}}}
\newcommand{\vshat}{\widehat{\mathbf{s}}}
\newcommand{\vxbar}{\overline{\mathbf{x}}}
\newcommand{\vxtil}{\widetilde{\mathbf{x}}}
\newcommand{\vxstar}{\mathbf{x}^*}
\newcommand{\vy}{\mathbf{y}}
\newcommand{\vyhat}{\widehat{\mathbf{y}}}

\newcommand{\vystar}{\mathbf{y}^*}
\newcommand{\vz}{\mathbf{z}}

\newcommand{\vdelta}{\mathbf{\Delta}}
\newcommand{\vdeltabar}{\overline{\mathbf{\Delta}}}
\newcommand{\vdeltatil}{\widetilde{\mathbf{\Delta}}}
\newcommand{\vdeltahat}{\widehat{\mathbf{\Delta}}}
\newcommand{\vdeltastar}{\mathbf{\Delta}^*}
\newcommand{\vdel}{\mathbf{\delta}}
\newcommand{\vdeltil}{\widetilde{\mathbf{\delta}}}
\newcommand{\vmu}{\mathbf{\mu}}

\newcommand{\ma}{\mathbf{A}}
\newcommand{\mahat}{\widehat{\mathbf{A}}}
\newcommand{\mabar}{\overline{\mathbf{A}}}

\newcommand{\mb}{\mathbf{B}}

\newcommand{\mc}{\mathbf{C}}
\newcommand{\md}{\mathbf{D}}
\newcommand{\me}{\mathbf{E}}
\newcommand{\mg}{\mathbf{G}}

\newcommand{\mh}{\mathbf{H}}
\newcommand{\mi}{\mathbf{I}}
\newcommand{\mk}{\mathbf{K}}
\newcommand{\mm}{\mathbf{M}}
\newcommand{\mmtil}{\widetilde{\mathbf{M}}}
\newcommand{\mn}{\mathbf{N}}
\newcommand{\matp}{\mathbf{P}}
\newcommand{\mpbar}{\overline{\mathbf{P}}}
\newcommand{\mptil}{\widetilde{\mathbf{P}}}
\newcommand{\mq}{\mathbf{Q}}
\newcommand{\mr}{\mathbf{R}}
\newcommand{\ms}{\mathbf{S}}
\newcommand{\msbar}{\overline{\mathbf{S}}}
\newcommand{\mstil}{\widetilde{\mathbf{S}}}
\newcommand{\mt}{\mathbf{T}}
\newcommand{\mttil}{\widetilde{\mathbf{T}}}
\newcommand{\mthat}{\widehat{\mathbf{T}}}
\newcommand{\mv}{\mathbf{V}}
\newcommand{\matu}{\mathbf{U}}
\newcommand{\mw}{\mathbf{W}}
\newcommand{\mwtil}{\widetilde{\mathbf{W}}}
\newcommand{\mwhat}{\widehat{\mathbf{W}}}
\newcommand{\mx}{\mathbf{X}}
\newcommand{\mxbar}{\overline{\mathbf{X}}}
\newcommand{\mxtil}{\widetilde{\mathbf{X}}}
\newcommand{\my}{\mathbf{Y}}
\newcommand{\myhat}{\widehat{\mathbf{Y}}}
\newcommand{\mmhat}{\widehat{\mathbf{M}}}
\newcommand{\mz}{\mathbf{Z}}
\newcommand{\mztil}{\widetilde{\mathbf{Z}}}
\newcommand{\mzhat}{\widehat{\mathbf{Z}}}
\newcommand{\mzero}{\mathbf{0}}

\DeclareMathOperator{\tr}{tr}

\usepackage{enumitem}

\def\eps{\epsilon} 

\begin{document}

\author{Mehrdad Ghadiri\footnote{Georgia Institute of Technology, \url{ghadiri@gatech.edu}} \and Richard Peng\footnote{Georgia Institute of Technology \& University of Waterloo, \url{y5peng@uwaterloo.ca}} \and Santosh S. Vempala\footnote{Georgia Institute of Technology, \url{vempala@gatech.edu}}}

\title{
The Bit Complexity of Efficient Continuous Optimization
}
\date{}
\maketitle
\begin{abstract}
We analyze the bit complexity of efficient algorithms for fundamental optimization problems, such as linear regression, $p$-norm regression, and linear programming (LP). State-of-the-art algorithms are iterative, and in terms of the number of arithmetic operations, they match the current time complexity of multiplying two $n$-by-$n$ matrices (up to polylogarithmic factors). However, previous work has typically assumed infinite precision arithmetic, and due to complicated inverse maintenance techniques, the actual running times of these algorithms are unknown.
To settle the running time and bit complexity of these algorithms, we demonstrate that a core common subroutine, known as \emph{inverse maintenance}, is backward-stable. Additionally, we show that iterative approaches for solving constrained weighted regression problems can be accomplished with bounded-error pre-conditioners.
Specifically, we prove that linear programs can be solved approximately in matrix multiplication time multiplied by polylog factors that depend on the condition number $\kappa$ of the matrix and the inner and outer radius of the LP problem. $p$-norm regression can be solved approximately in matrix multiplication time multiplied by polylog factors in $\kappa$. Lastly, linear regression can be solved approximately in input-sparsity time multiplied by polylog factors in $\kappa$.
Furthermore, we present results for achieving lower than matrix multiplication time for $p$-norm regression by utilizing faster solvers for sparse linear systems.

\let\thefootnote\relax\footnotetext{A preliminary version of this work that focuses only on the $p$-norm problem and the running time for sparse instances appears in \url{arxiv:2109.11537} \cite{ghadiri2021faster}.}
\end{abstract}

\newpage

\tableofcontents

\thispagestyle{empty}
\newpage
\pagenumbering{arabic}
\section{Introduction}

Over the past two decades,
many breakthroughs in algorithm design have relied on continuous algorithmic primitives~\cite{lee2013path,bubeck2018homotopy,AdilKPS19,CLS21,DBLP:conf/soda/Brand20,BrandLSS20}.
The increased attention on continuous methods has in turn
led to renewed interest and improved runtime bounds
for fundamental numerical routines. 
Many of these bounds, however, were initially claimed assuming
exact computations with infinite precision\footnote{This is unlike the development in the classic book~\cite{grotschel2012geometric}, where bit complexity analysis is carried out carefully when establishing polynomial time bounds.}.
Relaxing this assumption, of course, is crucial for claiming
truly fast(er) algorithms.
As an illustration, in the case of linear systems,
Krylov space methods such as the celebrated Conjugate Gradient algorithm
are known to take $O(n \cdot \nnz(\ma))$ arithmetic operations (where $\nnz$ denotes the number of nonzero entries),
but each involving intermediate numbers with $\Omega(n)$ bits in the worst case.
So, even though each iteration is simply a matrix-vector multiplication, the cost of each iteration is $O(n\cdot\nnz(\ma))$ and the overall cost is $O(n^2\cdot\nnz(\ma))$, which makes it considerably worse than ``direct methods" --- linear systems can be solved with bit complexity $\tilde{O}(n^\omega)$~\cite{storjohann2005shifted}. It was shown via a careful bit complexity analysis, that the block-Krylov method can be used to solve sufficiently sparse linear systems for poly-conditioned matrices faster than matrix multiplication \cite{PengVempala21,nie2022matrix}. 

For more general convex optimization problems such as regression and linear programming, fast iterative methods have been studied intensively for decades, with much of the focus on the interior-point method (IPM) for convex optimization. 
Since Karmarkar~\cite{Karmarkar84} and Vaidya's seminal papers~\cite{vaidya1989speeding}, maintaining the inverse of a matrix  modified by low-rank updates has been an important tool in fast algorithms for linear programming~\cite{DBLP:conf/stoc/CohenLS19,DBLP:conf/soda/Brand20,BrandLSS20,BLLSS0W21,jiang2020fasterb,song2021oblivious}, $p$-norm regression \cite{bubeck2018homotopy,AdilKPS19,AdilPS19,AdilS20,adil2022fast,adil2021unifying}, semi-definite programming~\cite{jiang2020faster,huang2022solving,jiang2022faster} and a host of dynamic optimization problems~\cite{van2019dynamica,van2019dynamicb,van2021fast,jiang2022dynamic}. Recent successes in achieving asymptotic complexity close to the time required for matrix multiplication all rely on solving a slowly-changing linear system in each iteration. As a result, they lend themselves to inverse maintenance, rather than solving the linear system from scratch in each iteration.

In the optimization literature, it is widely acknowledged that
the bit complexity analysis can be difficult.
For example, in the paper that presents his interior-point method~\cite{Renegar88}, Renegar makes the following remark about the
bit complexity of the original interior point algorithm due to Karmarkar:
\begin{displayquote}
``In the original version of this paper I wrote that I did not see how Karmarkar's
algorithm could be carried out with $O(L)$ bits of accuracy (assuming the number
of bits required to represent the original problem is $L$) as Karmarkar claimed in
his paper. Subsequently, Karmarkar convinced me that this could be done if one
does not rely on rank one updates, as the algorithm in the present paper does not.
The argument, embedded in our complexity analysis, relies on the fact that the
linear equations that need to be solved need only be solved approximately, and this can be done efficiently using Cholesky factorization and the fact that the condition
number of the corresponding matrices are bounded by $2^{O(L)}$.'' \cite{Renegar88}
\end{displayquote}
Here $L$ refers to the total bit complexity of the problem (i.e., the sum of the number of bits of all entries of $\ma, \vb, \vc$). Later works argued that $L$ can be replaced by the log of the maximum subdeterminant of the constraint matrix $\ma$ \cite{lee2013path,DBLP:conf/stoc/CohenLS19,DBLP:conf/soda/Brand20}. Since the latter is smaller, we adopt the latter definition for $L$ for the rest of the paper. Let us note right away that $L$ can be as large as $n\ell$ where $\ell$ is the number of bits used to represent any single entry of $\ma$. Moreover $L=\Omega(n)$, with high probability, on random matrices \cite{tao2005random}. This implies that in most instances, the actual
running time of these algorithms is a factor of $n$ more than the number of arithmetic operations. In this paper, we show that these algorithms (with proper modifications and roundings) can be carried out with a bit complexity depending on the logarithm of the condition number (which is $O(\log n)$, with high probability, on random matrices \cite{edelman1988eigenvalues,edelman1989eigenvalues}), and the logarithm of the ratio of the outer and inner radius of the LP. In another paper, that introduced his condition number \cite{renegar1996condition}, Renegar promotes the use of the conjugate-gradient method for solving the linear systems arising in each step of the IPMs. Note that this also leads to an extra factor of $n$.

Since matrix inverses are only computed approximately, it is important to show the number of bits of accuracy needed to guarantee the final target accuracy remains small despite the error accumulation during inverse maintenance. This property is closely related to what numerical analysts call {\em stability} and is carefully studied in many numerical algorithms, e.g., for the computation of eigenvalues~\cite{trefethen1997numerical}; however, it is not rigorously established for state-of-the-art results based on the IPM.  

The situation is more troubling for nonlinear optimization problems such as $p$-norm minimization, i.e., $\min \{\|\vx\|_p^p\, :\, \ma^\top\vx = \vb\}$. For any $p\ge 2$, there are iterative algorithms that need only 
$\Otil(n^{1/3})$ iterations (unlike LP, which currently needs $\sqrt{n}$ iterations). However, the analysis of these algorithms~\cite{AdilKPS19,AdilPS19,AdilS20,bubeck2018homotopy} assumes infinite bit precision, while using sophisticated variants of inverse maintenance. Another difficulty with $p$-norm minimization is that the bit complexity of the exact solution can be unbounded (since it can be irrational), while for linear systems (and linear regression, i.e., $p=2$), it can be bounded by $\Otil(n\ell)$ using rational number representations, where $\ell$ is the bit complexity of the entries of the input matrix. It has been shown that a linear system can be solved in time $\Otil(n^{\omega}\cdot \ell)$ \cite{storjohann2005shifted}; however, the bit complexity of algorithms even for solving linear regression problems in input-sparsity time, a widely studied and important problem \cite{clarkson2014sketching,NelsonN13,CohenLMMPS15}, is not established. 

In this paper, we address the bit complexity of state-of-the-art algorithms for linear regression, $p$-norm minimization, and linear programming. Our core technical results bound the bit complexity of general inverse maintenance and iterative algorithms for solving linear regression problems. We believe that these tools will be broadly useful for numerical algorithms in continuous optimization. 

The impractical nature of matrix multiplication algorithms with exponents below $2.5$ means the fastest bit complexity bounds we state are only of theoretical interest.
However, our results apply to all matrix multiplication algorithms up to the current fastest one, with matrix exponent $\omega\approx 2.372$ \cite{AW21}.
In particular, they apply to algorithms with $O(n^3)$ and $O(n^{2.808})$ (Strassen's algorithm \cite{strassen1969gaussian}) running times. Moreover, the iterative algorithms we consider are remarkably effective in practice.
For example, our analyses in Section~\ref{sec:ipm} is based on
the interior point method used in the Gurobi library~\footnote{\url{https://www.gurobi.com/wp-content/plugins/hd_documentations/documentation/9.0/refman.pdf}};
iterative refinement approaches for $p$-norm regression have shown promising results in practice~\cite{AdilPS19};
and iterative approaches for linear regression have resulted in theoretically and practically faster algorithms for tensor decomposition~\cite{fahrbachsubquadratic}.

\subsection{Results}
\label{subsec:results}

We start with some definitions related to the stability and bit complexity of algorithms. The condition number of a function $f:\R^{n}\rightarrow\R^m$ is defined as the smallest nonnegative real number $\kappa_f$ such that
\[
\frac{\norm{f(\vx+\delta \vx) - f(\vx)}}{\norm{f(\vx)}} = \kappa_f \cdot \frac{\norm{\delta \vx}}{\norm{\vx}} + O \left( \left( \frac{\norm{\delta \vx}}{\norm{\vx}} \right)^2 \right),
\]
where $\vx,\delta \vx \in \R^n$.
For the inverse of matrices, this coincides with the condition number of the matrix defined as $\kappa(\ma) := \norm{\ma}_2 \cdot \norm{\ma^{-1}}_2$.
It has been shown that a recursive algorithm based on fast matrix multiplication is logarithmically stable \cite{DemmelDH07} in the following sense.
\[
\frac{\norm{g(\ma) - \ma^{-1}}}{\norm{\ma^{-1}}} \leq O(\veps) \kappa(\ma)^{\poly \log(n)} + O(\veps^2),
\]
where $g(\ma)\in\R^{n\times n}$ is the output of the algorithm for the inverse. Taking $\veps<\epsilon \norm{\ma^{-1}} \kappa(\ma)^{\poly \log(n)}$, we can guarantee that $\norm{g(\ma) - \ma^{-1}} \leq \epsilon$.
This implies that by using $\poly \log(n) \cdot \log(\kappa) + \log(1/\epsilon)$ bits, we can achieve $\norm{g(\ma) - \ma^{-1}} \leq \epsilon$. Therefore we consider the following definitions of numerical stability that are equivalent up to a $\kappa(\ma)$ factor.

\begin{definition}[Numerical Stability of Computing the Inverse]
Let $\ma\in \R^{n\times n}$ be an invertible matrix. Consider an algorithm that computes the matrix $\mm\in \R^{n\times n}$ as the inverse of $\ma$. Then the algorithm is forward stable if $\norm{\mm - \ma^{-1}}\leq \veps$, and it is backward stable if $\norm{\mm^{-1} - \ma} \leq \veps$.
\end{definition}
As we will discuss later, forward stability suffices for solving linear regression. However, for inverse maintenance guarantees, we require backward stability.

Our first result bounds the bit complexity of solving a linear regression problem in input-sparsity time.  We later extend this to certain weighted constrained regression problems that are used as a subprocedure for the $p$-norm regression problem. We will use \emph{running time} for the total time counting bit-level operations and \emph{bit complexity} to refer to the size of representations. We also use ``with high probability''  to mean with probability at least $1 - n^{-C}$ for any constant $C$.

\begin{restatable}[Linear Regression]{theorem}{thmLinReg}
\label{thm:lin_reg}
Let $\ma\in\R^{n\times d}$ be a full-rank matrix with $n\geq d$ and a condition number bounded by $\kappa$, and $\vb\in\R^{n}$ all with bit complexity of $\log(\kappa)$. Let $\vxstar = \argmin_{\vx} \norm{\ma \vx - \vb}_2=(\ma^\top \ma)^{-1} \ma^\top \vb$ and $0<\epsilon<1$. Then there is an algorithm that, with high probability, computes $\vxhat$ such that 
\[
\norm{\vxhat - \vxstar}_{\ma^\top \ma} \leq \epsilon \cdot \norm{\ma (\ma^\top \ma)^{-1} \ma^\top \vb}_2,
\]
or equivalently
\[
\norm{\ma \vxhat - \vb}_2 \leq \epsilon \cdot \norm{\ma (\ma^\top \ma)^{-1} \ma^\top \vb}_2 + \norm{(\mi - \ma (\ma^\top \ma)^{-1} \ma^\top) \vb}_2,
\]
in time $\Otil((d^{\omega}+ d^2 \cdot \log^2 (1/\epsilon) + \nnz(\ma) \cdot \log^2 (1/\epsilon))\cdot \log \kappa)$.
\end{restatable}
The $\Otil$ in the above result and the rest of the paper hides $\poly \log(nd)$ factors and $\poly \log \log(\kappa/\epsilon)$ factors.
For simplicity, in all of our results, we assume the matrix $\ma$ has full column rank. However, this is not a limitation. As we show in \cref{sec:low-rank}, low-rank matrices can be modified to matrices with full column rank by concatenating a small factor of the identity matrix that introduces an small error.

We next consider the bit complexity of approximately solving linear programs.
We consider LPs of the following form:
\[
\min_{\ma^\top \vx = \vb, \vx\geq 0} \vc^\top \vx ~~~~~ \text{(primal)} ~~~~~   \text{and}  ~~~~~ \max_{\ma \vy\leq \vc} ~~ \vb^\top \vy ~~~~~ \text{(dual)}.
\]

\noindent
We start by defining a few parameters.
\begin{definition}
\label{def:radius}
Let $\ma\in\R^{n\times d}, \vb\in\R^d, \vc\in\R^n$ with $n\geq d$. For a linear program of the form $\min_{\ma^\top \vx = \vb, \vx\geq 0} \vc^\top \vx$, we define the following quantities:
\begin{itemize}
    \item Inner radius $r$: There exists an $\vx$ such that $\ma^\top \vx = \vb$ and $\vx_i\geq r\ge 0$ for all $i\in[n]$.
    
    \item Outer radius $R$: For all $\vx \geq 0$ with $\ma^\top \vx = \vb$, $\norm{\vx}_2\leq R$.

\end{itemize}
\end{definition}

\noindent
The next theorem states that the robust IPM \cite{DBLP:conf/stoc/CohenLS19,DBLP:conf/soda/Brand20} only requires numbers with $\Otil( \log(\frac{\kappa R}{\epsilon \cdot r}))$ bits in fixed-point arithmetic. We note that directly utilizing algorithms of \cite{DBLP:conf/stoc/CohenLS19,DBLP:conf/soda/Brand20}, does not imply the time complexity of the following result. First, the bit complexity of inverse maintenance has to be bounded (with proper rounding at update steps --- see the data structure in Algorithm \ref{alg:proj-ds}) and second, the modifications made to the problem to find an initial feasible solution, should be made in a way that ensures the condition number of the constraint matrix does not change significantly. We adopt the initialization approach of \cite{lee2021tutorial} and show that the condition number of the resulting matrix stays the same up to polynomial factors in $n$.

\begin{restatable}[Robust IPM]{theorem}{RobustIPM}
\label{thm:robust_ipm}
Given $\ma\in\R^{n\times d}$ with full column-rank and condition number $\kappa$, $\vb\in\R^{d}$, $\vc\in\R^{n}$ all with bit complexity of $\log(\kappa)$, and an error parameter $0<\epsilon<1$, suppose the inner radius and outer radius of the linear program $\min_{\ma^\top \vx = \vb, \vx\geq 0} \vc^\top \vx$ is $r$ and $R$, respectively. Then there is an algorithm that computes $\vxhat\in\R^{n}$ such that
\[
\vc^\top \vxhat \leq \min_{\ma^\top \vx = \vb, \vx\geq 0} \vc^\top \vx + \epsilon    ~~~ \text{, and } ~~~ \norm{\ma^\top \vxhat - \vb}_2 \leq \epsilon,
\]
in time $\Otil\left(\left(n^{\omega} + n^{2.5-\omegadual/2} + n^{2+1/6}\right) \cdot \log(\frac{\kappa R}{\epsilon \cdot r}) \cdot \log(\frac{R}{\epsilon \cdot r})\right)$.
\end{restatable}

We only assume that the bit complexity of $\ma,\vb,\vc$ is bounded by $\log(\kappa)$ for ease of notation. If the bit complexity of them is $\ell$, the first log factor will be replaced by $\ell + \log(\frac{\kappa R}{\epsilon \cdot r})$.
Note that our bit complexity depends on $\log(\kappa)$ as opposed to the bit complexity stated in \cite{DBLP:conf/stoc/CohenLS19}, which is the logarithm of the maximum determinant over the square submatrices. Note that although both quantities are $\Omega(n)$ in the worst case, for random matrices, the latter is $\Omega(n)$ while the former is $O(\log n)$. This is because, for random matrices, the condition number is polynomially bounded \cite{edelman1988eigenvalues,edelman1989eigenvalues} while the determinant is exponentially large \cite{tao2005random} with high probability. Moreover, $\log(R/r)$ has shown to be $O(\log n)$ in the smoothed analysis of LPs \cite{blum2002smoothed}. Finally, note that we are concerned with approximate solutions to LPs. An exact solution might require the bit complexity proportional to the logarithm of the maximum determinant of square submatrices. The exponent of the third term above is recently improved to $2+\frac{1}{18}$ by using more complicated data structures \cite{jiang2020fasterb}.

The above approach is not always the fastest algorithm for solving LPs approximately. The next result is based on solving linear systems using shifted number systems \cite{storjohann2005shifted}, which avoids the $\log(\kappa)$ factor. This approach does not use inverse maintenance techniques.

\begin{restatable}[Inverse-free IPM]{theorem}{thmInverseFreeIPM}
\label{thm:inverse_free_ipm}
Given $\ma\in\R^{n\times d}$ with full column-rank, $\vb\in\R^{d}$, $\vc\in\R^{n}$ all with bit complexity of $\ell$, and an error parameter $0<\epsilon<1$, suppose the inner radius and outer radius of the linear program $\min_{\ma^\top \vx = \vb, \vx\geq 0} \vc^\top \vx$ is $r$ and $R$, respectively. Then there is an algorithm that finds $\vxhat\in\R^{n}$ such that
\[
\vc^\top \vxhat \leq \min_{\ma^\top \vx = \vb, \vx\geq 0} \vc^\top \vx + \epsilon    ~~~ \text{, and } ~~~ \norm{\ma^\top \vxhat - \vb}_2 \leq \epsilon,
\]
in time $\Otil\left(n^{\omega+0.5} \cdot \left( \ell + \log(\frac{R}{\epsilon \cdot r}) \right) \cdot \log(\frac{R}{\epsilon \cdot r})\right)$.
\end{restatable}

The algorithm of \cref{thm:inverse_free_ipm} is faster than \cref{thm:robust_ipm} by a factor of $n^{0.5}$ when $\ell=O(1)$, $\log(\kappa)=\Omega(n)$, and $\log(R/r)=O(\log n)$. We discuss such a case in \cref{sec:lp-prelim}. This highlights the fact that when we consider the actual running time of algorithms, algorithms with smaller number of arithmetic operations do not necessarily have the smallest running time. Our final result for LPs is presented in \cref{thm:ell2_ipm_sparse} and shows one can go below matrix multiplication time for $\omega>2.5$ and sparse poly-conditioned matrices.

We next turn to $p$-norm minimization problems for $p \geq 2$. All of our results can be extended to the case of $p \in (1,2]$ by considering the dual norm using the approach explained in Section 7 of \cite{AdilKPS19}. Our first result bounds the bit complexity of solving the $p$-norm problem in both sparse and dense cases. Since the only difference between the two cases is the data structure we use, we present both of them in a single theorem.

\begin{restatable}[$p$-norm minimization]{theorem}{denseSparsePNorm}
\label{thm:main-p-norm}

Let $\ma\in \mathbb{R}^{n\times d}$ be a matrix with condition number bounded by $\kappa$, and $\vb\in\R^{d}$ be a vector with the bit complexity bounded by $\log(\kappa)$. Let $\vxstar=\argmin_{\ma^\top \vx=\vb} \norm{\vx}_p^p$. For $p \geq 2$, there is an algorithm that computes $\vxhat$ such that $\norm{\pi_\ma(\vxhat - \vxstar)}_2 \leq \epsilon \norm{\pi_{\ma}\vxstar}_2$, and 
\[
\norm{\vxhat}_p^p \leq (1+\epsilon) \norm{\vxstar}_p^p
\]
in time 
\[
\Otil_p \left( \left( n^{\omega} + n^{7/3} \cdot \log(1/\epsilon) \right) \log^{1.5}(\kappa / \epsilon) \log^2(1/\epsilon) \right).
\]
Moreover, for sparse matrices, there is an algorithm that returns an output with the same guarantees, with high probability, in time
\begin{align*}
\Otil_p\left(n^{7/3} \cdot  \left( 1 + \nnz(\ma)^{\frac{\omega-7/3}{\omega-1}} \right) \log^{2.5}(\kappa/\epsilon) \log^3(1/\epsilon)\right).
\end{align*}

\end{restatable}

The subscript $p$ hides a function $f(p)$.
For any value of $\omega>7/3$, $\nnz(\ma)=o(n^{\omega-1})$, and $\log(\kappa/\epsilon)=\poly(n)$, the above gives a running time $o_p(n^{\omega})$. For example, for polyconditioned matrices with $\nnz(\ma)=O(n)$ and the current value of $\omega \approx 2.372$, the running time is $\Otil_p(n^{2.363} \cdot \log^{5.5}(\epsilon^{-1}))$. Note that the powers of $\log(1/\epsilon)$ and $\log(\kappa/\epsilon)$ are different for $p$-norm and linear programming. This is because of the number of iterations of the algorithms arising from ``guessing'' the optimal values in subprocedures of our $p$-norm regression algorithm. Moreover note that for $p$-norm problem, we also modify the matrices by concatenating a (gradient) vector. We prove that this only affects the condition number of the matrix by a polynomial factor --- see \cref{subsec:solve-res-problem}. We also emphasize that we use a different approach than \cite{AdilKPS19} for solving the constrained weighted regression problems, that are subprocedures of the algorithm, to be sure that the numbers we work with only have $\log(\kappa/\epsilon)$ bits --- see \cref{subsec:weighted-lin-reg}. Note that taking the powers of $p$ of the numbers in the algorithm only increases the bit complexity by a factor of $p$ which is absorbed in the $\Otil_p$ notation.

Our approach for solving the $p$-norm minimization problems is to solve a series of \emph{smoothed} $p$-norm minimization problems (see \cref{subsec:res-problem,subsec:solve-res-problem}) to constant factor approximation. The smoothed $p$-norm problem, which we also refer to as mixed $(2,p)$-norm minimization, is defined as follows.

\begin{definition}[Smoothed $p$-norm minimization problem]
Let $\ma\in \mathbb{R}^{n\times d}$ and $\vb\in\R^{d}$, with $n\geq d$. For $p\geq 2$, let
\[
\vxstar = \argmin_{\vx\in\R^n:\ma^\top \vx = \vb}\sum_{i=1}^n \gamma_p(\vt_i, \vx_i),
\]
where for $t\in\R_{\geq 0}$ and $x\in\R$,
\begin{align*}
\gamma_p(t, x) := \begin{cases}
\frac{p}{2} t^{p-2} x^2 & \text{ if } |x|\leq t,\\
|x|^p + (\frac{p}{2} - 1) t^p & \text{ otherwise.}
\end{cases}
\end{align*}
Then the smoothed $p$-norm problem asks for $\vxhat$ such that $\norm{\pi_\ma(\vxhat - \vxstar)}_2 \leq \epsilon \cdot \norm{\pi_{\ma}\vxstar}_2$ and 
\[
\sum_{i=1}^n \gamma_p(\vt_i, \vxhat_i) \leq (1+\epsilon) \cdot \sum_{i=1}^n \gamma_p(\vt_i, \vxstar_i).
\]
\end{definition}

We show that the following mixed $(2,\infty)$-norm minimization problem can be used as a proxy for such smoothed $p$-norm problems, but this leads to larger running times for solving the $p$-norm minimization problem. However, since mixed $(2,\infty)$-norm minimization is an important problem in its own right, we present a multiplicative weights update algorithm for it as well.

\begin{definition}[Mixed $(2,\infty)$-norm minimization problem]
Let $\ma\in \mathbb{R}^{n\times d}$, $\vb\in\R^{d}$, $\vr,\vs\in\R^{n}_{\geq 0}$, with $n\geq d$. Let
\[
\vxstar = \argmin_{\vx\in\R^n:\ma^\top \vx = \vb} \norm{\vx}_{\vr}^2 + \norm{\vs \odot\vx}_{\infty},
\]
where $\odot$ is the entrywise (Hadamard) product, $\norm{\vx}_{\vr}^2 = \vx^\top \mr \vx$, and $\mr$ is the diagonal matrix corresponding to $\vr$.
Then the mixed $(2,\infty)$-norm minimization problem asks for $\vxhat$ such that $\norm{\pi_\ma(\vxhat - \vxstar)}_2 \leq \epsilon \cdot \norm{\pi_{\ma}\vxstar}_2$ and 
\[
\norm{\vxhat}_{\vr}^2 + \norm{\vs \odot\vxhat}_{\infty} \leq (1+\epsilon) \cdot \left( \norm{\vxstar}_{\vr}^2 + \norm{\vs \odot\vxstar}_{\infty} \right).
\]
\end{definition}

\noindent
We provide a constant factor approximation algorithm for this problem.

\begin{restatable}[Mixed $(2,\infty)$-norm minimization]{theorem}{denseSparseInftyNorm}
\label{thm:main-mixed-2-infty}
Let $\ma\in\R^{n\times d}$ and $\vb\in \R^{d}$, $\vr,\vs\in\R^{n}_{\geq 0}$, $n\geq d$, such that the condition number of $\ma$ is less than $\kappa$ and the bit complexity of all of them is bounded by $\log(\kappa)$.
For $0<\epsilon<1$, there is an algorithm that outputs $\vxhat$ such that $\norm{\pi_{\ma} (\vxhat - \vxstar)}_2 \leq \epsilon \cdot \norm{\pi_{\ma}\vxstar}_2$ and
\begin{align}
    \norm{\vxhat}_{\vr}^2 + \norm{\vs \odot\vxhat}_{\infty} =O(1) \cdot(\norm{\vxstar}_{\vr}^2 + \norm{\vs \odot\vxstar}_{\infty}),
\end{align}
where $\vxstar = \argmin_{\vx:\ma^\top \vx = \vb} \norm{\vx}_{\vr}^2 + \norm{\vs \odot\vx}_{\infty}$,
in time
\[
\Otil_p( (n^{\omega}+n^{7/3} \cdot  \log^2(1/\epsilon)) \log(\alpha_2\kappa/\epsilon) \log(\alpha_1 \kappa) \log(\kappa/\epsilon)),
\]
where $\alpha_1=1/(\min_{i\in[n]} \vr_i+\vs_i^2)$ and $\alpha_2 = (\max_{i\in[n]} \vr_i + \max_{i\in[n]} \vs_i)/\min_{i\in[n]} \vr_i$.
Moreover, for sparse matrices, there is an algorithm that returns an output with the same guarantees 
with probability at least $1-n^{-10} \cdot \log(\alpha_2)$
in time 
\[
\Otil\left(n^{7/3} \cdot  \left( 1 + \nnz(\ma)^{\frac{\omega-7/3}{\omega-1}} \right)\log^2(1/\epsilon) \log(\alpha_2\kappa/\epsilon) \log(\alpha_1 \kappa)\log^2(\kappa/\epsilon))\right).
\]

\end{restatable}

We finally note that, while the bit complexity of these problems is known for Laplacians and graph problems such as maximum flow, it was not known for general matrices prior to our work. The main reasons for this difference are the use of inverse maintenance techniques for general matrices and the difficulty of establishing bounds on the condition number of such matrices.

\subsection{Techniques}

Inverse maintenance is an important technique that has been used in optimization algorithms since Karmarkar \cite{Karmarkar84}. It has since been utilized in many other algorithms, such as iterative refinement for $p$-norm minimization and dynamic algorithms. The following identity, which is used for inverse maintenance, has been extensively used (without stability and bit complexity analysis) in optimization literature to speed up a variety of different iterative algorithms.

\begin{fact}[Sherman-Morrison-Woodbury identity \cite{woodbury1950inverting}]\label{fact:Woodbury}
For an invertible $n \times n$ matrix $\mm$ and matrices $\matu\in \R^{n \times r},\mc \in \R^{r \times r},\mv \in \R^{r \times n}$, we have
\[
(\mm+\matu \mc \mv)^{-1} = \mm^{-1} - \mm^{-1} \matu (\mc^{-1} + \mv \mm^{-1} \matu)^{-1} \mv \mm^{-1}.
\]
\end{fact}

Since the exact inverse of a matrix cannot necessarily be represented with a finite number of bits in fixed-point arithmetic, we only can use approximate inverses. Then the question is how many bits are required to maintain a small error when we apply the Sherman-Morrison-Woodbury identity in order to guarantee the convergence of our iterative algorithms? Note that the required error for inverses determines the bit complexity of them.

Our first main technique is to show that inverse maintenance via the Sherman-Morrison-Woodbury identity is backward stable. We need to present our numbers with $\Otil(\log(\kappa/\epsilon))$ bits to have this guarantee.
The following lemma states that after applying the Woodbury identity, the backward error only increases additively in each iteration. Therefore, if we apply this method for $\text{poly}(n)$ iterations, the error only increase by $\text{poly}(n,\kappa)\cdot \veps$. Therefore by picking $\veps$ to be appropriately small, we can guarantee that the inverse has small error over the course of an algorithm with $\text{poly}(n)$ iterations, such as interior point methods~\cite{DBLP:conf/stoc/CohenLS19,DBLP:conf/soda/Brand20} and multiplicative weights update methods~\cite{AdilKPS19}.

\begin{restatable}[Backward Stability of Inverse Maintenance]{lemma}{InvMainAsymm}
\label{lemma:smw-stability-asymm-fro}
Let $\mz \in \R^{n\times n}, \mztil \in \R^{n\times n}, \mc \in \R^{m\times m}$ be invertible matrices. Moreover let $\matu,\mv\in\R^{n \times m}$ such that $\mz+\matu \mc \mv^\top$ is invertible. Let $\kappa > m+n$ such that
\[
\norm{\matu}_{\fro},\norm{\mv}_{\fro},\norm{\mc}_{\fro},\norm{\mc^{-1}}_{\fro},\norm{\mz}_{\fro},\norm{\mz^{-1}}_{\fro},\norm{\mz+\matu \mc \mv^\top}_{\fro}, \norm{(\mz+\matu \mc \mv^\top)^{-1}}_{\fro} \leq \kappa 
\]
and $0 < \veps_1,\veps_2 < 1$. Suppose 
\begin{align}
\label{eq:smw-assumption}
\norm{\mztil -\mz}_{\fro} \leq \veps_1.
\end{align}
If $\md\in\R^{m\times m}$ is an invertible matrix such that 
\[
\norm{ \md^{-1} - (\mc^{-1} + \mv^\top \mztil^{-1} \matu)^{-1}}_{\fro} \leq \veps_2,
\]
then
\[
\norm{(\mztil^{-1} - \mztil^{-1} \matu \md^{-1} \mv^\top \mztil^{-1})^{-1} - (\mz + \matu \mc \mv^\top)}_{\fro} \leq 512\kappa^{26} \veps_2 +\veps_1.
\]
\end{restatable}

In addition to inverse maintenance, for the $p$-norm minimization problem we need high-accuracy solutions given a constant factor spectral approximation as the preconditioner. We note that even in the cases where we only solve one static linear regression problem (as opposed to a series of dynamically changing linear regression problems like algorithms for $p$-norm regression), one might need to use an iterative approach based on preconditioning instead of a direct solve to obtain a high-accuracy solution in certain running times. An example of this is high-accuracy input-sparsity time algorithms for solving linear regression problems \cite{clarkson2014sketching,CohenLMMPS15}.
Another example is illustrated by \cite{fahrbachsubquadratic} in the context of tensor decompositions in which the algorithm requires a preconditioning approach to achieve a speed-up to subquadratic time.

\begin{restatable}[High-accuracy solutions for constrained weighted linear regression]{lemma}{constrainedRichardson}
\label{lemma:constrained-richardson}
Let $\ma\in\R^{n\times d}$ have full column rank, $\vb\in\R^d$, and $\mw\in\R^{n\times n}$ be a diagonal matrix with $R \mi\succeq \mw\succeq \mi$. Moreover let $\vxstar = \argmin_{\vx: \ma^\top \vx = \vb} \frac{1}{2} \norm{\vx}_{\mw}^2$. Then
\[
\vxstar = \mw^{-1} \ma (\ma^\top \mw^{-1} \ma)^{-1} \vb.
\]
Moreover given a matrix $\mmtil^{-1}$ such that there exists matrix $\mm$ with $\norm{\mmtil^{-1} - \mm^{-1}}_{\fro} \leq \frac{\veps}{d \cdot \lambda\cdot \norm{\ma^\top \mw^{-1} \ma}_2}$ and $\ma^\top \mw^{-1} \ma \preceq \mm \preceq \lambda \ma^\top \mw^{-1} \ma$ with a constant $\lambda\geq 1$, there is an algorithm that finds $\vxhat$ such that
\[
\norm{\vxhat - \vxstar}_2 \leq ~ \epsilon \cdot \norm{\vxstar}_{2} , ~~
\norm{\vxhat}_{\mw} \leq (1+\epsilon) \norm{\vxstar}_{\mw} ~,  ~~ \text{and } ~~  \norm{\pi_{\ma}(\vxhat - \vxstar)}_{2} \leq \epsilon \norm{\pi_{\ma}\vxstar}_{2},
\]
where $\pi_{\ma}$ is the projection matrix of matrix $\ma$, in $O((d^2+\nnz(\ma))\cdot \log(\kappa(\ma)\cdot R) \cdot \log^2(\frac{R}{\epsilon}))$ time.
\end{restatable}

Note that since $\norm{\mm^{-1}} \leq \poly(n \kappa/R)$, we can take $\mmtil^{-1}$ to be a matrix with $O(\log(n R \kappa/\epsilon))$ bit complexity to satisfy the condition $\norm{\mmtil^{-1} - \mm^{-1}}_{\fro} \leq \frac{\veps}{d \cdot \lambda\cdot \norm{\ma^\top \mw^{-1} \ma}_2}$. A complication in \cref{lemma:constrained-richardson} is that we require a vector $\vxhat$ that is close to $\vxstar$ in two different norms: one induced by $\mw$ and the other induced by $\pi_{\ma}$. Interestingly, as we show, one does not need to take $\log(\kappa)$ iterations to achieve this.

\subsection{Discussion}
\label{sec:discussion}

Although the running times of optimization algorithms in terms of number of arithmetic operations have been extensively studied in the past decades, in many recent works, the bit complexity is left unanalyzed. \cite{DBLP:conf/stoc/CohenLS19} and \cite{AdilKPS19} present algorithms solving linear programs and $p$-norm minimization problems respectively with running times that match the matrix multiplication time $n^{\omega}$ up to polylogarithmic factors. However even solving one linear system under fixed-point arithmetic, by computing the inverse and applying it to the vector, requires bit complexity of $\Omega(\log(\kappa/\epsilon))$ even if the bit complexity of the original linear system is $O(1)$. This is exemplified by the following matrix that has a condition number of larger than $2^{n-1}$, by testing the vectors 
$\begin{bmatrix}
1 & 0 & 0 & \cdots & 0
\end{bmatrix}$ and $\begin{bmatrix}
(-1/2)^{n-1} & (-1/2)^{n-2} & \cdots & -1/2 & 1
\end{bmatrix}$ for the largest and smallest singular value, respectively.
\begin{align*}
\begin{bmatrix}
1 & 0 & 0 & 0 & \cdots & 0 & 0 \\
2 & 1 & 0 & 0 & \cdots & 0 & 0 \\
0 & 2 & 1 & 0 & \cdots & 0 & 0 \\
0 & 0 & 2 & 1 & \cdots & 0 & 0 \\
\vdots & \vdots & \vdots & \vdots & \ddots & \vdots & \vdots \\
0 & 0 & 0 & 0 & \cdots & 1 & 0 \\
0 & 0 & 0 & 0 & \cdots & 2 & 1 \\
\end{bmatrix} \in \R^{d\times d}.
\end{align*}

It is not a priori clear what bit complexity is required to guarantee convergence when we need to solve a series of dynamically changing linear systems as required by iterative approaches for solving $p$-norm minimization and LPs. Although the forward stability of the inverse maintenance processes has been considered \cite{yip1986note}, such bounds are not enough for algorithms that need $\poly(n)$ iterations.

Note that as illustrated by the above example, an algorithm with $O(n^{\omega})$ arithmetic operations and bit complexity of $\log(\kappa)$, in the worst case, has a running time of $O(n^{\omega+1})$. Therefore it is crucial to determine the right values for the power of $\log(\kappa)$ factor. Additionally as illustrated by \cref{thm:inverse_free_ipm}, an algorithm with smaller number of arithmetic operations does not necessarily have the best overall running time.

\paragraph{Outline.} 
We start by presenting our result on input-sparsity time linear regression in \cref{sec:lin_reg}.
We then present our numerically stable inverse maintenance in \cref{sec:inv-main}. 
Our data structures that use this inverse maintenance procedure for dense and sparse matrices are presented in \cref{sec:ds-bit}.
Equipped with these, we present our results on solving linear programs in \cref{sec:ipm}. We first discuss our overall algorithm and how to find the initial feasible solution in \cref{sec:lp-prelim}. We then present our LP solvers that uses robust IPM in \cref{sec:robust_ipm}, our LP solver based on shifted numbers in \cref{sec:inverse_free_ipm}, and our results for sparse LP and $\omega>2.5$ in \cref{sec:sparse_ipm}.

We introduce the outer loop of our algorithm for solving the $p$-norm minimization problem that uses a series of solutions to residual problems in \cref{subsec:res-problem}. Then in \cref{subsec:solve-res-problem}, we discuss how the residual problem can be solved effectively by solving instances of smoothed $p$-norm minimization problems and how the mixed $(2,\infty)$-norm minimization can be used as a proxy.
We then present our multiplicative weights update (MWU) algorithm to solve a mixed $(2,\infty)$-norm minimization problem in \cref{subsec:two-inf-norm}. Finally, in \cref{sec:two-p-norm}, we present our MWU algorithm for solving the smoothed $p$-norm minimization problem.

\subsection{Notation and Preliminaries}
\label{sec:notation_and_prelim}

\textbf{Linear algebra notations.}
We denote the Hadamard (entrywise product) with $\odot$. For a vector $\vx$, let $\abs{\vx}$ be a vector of same size such that $(\abs{\vx})_i = \abs{\vx_i}$ for all $i$; and $\vx^p$ denotes the vector with its $i$th entry equal to the $i$th entry of $\vx$ to the power of $p$, i.e., $(\vx^p)_i=(\vx_i)^p$. Similarly, for a diagonal matrix (or vector) $\mm$, $\sqrt{\mm}$ is a matrix where each entry is equal to the square root of the corresponding entry in $\mm$. 
For a matrix $\ma$ with $n$ rows and a subset $S\subseteq[n]$, let $\ma_S$ denote the matrix obtained by taking rows of $\ma$ with indices in $S$. $\ma_{:S}$ denote the matrix obtained by taking the columns of $\ma$ with indices in $S$. For a square matrix $\ma\in\R^{n\times n}$ and $S\subseteq [n]$, $\ma_{S,S}$ denotes the matrix obtained by taking entries of $\ma$ in $S\times S$. Note that, we apply these subindices before taking transpose, i.e., $\ma_S^\top = (\ma_S)^\top$. We denote the Moore-Penrose inverse (i.e., pseudoinverse) of $\ma$ with $\ma^{\dagger}$.

$\norm{\cdot}_{\fro}$ denotes the Frobenius norm. We denote the entrywise norm of matrices by $\vertiii{\cdot}$, e.g., $\vertiii{\ma}_{\infty}$ is the maximum magnitude over entries of $\ma$. For a matrix $\ma$, we denote its condition number by $\kappa(\ma):= \norm{\ma}_2 \norm{\ma^{\dagger}}_2$. In other words, the condition number of a matrix is its largest singular value divided by its smallest \emph{nonzero} singular value. We denote the orthogonal projection matrix of $\ma$ with $\pi_{\ma}$. In particular, if $\ma$ has full column rank, $\pi_{\ma}=\ma(\ma^\top \ma)^{-1} \ma^\top$.  Throughout the paper, to make the notation less cumbersome, we assume the bit complexity of the vector $\vb$ and matrix $\ma$ are at most $\log(\kappa)$. This means that the absolute value of each entry of $\vb$ and $\ma$ is either zero or in the interval $[\frac{1}{\kappa}, \kappa]$. This is without loss of generality since the factors of the running time depending on $\kappa$ can be replaced with $\log((\kappa\cdot 2^{\ell})/\epsilon)$, where $\ell$ is the bit complexity of the input.

When it is clear from the context, we denote the diagonal matrix corresponding to a vector with the capital letter of the vector, e.g., $\mw$ denotes the diagonal matrix corresponding to $\vw$. Also for $\vu,\vw\in\R^{n}$, we define
\[
\norm{\vu}_{\vw} = \norm{\vu}_{\mw} := \sqrt{\vu^\top \mw \vu},
\]
for $\vw\geq 0$. More generally for a symmetric positive semi-definite matrix $\mm$ we denote $\norm{\vu}_{\mm} = \sqrt{\vu^\top \mm \vu}$.
For vectors $\vu_1\in\R^{n_1},\ldots,\vu_{k}\in\R^{n_k}$, we denote by $(\vu_1,\ldots,\vu_k)\in\R^{n_1+\cdots+n_k}$, their concatenation. Note that $n_i$ could be equal to one, in which case $\vu_i$ is a number. For a number $t\in\R$, we denote the vector with all entries equal to $t$ with $\vec{t}$. The dimension of the vector will be clear from the context, e.g., if $\ma\in\R^{n\times d}$, $\begin{bmatrix}
\ma & \vec{0}
\end{bmatrix}$ denotes a matrix obtained by attaching a column of all zeros to the matrix $\ma$. For a function $f:\R\rightarrow \R$ and vector $\vu\in\R^n$, we define $(f(\vu))_i = f(\vu_i)$, i.e., we extend $f$ to $f:\R^n\rightarrow\R^n$.

\paragraph{Matrix multiplication.}
We denote the matrix multiplication exponent and its dual with $\omega$ and $\omegadual$, respectively. Moreover, we denote the cost of multiplying an $n$-by-$m$ matrix with an $m$-by-$d$ matrix with $\mtime(n,m,d)$, e.g., $\mtime(n,n,n)=n^{\omega}$, and $\mtime(n,n,n^{\omegadual})=n^{2+o(1)}$. We need the following lemma to bound the running time of rectangular matrix multiplication (for inverse maintenance) in our running time.
\begin{lemma}[\cite{gall2018improved,DBLP:conf/stoc/CohenLS19}]
\label{lemma:rectangular-matrix-mult}
Let $n\geq d$. Then multiplication of an $n\times d$ matrix with a $d\times n$ matrix or an $n\times n$ matrix with an $n\times d$ matrix can be performed in the following running time.
\[
n^{2+o(1)}+d^{(\omega-2)/(1-\alpha)} n^{2-\alpha\cdot (\omega-2)/(1-\alpha) + o(1)}.
\]
\end{lemma}
\noindent
For ease of notation, we drop $o(1)$ in the running time of matrix multiplication throughout the paper.

\paragraph{General assumptions.} We now state a few preliminary results to establish our assumptions in this paper.

\begin{remark}
\label{remark:polykappa-projection}
Let 
\[
\vxstar = \argmin_{\vx:\ma^\top \vx = \vb} \norm{\vx}_p^p.
\]
For $p\geq 2$, if $\norm{\vx^*}_p^p \leq \veps$, then $\norm{\vx^*}_2^p \leq d^{(p-2)/2} \veps$. Therefore without loss of generality, we can assume $\norm{\vx^*}_p^p > 1/\text{poly}(\kappa)$, since otherwise $\vx=\vec{0}$ will have a small error both in terms of the $p$-norm objective and in terms of $\norm{\ma^\top \vx - \vb}_2$.
\end{remark}

\noindent
The next lemma (proven in the appendix) states that we only need to focus on full column rank matrices.

\begin{restatable}{lemma}{LowRankLemma}
\label{lemma:low-rank}
Let $p\geq 2$, $\ma\in\R^{n\times d}$, $\vb\in\R^{d}$, $n\geq d$, such that the smallest nonzero singular value of $\ma$ is equal to $\sigma>0$. Moreover let $0<\veps_1<1$ and  $\veps_2=\veps_1 \cdot \frac{\sigma}{2\cdot d^{(p-2)/2p}}$. Let
\[
\mabar = \begin{bmatrix}
\ma \\ \veps_2\mi
\end{bmatrix}.
\]
Moreover let $0<\veps_3<1$, $\vxstar\in\R^{n}$ and $\vxhat\in\R^{n+d}$ such that
\[
\vxstar = \argmin_{\vx:\ma^\top \vx = \vb} \norm{\vx}_p^p ~~ \text{, }~~ \norm{\vxhat}_p^p \leq (1+\veps_3) \min_{\vx:\mabar^\top \vx = \vb} \norm{\vx}_p^p ~~ \text{, and } ~~ \norm{\mabar^\top \vxhat - \vb}_2 \leq \veps_3.
\]
Let $\vxtil\in\R^{d}$ be a vector with entries equal to the first $n$ entries of $\vxhat$. Then
\[
\norm{\ma^\top \vxtil - \vb}_2 \leq \veps_3 + \veps_1 \cdot \norm{\vb}_2  ~~ \text{, and }~~ \norm{\vxtil}_p^p \leq  (1+\veps_3)\norm{\vxstar}_p^p.
\]
\end{restatable}

\begin{remark} 
$\mabar$ in \cref{lemma:low-rank} has full column rank. Moreover, to achieve an error of $\epsilon$, we can pick $\veps_3 = \epsilon/2$ and $\veps_1 = \epsilon/(2\norm{\vb}_2)$. Therefore $\kappa(\mabar)\leq 4\cdot d^{(p-2)/2p}\kappa(\ma)\norm{\vb}_2$. Thus solving the problem with $\mabar$ only needs a polylogarithmic factor increase in bit complexity. Also since $n+d\leq 2n$, the polynomial factors in $n$ of running time only increase by constant factors. Therefore for the rest of the paper, without loss of generality, we assume the matrix $\ma$ has full column rank.
\end{remark}

\begin{remark}
Since $\ma$ has full column rank if $n=d$, then $\ma^\top \vx = \vb$ has a unique solution $\vxstar$, and we can compute a vector $\vxhat$ (by solving $\ma^\top \vx = \vb$) that is close to $\vxstar$ and with appropriate accuracy and bit complexity, we can guarantee both $\norm{\ma^\top \vxhat - b} \leq \epsilon$ and $\norm{\vxhat}_p^p \leq (1+\epsilon) \norm{\vxstar}_p^p$. Therefore for the rest of the paper, we assume $n>d$.
\end{remark}

\paragraph{Inverse maintenance.} The following directly follows from \cref{fact:Woodbury} and is one of the main tools for the robust IPM \cite{DBLP:conf/stoc/CohenLS19,DBLP:conf/soda/Brand20} to obtain a solution with about $n^{\omega}$ arithmetic operations.

\begin{corollary}
\label{cor:woodbury-proj-maintenance}
Let $\matp=\ma (\ma^\top (\mv+\mq) \ma)^{-1} \ma^\top$ where $\mv$ is a diagonal matrix, and $\mq$ be a sparse diagonal matrix with $T=\text{supp}(\mq)$. Then,
\[
\ma (\ma^\top (\mv + \mq) \ma)^{-1} \ma^\top = \matp - \matp_{:T} (\mq_{T,T}^{-1} + \matp_{T,T})^{-1} \matp_{:T}^\top.
\]
\end{corollary}
\begin{proof}
We have $\ma^\top \mq \ma = \ma_T^\top \mq_{T,T} \ma_T$ because the only nonzero entries of $\mq$ are the ones with indices in $T$. 
By \cref{fact:Woodbury}, we have
\begin{align*}
\ma (\ma^\top (\mv + \mq) \ma)^{-1} \ma^\top & = \ma (\ma^\top \mv \ma + \ma^\top \mq \ma)^{-1} \ma^\top \\ & =
\ma (\ma^\top \mv \ma + \ma_T^\top \mq_{T,T} \ma_T)^{-1} \ma^\top \\ & =
\ma \left( (\ma^\top \mv \ma)^{-1} - (\ma^\top \mv \ma)^{-1} \ma^\top_T (\mq_{T,T}^{-1} + \ma_T (\ma^\top \mv \ma)^{-1} \ma_T^\top)^{-1} \ma_T (\ma^\top \mv \ma)^{-1} \right) \ma^\top.
\end{align*}
The result follows by observing that
$\ma_T (\ma^\top \mv \ma)^{-1} \ma_T^\top = \matp_{T,T}$ and $\ma (\ma^\top \mv \ma)^{-1} \ma^\top_T =\matp_{:T}$.
\end{proof}

\section{Linear Regression}
\label{sec:lin_reg}

For a linear regression problem $\min_{\vx} \norm{\ma \vx - \vb}_2$ with $\vxstar = \argmin_{\vx} \norm{\ma \vx - \vb}_2$, we might want to make $\norm{\vx - \vxstar}_2$ or $\norm{\ma (\vx - \vxstar)}_2 = \norm{\vx - \vxstar}_{\ma^\top \ma}$ small. These are different for several reasons. For example, $\vxstar$ might not be unique (if $\ma$ is not full-rank), but $\ma \vxstar$ is unique. Even in the case where $\vxstar$ is unique $\norm{\vx - \vxstar}$ might be large while $\norm{\vx - \vxstar}_{\ma^\top \ma}$ is small, e.g., when $\vx - \vxstar$ is in the direction of the right singular vector of $\ma$ corresponding to the smallest singular value of $\ma$. However note that $\sigma_{\min}^2 \mi \preceq \ma^\top \ma \preceq \sigma_{\max}^2 \mi$, where $\sigma_{\min}$ and $\sigma_{\max}$ are the smallest and largest singular values of $\ma$. Therefore assuming $\sigma_{\min} \leq 1 \leq \sigma_{\max}$ (which can be achieved by scaling), $\norm{\vx - \vxstar}_2$ is within a $\kappa(\ma)^2$ factor of $\norm{\vx - \vxstar}_{\ma^\top \ma}$. 

In many applications, the goal is to bound $\norm{\vx - \vxstar}_{\ma^\top \ma}$ directly. For example, see \cref{cor:lp-projection-regression} which is a vector that is computed in each iteration of interior-point methods for solving linear programs. This then can be achieved by making sure $\norm{\vx - \vxstar}_2 \leq \frac{\epsilon}{\kappa(\ma)}$ or directly bounding $\norm{\vx - \vxstar}_{\ma^\top \ma}$. Even irrespective of bit complexity, the former might need $\log(\kappa/\epsilon)$ iterations. In this section, we show that an iterative approach can achieve the latter in $O(\log(1/\epsilon))$ iterations with an error-bounded precondition, avoiding the $\log(\kappa)$ factor in the number of iterations. We then use our approach to bound the bit complexity of solving a linear regression problem in input-sparsity time to high accuracy by using oblivious sketching approaches (see \cite{clarkson2014sketching,NelsonN13}) that find a spectral approximation of the matrix.

\begin{remark}
Note that a bound on $\norm{\vx^{(k)} - \vxstar}_{\ma^\top \ma}$ does not imply a multiplicative error bound on $\norm{\ma \vx - \vb}_2$. It only gives an additive error bound of the following form.
\[
\norm{\ma \vx - \vb}_2 \leq \norm{\ma (\vx - \vxstar)}_2 + \norm{\ma \vxstar - \vb}_2 = \norm{\vx - \vxstar}_{\ma^\top \ma} + \norm{(\mi - \ma (\ma^\top \ma) \ma^\top) \vb}_2.
\]
The second term of the right-hand side might be zero, in which case we have $\min_{\vx} \norm{\ma \vx - \vb}_2 = 0$. However, it is not necessarily possible to achieve a zero error even if the optimal solution has zero error (at least not with numbers represented in fixed-point arithmetic).
\end{remark}

\begin{lemma}[Bit complexity of Richardson's iteration]
\label{lemma:richardson}
Let $\ma\in\R^{n\times d}$ be a full-rank matrix, $n\geq d$.
Let $\lambda\geq 1$, 
 and $\mm,\mmtil\in\R^{d\times d}$ be symmetric matrices such that $\ma^\top \ma \preceq \mm \preceq \lambda \cdot \ma^\top \ma$ and $\norm{\mmtil^{-1} - \mm^{-1}}_{\fro} \leq \frac{\veps}{d \cdot \lambda\cdot \norm{\ma^\top \ma}_2}$. Let $\vx^{(k+1)}=\vx^{(k)} - \mmtil^{-1} (\ma^\top \ma \vx^{(k)}- \ma^\top \vb)$. Then we have
\[
\norm{\vx^{(k)}-\vxstar}_{\mm}
\leq
\left(1-\frac{1}{\lambda} + \veps\right)^k
\norm{\vx^{(0)} - \vxstar}_{\mm},
\]
where $\vxstar=\argmin_{\vx} \norm{\ma \vx-\vb}_2$.
\end{lemma}
\begin{proof}
We have $\vxstar = (\ma^\top \ma)^{-1} \ma^\top \vb$, and
\begin{align*}
\vx^{(k+1)} - \vxstar
& = 
\vx^{(k)} - \mmtil^{-1} (\ma^\top \ma \vx^{(k)}- \ma^\top \vb) - \vxstar
\\ & = 
\vx^{(k)} - \mmtil^{-1} (\ma^\top \ma \vx^{(k)}- \ma^\top \ma \vxstar) - \vxstar
\\ & = 
(\mi - \mmtil^{-1} \ma^\top \ma) (\vx^{(k)} - \vxstar)
\end{align*}
Therefore
\begin{align*}
\norm{\vx^{(k+1)} - \vxstar}_{\mm}
& = 
\norm{(\mi - \mmtil^{-1} \ma^\top \ma) (\vx^{(k)} - \vxstar)}_{\mm}
\\ & = 
\norm{(\mi - \mm^{-1} \ma^\top \ma + \mm^{-1} \ma^\top \ma - \mmtil^{-1} \ma^\top \ma) (\vx^{(k)} - \vxstar)}_{\mm}
\\ & \leq 
\norm{(\mi - \mm^{-1} \ma^\top \ma) (\vx^{(k)} - \vxstar)}_{\mm} + \norm{(\mm^{-1} \ma^\top \ma - \mmtil^{-1} \ma^\top \ma) (\vx^{(k)} - \vxstar)}_{\mm}
\end{align*}
Now we have 
\begin{align*}
\norm{(\mi - \mm^{-1} \ma^\top \ma) (\vx^{(k)} - \vxstar)}_{\mm}^2
& =
(\vx^{(k)} - \vxstar)^\top (\mi -  \ma^\top \ma \mm^{-1}) \mm (\mi - \mm^{-1} \ma^\top \ma) (\vx^{(k)} - \vxstar)
\end{align*}
Defining $\mh = \mm^{-1/2} \ma^\top \ma \mm^{-1/2}$, we have
\[
\mm^{1/2} (\mi - \mh)^2 \mm^{1/2} = (\mi -  \ma^\top \ma \mm^{-1}) \mm (\mi - \mm^{-1} \ma^\top \ma).
\]
Moreover, 
\[
\frac{1}{\lambda} \mi = \frac{1}{\lambda} \mm^{-1/2} \mm \mm^{-1/2} \preceq \mh \preceq \mm^{-1/2} \mm \mm^{-1/2} = \mi.
\]
Therefore $0\preceq \mi - \mh \preceq (1-\frac{1}{\lambda}) \mi$, which implies
\[
0 \preceq (\mi - \ma^\top \ma \mm^{-1}) \mm (\mi - \mm^{-1} \ma^\top \ma) \preceq (1-\frac{1}{\lambda})^2 \mm.
\]
Hence
\[
\norm{(\mi - \mm^{-1} \ma^\top \ma) (\vx^{(k)} - \vxstar)}_{\mm}^2 \leq  (1-\frac{1}{\lambda})^2 \norm{(\vx^{(k)} - \vxstar)}_{\mm}^2.
\]
Now we have
\[
\norm{(\mm^{-1} \ma^\top \ma - \mmtil^{-1} \ma^\top \ma) (\vx^{(k)} - \vxstar)}_{\mm}^2 = (\vx^{(k)} - \vxstar)^\top \ma^\top \ma (\mm^{-1} - \mmtil^{-1}) \mm (\mm^{-1} - \mmtil^{-1})\ma^\top \ma (\vx^{(k)} - \vxstar).
\]
Defining $\mg = \mm^{1/2} \mmtil^{-1} \mm^{1/2}$, we have
\[
(\mm^{-1} - \mmtil^{-1}) \mm (\mm^{-1} - \mmtil^{-1}) = \mm^{-1/2} (\mi - \mg)^2 \mm^{-1/2}
\]
Now note that 
\begin{align*}
\mi - \mg & = \mi - \mm^{1/2}(\mmtil^{-1} - \mm^{-1} + \mm^{-1}) \mm^{1/2}
\\ & =
\mi - \mm^{1/2}(\mmtil^{-1} - \mm^{-1}) \mm^{1/2} - \mm^{1/2} \mm^{-1} \mm^{1/2}
\\ & =
- \mm^{1/2}(\mmtil^{-1} - \mm^{-1}) \mm^{1/2}
\end{align*}
Therefore we have
\[
\norm{\mi - \mg}_2 \leq 
\norm{\mi - \mg}_F \leq \norm{\mm^{1/2}}_F^2 \norm{\mmtil^{-1} - \mm^{-1}}_F = \tr(\mm) \norm{\mmtil^{-1} - \mm^{-1}}_F \leq \frac{d\cdot \lambda \cdot \norm{\ma^\top \ma}_2 \cdot \veps}{d\cdot \lambda \cdot\norm{\ma^\top \ma}_2} = \veps
\]
Therefore we have
\begin{align*}
\norm{(\mm^{-1} \ma^\top \ma - \mmtil^{-1} \ma^\top \ma) (\vx^{(k)} - \vxstar)}_{\mm}^2 
& =
(\vx^{(k)} - \vxstar)^\top \ma^\top \ma \mm^{-1/2} (\mi - \mg)^2 \mm^{-1/2}\ma^\top \ma (\vx^{(k)} - \vxstar)
\\ & =
\norm{(\mi - \mg) \mm^{-1/2}\ma^\top \ma (\vx^{(k)} - \vxstar)}_2^2
\\ & \leq 
\veps^2 \cdot \norm{ \mm^{-1/2}\ma^\top \ma (\vx^{(k)} - \vxstar)}_2^2
\\ & = 
\veps^2 \cdot (\vx^{(k)} - \vxstar)^\top \ma^\top \ma \mm^{-1}\ma^\top \ma (\vx^{(k)} - \vxstar)
\\ & \leq
\veps^2 \cdot (\vx^{(k)} - \vxstar)^\top \ma^\top \ma (\vx^{(k)} - \vxstar)
\\ & \leq
\veps^2 \cdot (\vx^{(k)} - \vxstar)^\top \mm (\vx^{(k)} - \vxstar)
\\ & \leq 
\veps^2 \cdot \norm{\vx^{(k)} - \vxstar}_{\mm}^2
\end{align*}
Therefore combining the above, we have
\[
\norm{\vx^{(k+1)} - \vxstar}_{\mm} \leq \left(1 - \frac{1}{\lambda} + \veps \right) \norm{\vx^{(k)} - \vxstar}_{\mm}.
\]
\end{proof}

\begin{remark}
\label{remark:have-mtil}
To guarantee that $\norm{\mmtil^{-1} - \mm^{-1}}_{\fro} \leq \frac{\veps}{d \cdot \lambda\cdot \norm{\ma^\top \ma}_2}$, we require to use $\log(\frac{d^3\cdot \lambda \cdot \norm{\ma^\top \ma}_2 \norm{\mm^{-1}}_2}{\veps})$ bits.
Since $\ma^\top \ma\preceq \mm$, we have $(\ma^\top \ma)^{-1} \succeq \mm^{-1}$, which implies the about bound only requires $\poly\log(d\lambda)\log(\frac{\kappa}{\epsilon})$ bit complexity. 
If we use $\mm = \ma^\top \ma$, i.e., $\lambda = 1$, then picking $\veps = \frac{\epsilon}{\norm{\vb}_2}$ and $\vx^{(0)} = \vec{0}$, we have
\[
\norm{\vx^{(1)} - \vxstar}_{\ma^\top \ma} \leq \epsilon,
\]
since $\norm{\vxstar}_{\ma^\top \ma} = \norm{\ma(\ma^\top \ma)^{-1} \ma^\top \vb}_2 \leq \norm{\vb}_2$. This would give a bit complexity of $\Otil(\log(\frac{\kappa \cdot\norm{\vb}_2}{\epsilon}))$ for the numbers we require for our computation. So assuming access to matrix $\mmtil$, we can find $\vxhat$ such that
\[
\norm{\ma \vxhat - \vb}_2 \leq \epsilon + \min_{\vx} \norm{\ma \vx - \vb}_2,
\]
in time $\Otil(\left(\nnz(\ma) + d^2 \right) \cdot \log(\frac{\kappa \cdot \norm{\vb}_2}{\epsilon}))$.
\end{remark}

\begin{remark}
\label{remark:richardson-bit-complexity}
If we use a spectral approximation, then in each iteration, we can only guarantee a decrease of $(1-\lambda^{-1})$ factor in the error, and we have to perform $\log(1/\epsilon)$ iterations of the Richardson's iteration, to achieve the desired error. Therefore because in each iteration, the bit complexity of the solution vector is additively increased by $O(\log\kappa)$, the bit complexity of the $k$'th iteration is $O(k \cdot\log\kappa)$. This implies a running time of 
\[
\Otil((\nnz(\ma) + d^2)\cdot \log \kappa \cdot \log^2 \frac{1}{\epsilon}).
\] 
\end{remark}

The following bounds the occurring error in computations we perform in each iteration of iterative algorithms for solving linear programs.

\begin{corollary}
\label{cor:lp-projection-regression}
Let $\ma\in\R^{n\times d},\mw\in\R^{n\times n}$ be full-rank matrices and $\vu = \sqrt{\mw} \ma (\ma^\top \mw \ma)^{-1} \ma^\top \sqrt{\mw} \vg$. Moreover let $\lambda\geq 1$, 
 and $\mm,\mmtil\in\R^{d\times d}$ be symmetric matrices such that $\ma^\top \mw \ma \preceq \mm \preceq \lambda \cdot \ma^\top \mw \ma$ and $\norm{\mmtil^{-1} - \mm^{-1}}_{\fro} \leq \frac{\veps}{d \cdot \lambda\cdot \norm{\ma^\top \mw \ma}_2}$. Let $\vx^{(0)}=\vec{0}$, $\vx^{(k+1)}=\vx^{(k)} - \mmtil^{-1} (\ma^\top \mw \ma \vx^{(k)}- \ma^\top \sqrt{\mw} \vg)$, and $\vutil = \sqrt{W} \ma \vx^{(k)}$. Then we have
 \[
 \norm{\vutil - \vu}_2 \leq \lambda \cdot \left(1-\frac{1}{\lambda} + \veps\right)^k \norm{\sqrt{\mw} \ma (\ma^\top \mw \ma)^{-1} \ma^\top \sqrt{\mw} \vg}_2.
 \]
\end{corollary}
\begin{proof}
Consider applying Richardson's iteration to the following linear regression problem $\min_{\vx} \norm{\sqrt{W} \ma \vx - \vg}_2$ and suppose $\vxstar = \argmin_{\vx} \norm{\sqrt{W} \ma \vx - \vg}_2$. Then by \cref{lemma:richardson}, we have
\[
\norm{\vx^{(k)}-\vxstar}_{\mm}
\leq
\left(1-\frac{1}{\lambda} + \veps\right)^k
\norm{\vx^{(0)} - \vxstar}_{\mm}.
\]
Therefore
\[
\norm{\vx^{(k)}-\vxstar}_{\ma^\top \mw \ma}
\leq
\lambda \cdot \left(1-\frac{1}{\lambda} + \veps\right)^k
\norm{\vxstar}_{\ma^\top \mw \ma}.
\]
Now note that
\begin{align*}
\norm{\vxstar}_{\ma^\top \mw \ma}^2 & = \vg^\top \sqrt{\mw} \ma (\ma^\top \mw \ma)^{-1} (\ma^\top \mw \ma) (\ma^\top \mw \ma)^{-1} \ma^\top \sqrt{\mw} \vg
\\ & =
\norm{\sqrt{\mw} \ma (\ma^\top \mw \ma)^{-1} \ma^\top \sqrt{\mw} \vg}_2^2
\\ & \leq
\norm{\vg}_2^2,
\end{align*}
where the inequality follows since $\sqrt{\mw} \ma (\ma^\top \mw \ma)^{-1} \ma^\top \sqrt{\mw}$ is a projection matrix. Moreover
\[
\norm{\vx^{(k)}-\vxstar}_{\ma^\top \mw \ma} = \norm{\sqrt{W} \ma (\vx^{(k)}-\vxstar)}_2 = \norm{\vutil - \vu}_2.
\]
Combining the above with the inequalities implies the result.
\end{proof}

\cref{lemma:richardson} and \cref{remark:have-mtil} give a bound on the bit complexity and running time of finding a solution assuming access to a matrix close enough to the inverse of $\ma^\top \ma$. However in many applications, we do not even have access to $\ma^\top \ma$. For example, in the case of input-sparsity time linear regression algorithm \cite{clarkson2014sketching,CohenLMMPS15} since computing $\ma^\top \ma$ takes $\Omega(d\cdot \nnz(\ma))$ arithmetic operations. Another recent example is for \emph{subquadratic} time Kronecker regression algorithms since the size of $\ma^\top \ma$ is $\Omega(d^2)$ and computing it prevents obtaining a subquadratic time algorithm.

Since \cref{lemma:richardson} only needs a matrix that is close to the inverse of $\ma^\top \ma$, we instead find a spectral approximation $\mm$ of $\ma^\top \ma$, and then using classic approaches \cite{DemmelDH07,DemmelDHK07}, we find a matrix $\mmtil$ that is close to the inverse of $\mm$, i.e., $\norm{\mmtil^{-1} - \mm^{-1}}_{\fro} \leq \frac{\veps}{d \cdot \lambda\cdot \norm{\ma^\top \ma}_2}$. To find the spectral approximation we use the following result.

\begin{lemma}[\cite{NelsonN13,kane2014sparser}]
\label{lemma:oblivious-sketching}
Let $\ma\in\R^{n\times d}$, $0<\veps,\delta<1$, and $n\geq d$. There is an algorithm to sample a matrix $\ms$ with $O(d \log^8 (d/\delta)/\veps^2)$ rows, $n$ columns, and $s=\theta(\log^3(d/\delta)/\veps)$ nonzero entries per columns, so that
\[
(1-\veps) \ma^\top \ma \preceq
\ma^\top \ms^\top \ms \ma \preceq (1+\veps) \ma^\top \ma,
\]
with probability at least $(1-\delta)$. Moreover all entries of $\ms$ are in $\{0,\frac{1}{\sqrt{s}},-\frac{1}{\sqrt{s}} \}$. Finally $\ms$ can be sampled and multiplied with $\ma$ in time 
\[
O(\nnz(\ma) \cdot \poly\log(d/\delta) /\veps^2).
\]
\end{lemma}

\noindent
We are now equipped to prove our result for input-sparsity time linear regression.

\begin{proof}[Proof of \cref{thm:lin_reg}]
Our approach is to first compute a matrix $\mm$ such that $\ma^\top \ma \preceq \mm \preceq \lambda\cdot \ma^\top \ma$ for some constant $\lambda > 1$. We do this using \cref{lemma:oblivious-sketching}. We pick $\delta=\poly(n)$. We also pick $s$ to be a power of four, so the bit complexity of $\ms$ is controlled. Moreover we can pick a constant $\veps_1$, in a way so that $\frac{1}{1-\veps_1}$ is one plus a power of two. Then, we have
\[
\ma^\top \ma \preceq \frac{1}{1-\veps_1} \cdot  \ma^\top \ms^\top \ms \ma = \mm \preceq \frac{1+\veps_1}{1-\veps_1} \cdot \ma^\top \ma = \lambda \cdot \ma^\top \ma.
\]
Note that by our choice of parameters, the bit complexity of $\mm$ is equal to the bit complexity of $\ma$ up to constant and $\poly\log(nd)$ factors. Moreover, the condition number of $\mm$ is the same as $\ma$ (up to constant factors) since $\mm$ is spectrally close to $\ma^\top \ma$. Note that by \cref{lemma:oblivious-sketching}, we can compute $\mm$ in time $\Otil(\nnz(\ma) + d^{\omega})$ since the number of rows of $\ms \ma$ is $\Otil(d)$.

We now can compute $\mmtil^{-1}$ such that $\norm{\mmtil^{-1} - \mm^{-1}}_{\fro} \leq \frac{\veps_2}{d \cdot \lambda\cdot \norm{\ma^\top \ma}_2}$, for some constant $0<\veps_2<1$ so that $1-\frac{1}{\lambda}+\veps_2 < 1$, in $\Otil(d^{\omega} \cdot \log(\kappa))$ time using standard approaches \cite{DemmelDH07,DemmelDHK07}.
We then use Richardson's iteration (\cref{lemma:richardson}) to compute a solution to the regression problem. The running time of this step is $\Otil((\nnz(\ma) + d^2)\cdot \log \kappa \cdot \log^2 (1/\epsilon))$ according to \cref{remark:richardson-bit-complexity}. This gives a total running time of
\[
\Otil((d^{\omega}+ d^2 \cdot \log^2 (1/\epsilon) + \nnz(\ma) \cdot \log^2 (1/\epsilon))\cdot \log \kappa).
\]

\end{proof}

\section{Inverse Maintenance Stability}
\label{sec:inv-main}

In this section, we prove the backward stability of the inverse maintenance technique through the Sherman-Morrison-Woodbury identity (see \cref{fact:Woodbury}). Our formulation is based on the Frobenius norm error of the inverse matrix which in turn determines the bit complexity of the computed inverse. Note that due to the equivalence of norms, this provides bit complexity results in other norms up to polylogarithmic factors in dimension. 
We show inverse maintenance is backward stable with a bit complexity of $\Otil(\log(\kappa/\epsilon))$.

\begin{lemma}[Forward-backward error connection]
\label{lemma:forward-backward-equiv}
Let $\mm,\mn\in\R^{n\times n}$ be invertible matrices and $\kappa>1$ such that $\norm{\mn}_{\fro},\norm{\mn^{-1}}_{\fro} \leq \kappa$. Suppose $\norm{\mm-\mn}_{\fro} \leq \veps < \frac{1}{2\kappa}$. Then $\norm{\mm^{-1}}_{\fro} \leq 2 \kappa$, and
\[
\norm{\mm^{-1}-\mn^{-1}}_{\fro} \leq 2 \kappa^2 \cdot \veps
\]
\end{lemma}
\begin{proof}
Let $\me = \mm-\mn$. Then by using the Woodbury identity we have
\[
\mm^{-1} = \mn^{-1} - \mn^{-1} \me (\mi + \mn^{-1}\me)^{-1}\mn^{-1}
\]
Now note that
\[
(\mi + \mn^{-1}\me)^{-1} = (\mi + \mn^{-1} \mm -\mi)^{-1}=\mm^{-1} \mn.
\]
Therefore
\begin{align*}
\norm{\mm^{-1}}_{\fro} & \leq \norm{\mn^{-1}}_{\fro} + \norm{\mn^{-1}\me\mm^{-1}}_{\fro}
\\ & \leq  \norm{\mn^{-1}}_{\fro} + \norm{\mn^{-1}}_{\fro} \norm{\me}_{\fro} \norm{\mm^{-1}}_{\fro}
\\ & \leq 
\kappa + \kappa \veps \cdot \norm{\mm^{-1}}_{\fro}.
\end{align*}
Therefore
\[
\norm{\mm^{-1}}_{\fro} \leq \frac{\kappa}{1-\kappa\veps} \leq 2\kappa.
\]
Hence,
\begin{align*}
\norm{\mm^{-1}-\mn^{-1}}_{\fro} 
& = 
\norm{\mn^{-1} \me \mm^{-1}}_{\fro} \leq 
\norm{\mn^{-1}}_F \norm{\me}_F \norm{\mm^{-1}}_F
\leq 
2\kappa^2 \veps.
\end{align*}
\end{proof}

We are now equipped to analyze the stability of the Sherman-Morrison-Woodbury formula for inverse maintenance. Note that there are two sources of error in this formulation. One is from the inverse of the original matrix, and the other is from the inverse of the smaller matrix we need to compute to make the low-rank update to the inverse.

\InvMainAsymm*

\begin{proof}
By triangle inequality and \cref{eq:smw-assumption}, we have
\begin{align}
\nonumber
\norm{(\mztil^{-1} - \mztil^{-1} \matu \md^{-1} \mv^\top \mztil^{-1})^{-1} - (\mz + \matu \mc \mv^\top)}_{\fro} 
& \leq 
\norm{(\mztil^{-1} - \mztil^{-1} \matu \md^{-1} \mv^\top \mztil^{-1})^{-1} - (\mztil + \matu \mc \mv^\top)}_{\fro}
+
\norm{\mztil - \mz}_{\fro}
\\ & \leq 
\label{eq:inv-main-stab-first-tri}
\norm{(\mztil^{-1} - \mztil^{-1} \matu \md^{-1} \mv^\top \mztil^{-1})^{-1} - (\mztil + \matu \mc \mv^\top)}_{\fro} + \veps_1.
\end{align}
Note that $\ms := \mc^{-1} + \mv^\top \mztil^{-1} \matu$ is a Schur complement of the following matrix.
\[
\mt:=
\begin{bmatrix}
\mc^{-1} & \mv^\top \\
\matu & -\mztil
\end{bmatrix} = \begin{bmatrix}
\mi & \mzero \\
\matu \mc & \mi
\end{bmatrix}
\begin{bmatrix}
\mc^{-1} & \mzero \\
\mzero & - \mztil - \matu \mc \mv^{\top} 
\end{bmatrix}
\begin{bmatrix}
\mi & \mc \mv^{\top} \\
\mzero & \mi
\end{bmatrix}
\]
Note that since $\mc^{-1}$ and $-\mztil-\matu \mc \mv^\top$ (the Schur complement) are invertible, $\mt$ is invertible and  
\[
\mt^{-1} = \begin{bmatrix}
\mi & -\mc \mv^{\top} \\
\mzero & \mi
\end{bmatrix}
\begin{bmatrix}
\mc^{-1} & \mzero \\
\mzero & - \mztil - \matu \mc \mv^{\top} 
\end{bmatrix}^{-1}
\begin{bmatrix}
\mi & \mzero \\
-\matu \mc & \mi
\end{bmatrix}.
\]
Since $\ms$ is the Schur complement of $\mt$,
\begin{align}
\label{eq:inv-main-stab-t}
\nonumber
\norm{\ms^{-1}}_{\fro} \leq \norm{\mt^{-1}}_{\fro} & \leq \left(\sqrt{n+m} +\norm{\matu \mc}_{\fro}\right) \left(\sqrt{n+m} +\norm{ \mc \mv^\top}_{\fro}\right) \left( \norm{\mc}_{\fro} + \norm{(\mztil + \matu \mc \matu^{\top})^{-1}}\right) \\ & \leq 2(m+n+\kappa^4)\kappa^2 \leq 4 \kappa^6.
\end{align}
Moreover
\[
\norm{\ms}_{\fro} \leq \norm{\mc^{-1}}_{\fro} +
\norm{\mv^\top}_{\fro} \norm{\mztil^{-1}}_{\fro} \norm{\matu}_{\fro} \leq 2 \kappa^3.
\]
Therefore since $\norm{ \md^{-1} - (\mc^{-1} + \mv^\top \mztil^{-1} \matu)^{-1}}_{\fro} \leq \veps_2$, by \cref{lemma:forward-backward-equiv}
\begin{align}
\label{eq:inv-main-stab-d}
\norm{ \md - (\mc^{-1} + \mv^\top \mztil^{-1} \matu)}_{\fro} \leq (4\kappa^6)^2 \veps_2 = 16 \kappa^{12} \veps_2.
\end{align}
Now let
\[
\mm := \begin{bmatrix}
\md - \mv^\top \mztil^{-1} \matu& \mv^\top \\
\matu & -\mztil
\end{bmatrix}.
\]
Then by \eqref{eq:inv-main-stab-d},
\begin{align}
\label{eq:inv-main-stab-mt}
\norm{\mm - \mt}_{\fro} \leq 16 \kappa^{12} \veps_2.
\end{align}
Moreover by triangle inequality $\norm{\mt}_{\fro} \leq \norm{\mc^{-1}}_{\fro} + \norm{\matu}_{\fro} + \norm{\mv}_{\fro} + \norm{\mztil}_{\fro} \leq 4\kappa$.
Combining this with \eqref{eq:inv-main-stab-t} and \eqref{eq:inv-main-stab-mt}, noting that $\mm^{-1}$ is invertible since $-\mztil$ and $\md-\mv^{\top} \mztil^{-1} \matu +\mv^{\top} \mztil^{-1} \matu = \md$ (the Schur complement) are invertible, and using \cref{lemma:forward-backward-equiv}, we have
\[
\norm{\mm^{-1}-\mt^{-1}} \leq (4 \kappa^{6})^2 \cdot 16 \kappa^{12} \veps_2 = 256 \kappa^{24} \veps_2.
\]
Now note that $-\mztil - \matu \mc \mv^\top$ and $-\mztil - \matu (\md - \mv^\top \mztil^{-1} \matu)^{-1} \mv^\top$ are the Schur complements (of the corresponding block) of $\mt$ and $\mm$, respectively. Therefore
\[
\norm{\left(\mztil + \matu (\md - \mv^\top \mztil^{-1} \matu)^{-1} \mv^\top\right)^{-1} - \left(\mztil + \matu \mc \mv^\top\right)^{-1}}_{\fro} \leq \norm{\mm^{-1}-\mt^{-1}}_{\fro} \leq 256 \kappa^{24} \veps_2.
\]
By Woodbury identity,
\begin{align*}
\left(\mztil + \matu (\md - \mv^\top \mztil^{-1} \matu)^{-1} \mv^\top\right)^{-1} 
& = 
\mztil^{-1} - \mztil^{-1} \matu (\md - \mv^\top \mztil^{-1} \matu + \mv^\top \mztil^{-1} \matu)^{-1} \mv^\top \mztil^{-1}
\\ & =
\mztil^{-1} - \mztil^{-1} \matu \md^{-1} \mv^\top \mztil^{-1}.
\end{align*}
Therefore
\[
\norm{\left(\mztil^{-1} - \mztil^{-1} \matu \md^{-1} \mv^\top \mztil^{-1}\right) - \left(\mztil + \matu \mc \mv^\top\right)^{-1}}_{\fro} \leq 256 \kappa^{24} \veps_2.
\]
Thus since $\norm{\mz+\matu \mc \mv^\top}_{\fro}, \norm{(\mz+\matu \mc \mv^\top)^{-1}}_{\fro} \leq \kappa$, by \cref{lemma:forward-backward-equiv},
\[
\norm{\left(\mztil^{-1} - \mztil^{-1} \matu \md^{-1} \mv^\top \mztil^{-1}\right) - \left(\mztil + \matu \mc \mv^\top\right)^{-1}}_{\fro} \leq (2\kappa^2) \cdot 256 \kappa^{24} \veps_2 = 512 \kappa^{26} \veps_2.
\]
The result follows from combining this with \eqref{eq:inv-main-stab-first-tri}.
\end{proof}

We proved the stability of the inverse maintenance process in this section. 
Equipped with this, we next develop inverse maintenance data structures for both sparse and dense matrices. We later use these data structures for iterative algorithms for solving $p$-norm minimization and LP.

\subsection{Bit Complexity of Dense Inverse Maintenance Data Structure}
\label{sec:ds-bit}

In this section, we present our data structures for dense matrices. For dense matrices, our data structure only relies on the stability of inverse maintenance as proven in \cref{lemma:smw-stability-asymm-fro}.

\RestyleAlgo{algoruled}
\IncMargin{0.15cm}
\begin{algorithm}[t]

\textbf{Variables:} $n,d\in\N$ with $n>d$, $\ma\in\R^{n\times d}$, $\mz^{-1}\in\R^{d\times d}$, $\vecv\in \R^{n}$, $1>\veps>0$.

 \SetKwProg{ddsInit}{Procedure}{$(\ma\in\R^{n\times d}, \vw\in\R^{n},\veps)$}{}
  \ddsInit{\initialize}{
  Set $\ma$, $\veps$, $n$, and $d$ according to the input\\
  Set $\vecv \leftarrow \vw$\\
  Compute the matrix $\mz^{-1}$ such that $\norm{\mz - \ma^\top \mv \ma}_{\fro} \leq \veps$ \DontPrintSemicolon\tcp*{\textcolor{blue}{for example using the approach of \cite{DemmelDH07,DemmelDHK07}}}
  }

\SetKwProg{ddsUpdate}{Procedure}{$(S\subseteq[n], \vu\in\R^{\abs{S}})$}{}
  \ddsUpdate{\update}{
  Set $\vq_S=\vu - \vecv_{S}$ where $\vq\in\R^{n}$\\
  Set $\vecv_{S} = \vu$\\
  Update $\mz^{-1}$ using Woodbury identity as the following
    $
    \mz^{-1} \leftarrow \mz^{-1} - \mz^{-1} \ma_S^{\top} \md^{-1} \ma_{S} \mz^{-1}
    $, where $\md^{-1} \in \R^{\abs{S}\times\abs{S}}$ is a matrix such that $\norm{\md^{-1}-(\mq^{-1}_{S,S} + \ma_S \mz^{-1} \ma_S^\top)^{-1}}_{\fro} \leq \veps$, and $\mq$ is the diagonal matrix corresponding to $\vq$.\\
    Round entries of $\mz^{-1}$ to numbers with $\ceil{\log(10 d^2  \cdot \kappa/\veps)}$ bits.
    }

\SetKwProg{ddsQuery}{Procedure}{$(\vbtil\in\R^d)$}{}
  \ddsQuery{\query}{
  Compute and return $\mz^{-1} \vbtil$.}
  
\caption{Dense inverse maintenance data structure $(\dds)$}
\label{alg:dense-ds}
\end{algorithm}

\begin{theorem}[Dense data structure --- $\dds$]
\label{thm:dense-ds}
Let $n\geq d$.
There exists a data structure with the following operations that maintain an explicit matrix $\mz^{-1}$.
\begin{itemize}
    \item $\initialize(\ma\in\R^{n\times d}, \vw\in\R^{n},  \veps)$: Sets $\vecv = \vw$, and computes a linear operation $\mz^{-1}$ such that \[\norm{\mz - \ma^\top \mv \ma}_{\fro} \leq \veps,\] where $\mv$ is the diagonal matrix corresponding to $\vecv$.
    
    \item $\update(S\subseteq[n], \vu\in\R^{\abs{S}})$: Sets $\vecv_S = \vu$ and updates $\mz^{-1}$ such that after $k$ calls to $\update$, we have
\[
\norm{\mz - \ma^\top \mv \ma}_{\fro} \leq (512 \cdot k \cdot \kappa^{26} + 2 \cdot k \cdot \kappa^2 + 1) \cdot \veps,
\]
where $\kappa>n+d$ is a number such that
\[
\norm{\ma}_{\fro},\norm{\mq_{S,S}}_{\fro},\norm{\mq_{S,S}^{-1}}_{\fro}, \norm{\ma^\top \mv \ma}_{\fro}, \norm{(\ma^\top \mv \ma)^{-1}}_{\fro} \leq \kappa,
\]
during all the updates and 
\[
\veps < \frac{1}{2\kappa \cdot (512 \cdot (k+1) \cdot \kappa^{26} + 2 \cdot k \cdot \kappa^2 + 1)}.
\]

    \item $\query(\vbtil\in\R^d)$: Computes and returns $\mz^{-1} \vbtil$.
\end{itemize}
The running time of initialize, update, and query are $\mtime(n,d,d) \cdot O(\log(\kappa/\veps))$, \[\left(\mtime(d,\abs{S},\abs{S}) +\mtime(d,d,\abs{S})+\abs{S}^{\omega}\right) \cdot O(\log(\kappa/\veps)),\] and $d^2\cdot O(\log(\kappa/\veps))$, respectively. Moreover the bit complexity of the matrix $\mz^{-1}$ is $\Otil(\log(\kappa/\veps))$.
\end{theorem}
\begin{proof}
We show that the data structure in Algorithm \ref{alg:dense-ds} achieves the desired result.
Computing the matrix $\ma^\top \mv \ma$ takes $\mtime(n,d,d) \cdot O(\log(\kappa))$ time, and computing the matrix $\mz^{-1}$ such that $\norm{\mz - \ma^\top \mv \ma}_{\fro} \leq \veps$ takes $d^\omega \cdot O(\log(\kappa)/\veps)$ takes. This gives the bound on the bit complexity of initialization follows from $d\leq n$.

Now note that at the end of each update procedure, we round the matrix to numbers with $\Otil(\log(\kappa/\epsilon))$ bits.
For update we need to compute the matrix $\md$ such that  $\norm{\md^{-1}-(\mq^{-1}_{S,S} + \ma_S \mz^{-1} \ma_S^\top)^{-1}}_{\fro} \leq \veps$ which can be done in time $\left(\mtime(\abs{S},d,d)+\mtime(\abs{S},d,\abs{S}) + \abs{S}^{\omega}\right) \cdot O(\log(\kappa/\veps))$. Given the matrix $\md^{-1}$, updating the matrix $\mz^{-1}$ can be done in time $\left(\mtime(d,\abs{S},\abs{S}) +\mtime(d,d,\abs{S})\right) \cdot O(\log(\kappa/\veps))$.
The cost of the query is a matrix-vector multiplication which is $d^2\cdot O(\log(\kappa/\veps))$.

We now need to bound the error of our matrix after $k$ updates.  We prove this by induction. The base case trivially follows from the condition in the initialization. Now suppose after $k$ updates 
\[
\norm{\mz - \ma^\top \mv \ma}_{\fro} \leq (512 \cdot k \cdot \kappa^{26} + 2 \cdot k \cdot \kappa^2 + 1) \cdot \veps,
\]
and $\mq$ is the diagonal matrix corresponding to the $k+1$'st update. Let 
\[
\mztil^{-1} = \mz^{-1} - \mz^{-1} \ma_S^{\top} \md^{-1} \ma_{S} \mz^{-1} ~~ , \text{with} ~~  \norm{\md^{-1}-(\mq^{-1}_{S,S} + \ma_S \mz^{-1} \ma_S^\top)^{-1}}_{\fro} \leq \veps,
\]
and $\mzhat^{-1}$ is the matrix obtained by rounding the entries of $\mztil^{-1}$ to numbers with $\ceil{\log(10 d^2  \cdot \kappa/\veps)}$ bits. \cref{lemma:smw-stability-asymm-fro} directly gives 
\[
\norm{\mztil - \ma^\top (\mv+\mq) \ma}_{\fro} \leq (512 \cdot (k+1) \cdot \kappa^{26} + 2 \cdot k \cdot \kappa^2  + 1) \cdot \veps.
\]
Moreover since by assumption $\norm{\ma^\top (\mv+\mq) \ma}_{\fro},\norm{(\ma^\top (\mv+\mq) \ma)^{-1}}_{\fro} \leq \kappa$, and 
\[
\norm{\mztil - \ma^\top (\mv+\mq) \ma}_{\fro} \leq (512 \cdot (k+1) \cdot \kappa^{26} + 2 \cdot k \cdot \kappa^2 + 1) \cdot \veps < \frac{1}{2\kappa},
\]
by \cref{lemma:forward-backward-equiv}, we have $\norm{\mztil^{-1}}_{\fro} \leq 2\kappa$. Therefore the number of bits before decimal for the entries $\mztil^{-1}$ is bounded by $\log_2(2\kappa) + 1$. Therefore the rounding only introduces error in the bits after the decimal point. Therefore
\[
\norm{\mzhat^{-1} - \mztil^{-1}} \leq \veps.
\]
Invoking \cref{lemma:forward-backward-equiv} and because $\veps < \frac{1}{4\kappa}$ (by assumption), $\norm{\mztil^{-1}}_{\fro} \leq 2\kappa$, and 
\[
\norm{\mztil}_{\fro} \leq \norm{\mztil - \ma^\top (\mv+\mq) \ma}_{\fro} + \norm{\ma^\top (\mv+\mq) \ma}_{\fro} \leq \frac{1}{2\kappa} + \kappa \leq 2 \kappa,
\]
we have $\norm{\mzhat-\mztil}_{\fro} \leq 2\kappa^2 \cdot \veps$. Therefore by triangle inequality,
\[
\norm{\mzhat - \ma^\top (\mv+\mq) \ma}_{\fro} \leq \norm{\mzhat - \mztil}_{\fro} +
\norm{\mztil - \ma^\top (\mv+\mq) \ma}_{\fro}
\leq (512 \cdot (k+1) \cdot \kappa^{26} + 2\cdot (k+1) \kappa^2 + 1) \cdot \veps.
\]
Therefore the desired error bound holds.
\end{proof}

\begin{remark}
Note that if we do not perform the rounding step in the $\update$ procedure of $\dds$, after $k$ updates, the bit complexity increases by a factor of $k$ since the update involves multiplying matrices. However, because of the upper bound assumption on $\veps$, after the rounding, the bit complexity of the resulting matrix is $\Otil(\log(k \cdot \kappa/\veps))$. As we see later the number of iterations in our algorithms is of the form $\poly(n) \log(f(n))$, where $f(n)$ is at most an exponential function in $n$ (e.g., for LPs, the number of iterations is $\Otil(n^{0.5}\log(\frac{R}{r\cdot \epsilon}))$). Therefore $\Otil(\log(k \cdot \kappa/\veps))$ in our algorithms is $\Otil(\log(\kappa/\veps))$.
\end{remark}

\begin{remark}
\cref{thm:dense-ds} requires $\norm{\mq_{S,S}}_{\fro},\norm{\mq_{S,S}^{-1}}_{\fro} \leq \kappa$, where $\mq$ is the diagonal matrix corresponding to the difference of weights from one iteration to the next. Although we usually update the inverse only for weights that have changed significantly in our algorithms, even if the difference is small, this condition does not impose a limitation on our data structure since we can implement each update as two updates. For example, if $\vq_i>0$ is small, we first make an update with $\vqtil_i^{(1)} = \vq_i+1$ and then we make a second update with $\vqtil_i^{(2)} = -1$. Note that when doing this a number should be added and subtracted so that the Frobenius norm of $\ma^\top \mv \ma$ and $(\ma^\top \mv \ma)^{-1}$ also stay small when we perform the auxiliary update step.
\end{remark}

\subsection{Bit Complexity of Sparse Inverse Maintenance Data Structure}

We now turn to the sparse case. We first give an algorithm and bit complexity bounds for computing a linear operator for the inverse of a sparse matrix based on the block-Krylov approach \cite{eberly2006solving,eberly2007faster}. Our algorithm builds on \cite{PengVempala21} but has some differences from the algorithm presented in that work since we use the improved analysis presented in \cite{nie2022matrix}.

\begin{restatable}[\cite{PengVempala21,nie2022matrix}]{theorem}{SparseInverse}
\label{thm:sparse_inverse}
Given a sparse $n\times d$ matrix $\ma$ with max entry-wise magnitude at most $1$,
a diagonal $n\times n$ matrix $\mw$ with entry-wise magnitude at most $1$
and $m\leq d^{(\omega-2)/(\omega-1)}$, along with $\kappa$ that upper bounds the condition numbers of $\ma$ and $\mw$, with probability at least $1-n^{-20}$, 
we can obtain in time 
\[
\Otil\left(\left(d\cdot \nnz\left(\ma\right)\cdot m + \left(\frac{d}{m}\right)^\omega m^2\right)\log\left(\kappa\right)\right)
\]
a linear operator $\mz_{\ma^\top \mw \ma}$ such that
\[
\norm{\mz_{\ma^\top \mw \ma}
- \left(\ma^\top \mw \ma\right)^{-1}
}_{\fro}
\leq
\kappa^{-10} n^{-10}.
\]
Moreover, for a $d\times r$ matrix $\mb$, where $r\leq d/m$, $\mz_{\ma^\top \mw \ma} \mb$ can be computed in time $\Otil((r \cdot \nnz(\ma)\cdot m^2 + d^{2} r^{\omega-2})\log(\kappa/\epsilon))$.
\end{restatable}

\begin{proof}
The statements below are closely based on the top-level
claims in~\cite{PengVempala21}\footnote{Version 2, \url{https://arxiv.org/pdf/2007.10254v2.pdf}}. 
Since $\ma^\top \mw \ma$ is already symmetrized,
we can ignore the outer step involving a multiplication
by the transpose of an asymmetric matrix.
So we will show how to give access to an
operator $\mz_{\ma^\top \mw \ma}$ such that
\begin{align}
\norm{
\mz_{\ma^\top \mw \ma}
-
\left( \ma^\top \mw \ma \right)^{-1}
}_{\fro}
\leq
\veps
\end{align}

The algorithm that computes access to this $\mz$ was given in Section 7
of~\cite{PengVempala21}.
\begin{enumerate}
    \item Perturb with random Gaussian $\mr$ to form the perturbed matrix
    \[
    \mahat
    =
    \ma^\top \mw \ma
    +
    \mr
    \]
    \item Generate Krylov space with $\Otil(m)$ extra columns,
    \[
    K
    = 
    \left[
    \begin{array}{ccccc}
    \mg^{s}
    &
    \mahat \mg^{s}
    &
    \mahat^2 \mg^{s}
    &
    \ldots
    &
    \mahat^{m - 1} \mg^{s}
    \end{array}
    \right],
    \]
    where $s=d/m$, and $\mg^{s}$ is a sparse Gaussian matrix with $s$ columns and $\nnz(\mg^{s}) = \Otil(n)$.
    \item Compute the inverse of the block-Hankel matrix $\mh=\mk^\top \mahat \mk$.
\end{enumerate}

Specifically, for step (3),
the $\mz_\mh$ generated by the block-Hankel solver
is the product of two explicit matrices, each with
$\Otil(m \log(\kappa))$ bits,
\[
\mz_\mh
=
\mx_{\mh} \my_{\mh}^\top
\]
such that the cost of 
computing $\mx_{\mh} \mb$, $\my_{\mh} \mb$, $\mx_{\mh}^{\top} \mb$, $\my_{\mh}^{\top} \mb$
for some $d$-by-$r$ matrix $\mb$ with up to
$\Otil(m \log{\kappa})$ bits per entry is
$\Otil(m^2 \cdot
\mtime(\frac{d}{m}, \frac{d}{m}, r) \cdot \log{\kappa})$
by Lemma 6.6 of~\cite{PengVempala21}\footnotemark[1].
The cost of multiplying $\mz_{\ma^\top \mw \ma}$ against a $d$-by-$r$ matrix
$\mb$ is then broken down into three parts:
\begin{enumerate}
    \item The cost of performing the multiplication $\mk^\top \mb$ which takes $\Otil(\nnz(A)\cdot m^2 \cdot r \cdot \log^2(\kappa/\veps))$ time.
    
    \item The cost of multiplying $\my_{\mh}^\top$ and $\mx_{\mh}$ against a $d$-by-$r$ matrix: by Lemma 6.6 of~\cite{PengVempala21}\footnotemark[1],
    this takes time $\Otil(\mtime(\frac{d}{m},\frac{d}{m},r) \cdot m^2 \log{\kappa})$ by performing fast Fourier transform on the blocks of $\mx_{\mh}$ and $\my_{\mh}$.
    
    \item The cost of multiplying $\mk$ with a $d$-by-$r$ matrix from right which takes $\Otil(\nnz(A)\cdot m^2 \cdot r \cdot \log^2(\kappa/\veps))$ time.
\end{enumerate}

\end{proof}

\noindent
We are now equipped to present our sparse data structure and bound the bit complexity of its operations.

\RestyleAlgo{algoruled}
\IncMargin{0.15cm}
\begin{algorithm}[t]

\textbf{Variables:} $n,d,m\in\N$ with $n>d$, $m \leq n^{0.25}$, $\ma\in\R^{n\times d}$, linear operator $\mz\in\R^{d\times d}$, explicit matrix $\mt \in\R^{d\times d}$, $\vecv\in \R^{n}$, $1>\veps>0$.

 \SetKwProg{sdsInit}{Procedure}{$(\ma\in\R^{n\times d}, \vw\in\R^{n},\veps)$}{}
  \sdsInit{\initialize}{
  Set $\ma$, $\veps$, $n$, and $d$ according to the input\\
  Set $\vecv \leftarrow \vw$\\
  Compute the linear operator $\mz$ such that $\norm{\mz^{-1} - \ma^\top \mv \ma}_{\fro} \leq \veps$ \DontPrintSemicolon\tcp*{\textcolor{blue}{using \cref{thm:sparse_inverse} and setting the error bounds small enough according to \cref{lemma:forward-backward-equiv}}}
  Set $\mt$ to the matrix of all zeros
  }

\SetKwProg{sdsUpdate}{Procedure}{$(S\subseteq[n], \vu\in\R^{\abs{S}})$}{}
  \sdsUpdate{\update}{
    \If{$\abs{S} \geq \frac{n}{m}$}{
     Set $\vw\in\R^n$ to a vector with $\vw_i=\vu_i$, if $i\in S$, and $\vw_i = \vecv_i$, if $i\in[n]\setminus S$. \\
     \initialize($\ma,\vw, \veps$)
    } \Else {
    Set $\vq_S=\vu - \vecv_{S}$ where $\vq\in\R^{n}$\\
  Set $\vecv_{S} = \vu$\\
Update the matrix $\mt$ as the following
    \begin{align}
    \label{eq:sparse-ds-query}
    \mt \leftarrow \mt - (\mz + \mt)^\top \ma_S^{\top} \md^{-1} \ma_{S} (\mz+\mt),
    \end{align}
    where $\md^{-1} \in \R^{\abs{S}\times\abs{S}}$ is a matrix such that $\norm{\md^{-1}-(\mq^{-1}_{S,S} + \ma_S (\mz + \mt) \ma_S^\top)^{-1}}_{\fro} \leq \veps$, and $\mq$ is the diagonal matrix corresponding to $\vq$.\\
  
  Round entries of $\mt$ to numbers with $\ceil{\log(10 d^2  \cdot \kappa/\veps)}$ bits.
    }
    }

\SetKwProg{sdsQuery}{Procedure}{$(\vbtil\in\R^d)$}{}
  \sdsQuery{\query}{
  Compute and return $\mz  \vbtil + \mt \vbtil$. \DontPrintSemicolon\tcp*{\textcolor{blue}{$\mz  \vbtil$ is computed according to \cref{thm:sparse_inverse}}}}
  
\caption{Sparse inverse maintenance data structure $(\sds)$}
\label{alg:sparse-ds}
\end{algorithm}

\begin{theorem}[Sparse data structure --- $\sds$]
\label{thm:sparse-ds}
Let $n\geq d$ and $m\leq n^{1/4}$.
There exists a data structure with the following operations that maintain an inverse operator as the sum of an explicit matrix $\mt$ and a block-Krylov-based inverse (as represented in \cref{thm:sparse_inverse}).
\begin{itemize}
    \item $\initialize(\ma\in\R^{n\times d}, \vw\in\R^{n}, \veps)$: Sets $\vecv = \vw$, and initializes the explicit matrix $\mt\in\R^{d\times d}$ and a linear operator $\mz$ (see \cref{thm:sparse_inverse}) such that $\norm{(\mz+\mt)^{-1} - \ma^\top \mv \ma}_{\fro} \leq \veps$, where $\mv$ is the diagonal matrix corresponding to $\vecv$.
    
    \item $\update(S\subseteq[n]$, $\vu\in\R^{\abs{S}})$: Sets $\vecv_S = \vu$ and updates $\mz$ and $\mt$ such that after $k$ calls to $\update$, we have
\[
\norm{(\mz+\mt)^{-1} - \ma^\top \mv \ma}_{\fro} \leq (512 \cdot k \cdot \kappa^{26} + 2 \cdot k \cdot \kappa^2 + 1) \cdot \veps,
\]
where $\kappa>n+d$ is a number such that
\[
\norm{\ma}_{\fro},\norm{\mq_{S,S}}_{\fro},\norm{\mq_{S,S}^{-1}}_{\fro}, \norm{\ma^\top \mv \ma}_{\fro}, \norm{(\ma^\top \mv \ma)^{-1}}_{\fro} \leq \kappa,
\]
during all the updates and 
\[
\veps < \frac{1}{2\kappa \cdot (512 \cdot (k+1) \cdot \kappa^{26} + 2 \cdot k \cdot \kappa^2 + 1)}.
\]

    \item $\query(\vbtil)$: Computes and returns $(\mz + \mt) \vbtil$.
\end{itemize}
The running time of initialize, and query are $\Otil\left(\left(d\cdot \nnz\left(\ma\right)\cdot m + \left(\frac{d}{m}\right)^\omega m^2\right)\log^2\left(\kappa/\veps\right)\right)$, and $\Otil(\nnz(\ma)\cdot m^2+d^2\cdot\log^2(\kappa/\veps))$, respectively. The running time of updates is equal to initialize if $\abs{S}\geq n/m$, and is equal to
\[
\left(\nnz(\ma)\cdot m^2\cdot \abs{S} + d^2 \cdot \abs{S}^{\omega-2}+\mtime(d,\abs{S},\abs{S}) + \abs{S}^\omega\right) \cdot O(\log^2(\kappa/\veps)),
\]
otherwise.
\end{theorem}
\begin{proof}
We show that the data structure in Algorithm \ref{alg:sparse-ds} achieves the desired result.
The running time for the initialization and update when $\abs{S} \geq n/m$ follow directly from \cref{thm:sparse_inverse}. The running time of query follows by invoking the second part of \cref{thm:sparse_inverse} for a matrix with one column.

For updates with $\abs{S} < n/m$, we first need to compute $\mz^\top \ma_S^\top$ which by \cref{thm:sparse_inverse} can be done in time $\Otil((\nnz(\ma)\cdot m^2\cdot \abs{S} + d^2 \cdot \abs{S}^{\omega-2})\log^2\left(\kappa/\veps\right))$. After this multiplication, the number of bits of the resulting matrix can be reduced to $\Otil(\log(\kappa/\veps))$ because the condition numbers of $\ma^\top \mv \ma$ is bounded by $\kappa^{O(1)}$. Note that this rounding error can be counted as the error of the linear operator of the inverse.
Then with an extra cost of $\Otil(d^2 \log(\kappa/\epsilon))$, we can compute $(\mz+\mt)^\top \ma_S^{\top}$. Therefore $\ma_S (\mz+\mt) \ma_S^\top$ can be computed in time $\Otil(\mtime(\abs{S},d,\abs{S})\cdot \log(\kappa/\veps))$. Now $\md^{-1}$ can be computed in time $\Otil(\abs{S}^{\omega}\cdot\log(\kappa/\veps))$. Finally since we already have computed $(\mz + \mt)^\top \ma_S^\top$, $\mt$ can be updated in time $\Otil((\mtime(d,\abs{S},\abs{S})+\mtime(d,d,\abs{S}))\cdot \log(\kappa/\veps))$.

We now need to bound the error of our matrix after $k$ updates.  We prove this by induction. The base case trivially follows from the condition in the initialization and \cref{lemma:forward-backward-equiv}. Now suppose after $k$ updates 
\[
\norm{(\mz + \mt)^{-1} - \ma^\top \mv \ma}_{\fro} \leq (512 \cdot k \cdot \kappa^{26} + 2 \cdot k \cdot \kappa^2 + 1) \cdot \veps,
\]
and $\mq$ is the diagonal matrix corresponding to the $k+1$'st update. Let 
\[
\mttil = \mt - (\mz + \mt)^\top \ma_S^{\top} \md^{-1} \ma_{S} (\mz+\mt) ~~ , \text{with} ~~  \norm{\md^{-1}-(\mq^{-1}_{S,S} + \ma_S (\mz + \mt) \ma_S^\top)^{-1}}_{\fro} \leq \veps,
\]
and $\mthat$ is the matrix obtained by rounding the entries of $\mttil$ to numbers with $\ceil{\log(10 d^2  \cdot \kappa/\veps)}$ bits. Also let $\mztil := \mz + \mttil$ and $\mzhat := \mz + \mthat$.
\cref{lemma:smw-stability-asymm-fro} directly gives 
\[
\norm{\mztil^{-1} - \ma^\top (\mv+\mq) \ma}_{\fro} \leq (512 \cdot (k+1) \cdot \kappa^{26} + 2 \cdot k \cdot \kappa^2  + 1) \cdot \veps.
\]
Moreover since by assumption $\norm{\ma^\top (\mv+\mq) \ma}_{\fro},\norm{(\ma^\top (\mv+\mq) \ma)^{-1}}_{\fro} \leq \kappa$, and 
\[
\norm{\mztil^{-1} - \ma^\top (\mv+\mq) \ma}_{\fro} \leq (512 \cdot (k+1) \cdot \kappa^{26} + 2 \cdot k \cdot \kappa^2 + 1) \cdot \veps < \frac{1}{2\kappa},
\]
by \cref{lemma:forward-backward-equiv}, we have $\norm{\mztil}_{\fro} \leq 2\kappa$. Therefore the number of bits before decimal for the entries $\mztil$ is bounded by $\log_2(2\kappa) + 1$. Therefore the rounding only introduces error in the bits after the decimal point. Therefore
\[
\norm{\mzhat - \mztil}_{\fro} = \norm{\mthat - \mttil}_{\fro}\leq \veps.
\]
Invoking \cref{lemma:forward-backward-equiv} and because $\veps < \frac{1}{4\kappa}$ (by assumption), $\norm{\mztil}_{\fro} \leq 2\kappa$, and 
\[
\norm{\mztil^{-1}}_{\fro} \leq \norm{\mztil^{-1} - \ma^\top (\mv+\mq) \ma}_{\fro} + \norm{\ma^\top (\mv+\mq) \ma}_{\fro} \leq \frac{1}{2\kappa} + \kappa \leq 2 \kappa,
\]
we have $\norm{\mzhat^{-1}-\mztil^{-1}}_{\fro} \leq 2\kappa^2 \cdot \veps$. Therefore by triangle inequality,
\[
\norm{\mzhat^{-1} - \ma^\top (\mv+\mq) \ma}_{\fro} \leq \norm{\mzhat^{-1} - \mztil^{-1}}_{\fro} +
\norm{\mztil^{-1} - \ma^\top (\mv+\mq) \ma}_{\fro}
\leq (512 \cdot (k+1) \cdot \kappa^{26} + 2\cdot (k+1) \kappa^2 + 1) \cdot \veps.
\]
Therefore the desired error bound holds.
\end{proof}

\section{Linear Programmming Using Interior Point Methods (IPM)}
\label{sec:ipm}

In this section, we consider linear programming problems of the following form.
\[
\min_{\vx:\ma^\top \vx =\vb, \vx \geq 0} \vc^\top \vx,
\]
where $\ma\in\R^{n\times d}$, $\vb\in\R^d$ and $\vc\in\R^n$. We consider a variety of interior point methods for this problem. Our main result is the following that bounds the bit complexity of the algorithm of \cite{DBLP:conf/soda/Brand20}, which is the derandomized version of \cite{DBLP:conf/stoc/CohenLS19} --- see \cref{sec:robust_ipm}. A main difference between our algorithm and that of \cite{DBLP:conf/stoc/CohenLS19,DBLP:conf/soda/Brand20} is the choice of initial feasible solution. Inspired by \cite{lee2021tutorial} and in contrast with \cite{DBLP:conf/stoc/CohenLS19,DBLP:conf/soda/Brand20}, we select the initial feasible solution so that the condition number and $R/r$ stay the same up to polynomial factors. Recall $r$ and $R$ are inner and outer radius of the LP (see \cref{def:radius}).

\RobustIPM*

As discussed in \cref{sec:discussion}, $\log(\kappa)$ can be $\Omega(n)$ even for matrices with bit complexity $O(1)$. Moreover, as we discuss in \cref{sec:lp-prelim}, $\log(R/r)$ can be $\Omega(n)$ as well. This gives a total running time of $O(n^{\omega+2})$ for algorithms of \cite{DBLP:conf/stoc/CohenLS19,DBLP:conf/soda/Brand20}. Note that there are instances in which $\log(R/r)=O(1)$ while $\kappa = \Omega(n)$. Motivated by this, we present the following algorithm based on solving linear systems using shifted numbers \cite{storjohann2005shifted} that replaces the $\log(\kappa)$ factor with $n^{0.5}$. In instances with $\log(R/r)=O(1)$, $\kappa = \Omega(n)$, this approach is faster than \cref{thm:robust_ipm} by a factor of $n^{0.5}$.

\thmInverseFreeIPM*

We use the classic IPM that uses the $2$-norm as its potential function for the above result. A similar approach combined with the sparse solver can be used to improve the running time of solving linear programs beyond matrix multiplication for sparse instances when $\omega >2.5$ (for example, algorithms based on the Strassen algorithm with $\omega\approx 2.808$ \cite{strassen1969gaussian}). Note that such matrix multiplication algorithms are the ones that are used in practice.

\begin{restatable}{theorem}{SparseIPM}[$\ell_2$-IPM for sparse matrices]
\label{thm:ell2_ipm_sparse}
Let the matrix multiplication exponent $\omega>2.5$.
Given $\ma\in\R^{n\times d}$ with full column-rank, $\vb\in\R^{d}$, $\vc\in\R^{n}$, $1\leq m\leq n^{1/4}$ an error parameter $0<\epsilon<1$, and a linear program $\min_{\ma^\top \vx = \vb, \vx\geq 0} \vc^\top \vx$ with inner radius $r$ and outer radius $R$, there exists an algorithm that finds $\vxhat\in\R^{n}$ such that
\[
\vc^\top \vxhat \leq \min_{\ma^\top \vx = \vb, \vx\geq 0} \vc^\top \vx + \epsilon \cdot \norm{\vc}_{\infty}R   ~~~ \text{, and } ~~~ \norm{\ma^\top \vxhat - \vb}_2 \leq \epsilon \cdot \norm{\vb}_2,
\]
in time $\Otil\left(\left(\nnz(\ma)\cdot m^2\cdot n + \frac{n^{\omega}}{m^{\omega-2.5}} +n^{2.5}\right)\cdot \log^2(\frac{\kappa+\norm{\vb}_2}{\veps}) \cdot \log(\frac{n \cdot R}{\epsilon \cdot r})\right)$ with high probability.
\end{restatable}

For the case of $\nnz(\ma)=O(n)$, if we use the Strassen algorithm and $\kappa/\epsilon$ and $R/r$ are polynomials in $n$, then the above result implies a running time of $\Otil(n^{2.756})$. Moreover, for any $\omega>2.5$ and $\ma$ with $\nnz(\ma) = o(n^{\omega-1})$, there exits an $m$ such that the above running time is smaller than $n^{\omega}$. 

In \cref{sec:lp-prelim}, we discuss some definitions, parameters, the general IPM approach for solving LPs and our choice of initial feasible solutions. We then prove \cref{thm:robust_ipm,thm:inverse_free_ipm,thm:ell2_ipm_sparse} in \cref{sec:robust_ipm,sec:inverse_free_ipm,sec:sparse_ipm}, respectively.

\subsection{LP Preliminaries and Initial Feasible Point}
\label{sec:lp-prelim}

We start by defining the central path. The interior point method first finds a feasible solution on the central path and then following the central path to get close to the optimal solution.
\begin{definition}
A point $\vx\in\R^{n}_{\geq 0}$ is on the central path if there exist $\vs\in\R^{n}_{\geq 0},t\in\R_{\geq 0}$ such that
\begin{align*}
\vx \odot \vs & = \vec{t},
\\
\ma^\top \vx & = \vb, 
\\
\ma \vy + \vs & = \vc,
\end{align*}
Note that $\vx$ is an optimal solution if there exists $\vs\in\R^{n}_{\geq 0}$ such that $\vx \odot \vs = \vec{0}$, and the other two constraints are also satisfied.
\end{definition}
The first step of solving linear programs using IPMs is to find an initial feasible solution on the central path. This is achieved by modifying the linear program so that a feasible solution of the modified program is known.

\begin{definition}[Modified linear program]
\label{def:modified-LP}
Consider a linear program $\min_{\vx: \ma^\top \vx = \vb,\vx\geq 0} \vc^\top \vx$, with inner radius $r$ and outer radius $R$. For any $\overline{R}\geq 10R$, $t\geq 8 \norm{\vc}_{\infty}\overline{R}$, we define the modified primal linear program by
\[
\min_{(\vx^+,\vx^-,\vx^\theta)\in\ptope_{\overline{R},t}} \vc^\top \vx^+ + \vctil^\top \vx^-,
\]
where 
\[
\ptope_{\overline{R},t} = \{(\vx^+,\vx^-,\vx^\theta)\in\R^{2n+1}_{\geq 0} :\ma^\top (\vx^+ - \vx^-) = \vb, \sum_{i=1}^n \vx^+_i +\vx^{\theta}=\vbtil \},
\]
with $\vx_{\vc}^+=\frac{t}{\vc+\vec{t}/\overline{R}}$, $\vx_{\vc}^-=\vx_{\vc}^+ - \ma (\ma^\top \ma)^{-1} \vb$, $\vctil =t/\vx^-_{\vc}$, $\vbtil=\sum_{i=1}^n \vx_{\vc,i}^+ + \overline{R}$. We define the corresponding dual polytope by
\[
\dualptope_{\overline{R},t} = \{(\vs^+,\vs^-,\vs^\theta)\in\R^{2n+1}_{\geq 0}:\ma \vy + \lambda \vec{1}  +\vs^+ =\vc, -\ma \vy +\vs^-=\vctil, \lambda +\vs^\theta =0 \text{ for some } \vy\in\R^d \text{ and } \lambda \in \R\}.
\]
Note that defining
\[
\mabar = \begin{bmatrix}
\ma & \vec{1} \\
-\ma & \vec{0} \\
\vec{0}^\top & 1
\end{bmatrix}, ~~ \vbbar = \begin{bmatrix}
\vb \\ \vbtil
\end{bmatrix}, ~~ \text{ and } ~~ \vcbar = \begin{bmatrix}
\vc \\ \vctil \\ 0
\end{bmatrix},
\]
the modified primal problem is $\min_{\vxbar:\mabar^\top \vxbar = \vbbar, \vxbar\geq 0}  \vcbar^\top \vxbar$.
\end{definition}

The next lemma states that an initial feasible solution of the modified linear program is known. Moreover starting from that feasible solution, if we decrease the centrality (entries of the vector $\vx \odot \vs$) by an appropriate amount, we can reach a point close to the central path of the original linear program.

\begin{lemma}[Theorem 11 on \cite{lee2021tutorial}]
\label{lemma:init-feasible}
Given a linear program $\min_{\vx: \ma \vx = \vb,\vx\geq 0} \vc^\top \vx$, with inner radius $r$, andouter radius $R$. For any $0\leq \veps\leq 0.5$, the modified linear program (\cref{def:modified-LP}), with $\overline{R} = \frac{5}{\veps} R$ and $t = 2^{16} \veps^{-3} n^2 \frac{R}{r} \cdot \norm{\vc}_{\infty}R$ has the following properties:
\begin{enumerate}
    \item The point $(
    \vx_{\vc}^+, \vx_{\vc}^-, \overline{R}
    )$ (as defined in \cref{def:modified-LP}) is on the central path of the modified linear program with $(\vs^+,\vs^-,\vs^\theta)$ and $t$, where $\vs^+=\vec{t}/\vx^+$, $\vs^- = \vec{t}/\vs^{-}$, and $\vs^\theta = t/\vx^{\theta}$.
    
    \item For any feasible primal point $\vxbar=(
    \vx^+, \vx^-, \vx^{\theta}
    ) \in \ptope_{\overline{R}, t}$ and dual $\vsbar = (\vs^+, \vs^-, \vs^{\theta})\in \dualptope_{\overline{R}, t}$ such that $\frac{5}{6} \norm{\vc}_{\infty}R \leq \vx_i \vs_i \leq \frac{7}{6} \norm{\vc}_{\infty}R$, we have that $(\vx^+ -\vx^-,\vs^+ - \vec{\vs}^{\theta}) \in \ptope \times \dualptope$. In addition, $\vx_i^-  \leq \veps \vx_i^+$ and $\vec{\vs}^\theta \leq \veps \vs_i^+$ for all $i\in[n]$.
\end{enumerate}
\end{lemma}

As we will show for all of our interior point methods, we can take steps of the form $(1-\frac{1}{C\sqrt{n}}) t$ for some constant $C$. Therefore starting from the initial feasible solution of the modified linear program, we can reach a point close to a feasible solution of the original linear program in $O(\sqrt{n} \log(\frac{n\cdot R}{\veps\cdot r}))$ iterations. We then can run our interior point algorithms on that point to reach a point that is $\epsilon$ close to the optimal. This can be performed in $O(\sqrt{n}\log(\frac{n \cdot \norm{\vc}_{\infty}R}{\epsilon}))$ additional iterations. This is illustrated in Algorithm \ref{alg:ipm-init}. In this algorithm we denote the IPM algorithms by $\genericipm$ since we use different IPMs in Sections \ref{sec:robust_ipm}, \ref{sec:inverse_free_ipm}, and \ref{sec:sparse_ipm}. Essentially the differences between IPMs is the way they measure the closeness to the central path, the linear systems they solve in each iteration (which is characterized by approximations of the gradient vector and vectors $\vx$ and $\vs$ that are used), and the way these linear systems are solved. The former is formalized in the following  definition.

\begin{definition}
We consider an algorithm $\genericipm(\ma,\vx^{(0)},\vs^{(0)}, t^{(0)}, t^{(\final)}, \veps)$, a generic interior point method, if for a potential function $f$, a function $g$ depending on $n$, and given $\vx^{(0)},\vs^{(0)}\in\R^n$, $t\in\R$, such that $f(\vx^{(0)},\vs^{(0)},t^{(0)})\leq g(n)$, it returns $\vx^{(\final)},\vs^{(\final)}$ such that $f(\vx^{(\final)},\vs^{(\final)},t^{(\final)})\leq g(n)$, and $\norm{\ma^\top (\vx^{(\final)} - \vx^{(0)})} \leq \veps$. For IPMs based on the $2$-norm, $f(\vx,\vs,t) = \norm{(\vx \odot \vs - \vec{t})/t}_2$, and $g(n) = 0.01$. For robust IPMs, $f(\vx,\vs,t)=\phil((\vx \odot \vs - \vec{t})/t)$, where $\phil(\vu)=\sum_{i=1}^n \cosh(\lambda \vu_i)$, $\lambda>0$ is a parameter, and $g(n) = 16n$. 
\end{definition}

An IPM algorithm updates primal and slack vectors $\vx$ and $\vs$, in each iteration, by solving the following linear system and setting $\vx=\vx+\vdeltil_{\vx}$ and $\vs=\vs+\vdeltil_{\vs}$,

\begin{align*}
\mxbar \vdeltil_{\vs} + \msbar \vdeltil_{\vx} & = \vdeltil_{\vmu}, \\
\ma^\top \vdeltil_{\vx} & = 0, \nonumber \\
\ma \vdeltil_{\vy} + \vdeltil_{\vs} & = 0, \nonumber
\end{align*}
where $\vxbar, \vsbar, \vdeltil_{\vmu}$ are vectors close (in some norm) to $\vx,\vs,\vdel_{\vmu}$, and $\vdel_{\mu}$ is a vector function of the gradient of the potential function $f$. Note that we solve these linear systems approximately, but because the error is additive (see \cref{cor:lp-projection-regression}), the total feasibility error of the algorithm can be bounded.

We use a robust IPM in \cref{sec:robust_ipm}, and IPMs based on the $2$-norm in Sections \ref{sec:inverse_free_ipm} and \ref{sec:sparse_ipm}. In \cref{sec:robust_ipm}, the linear systems are solved by multiplication with an inverse initially obtained by divide-and-conquer algorithms and fast matrix multiplication \cite{DemmelDH07,DemmelDHK07}, and maintained by the Woodbury identity under low-rank updates --- see \cref{cor:woodbury-proj-maintenance}. In \cref{sec:inverse_free_ipm}, the linear systems are solved using shifted-number representation \cite{storjohann2005shifted} --- see \cref{thm:storjohann-lin-sys}. In \cref{sec:sparse_ipm}, the linear systems are solved by multiplication by representation of inverses obtained by block Krylov method \cite{eberly2006solving,eberly2007faster,PengVempala21,nie2022matrix}, and maintained by the Woodbury identity under low-rank updates. The running time of linear system solvers in Sections \ref{sec:robust_ipm} and \ref{sec:sparse_ipm} depend on the condition number of the corresponding matrix. Since the modified linear program changes the matrix, we need to argue that its condition number does not blow up compared to the original matrix.

\RestyleAlgo{algoruled}
\begin{algorithm}[t]

\textbf{Assumption:} The linear program has inner radius $r$ and outer radius $R$\\

\textbf{Input:} Full column rank matrix $\ma \in \R^{n\times d}$ and vectors $\vb\in\R^{d},\vc\in\R^n$; Error parameters $0 < \epsilon_1,\epsilon_2 < 1$. \\

\textbf{Output:} $\vxhat\in\Q^{n}_{\geq 0}$ such that $\norm{\ma^\top \vxhat - \vb}_2\leq \epsilon_2 \norm{\vb}_2$ and $\vc^\top \vxhat \leq \min_{\vx:\ma^\top \vx = \vb,\vx\geq 0} \vc^\top \vx + \epsilon_1$. \\

Let $\veps = 1/(100\sqrt{n})$, $\overline{R} = \frac{5}{\veps} R$, $t = 2^{16} \veps^{-3} n^2 \frac{R}{r} \cdot \norm{\vc}_{\infty}R$.\\

Let $\mabar,\vbbar,\vcbar,\vx_{\vc}^+,\vx_{\vc}^-$ be as defined in \cref{def:modified-LP} for the modified linear programming problem.\\

Let $\vx_{\vc}^{\theta} = \overline{R}$, $\vxbar^{(0)} = (\vx_{\vc}^+,\vx_{\vc}^-,\vx_{\vc}^{\theta})$, and $\vsbar^{(0)} = t/\vxbar^{(0)}$\\

Let $(\vxbar^{(\final)},\vsbar^{(\final)}) = \genericipm(\mabar,\vxbar^{(0)},\vsbar^{(0)}, t, \norm{\vc}_{\infty}R, \epsilon_2)$ \label{alg-step:ipm-init-modified-initial}\\

Set $\vx^{(0)}=\vx^+ - \vx^-$ and $\vs^{(0)}=\vs^+ - \vs^-$ where $\vxbar^{(\final)} = (\vx^+,\vx^-,\vx^{\theta})$ and $\vsbar^{(\final)} = (\vs^+,\vs^-,\vs^{\theta})$ \\

Let $(\vx^{(\final)},\vs^{(\final)}) = \genericipm(\ma,\vx^{(0)},\vs^{(0)},\norm{\vc}_{\infty}R, \epsilon /2n,\epsilon_2)$ \label{alg-step:ipm-init-original-initial}.

\caption{Path following interior point method (IPM)}
\label{alg:ipm-init}
\end{algorithm}

\begin{lemma}
Condition number of $\mabar$ (as defined in \cref{def:modified-LP}) is less than $8 \cdot(\kappa(A)+\log(n))^7$.
\end{lemma}
\begin{proof}
First note that the condition number of $\ma$ and $\mb:=\begin{bmatrix}
\ma^\top & - \ma^\top & \vec{0}
\end{bmatrix}^\top$ are the same. Therefore setting $\vg = \begin{bmatrix}
\vec{1}^\top & \vec{0}^\top & 1
\end{bmatrix}^\top$, since $\norm{\vg}_2 \leq \sqrt{n+1}$, and
\[
\norm{(\mi-\mb(\mb^{\top} \mb)^{-1} \mb^\top)\vg}_2 \geq 1,
\]
by \cref{lemma:cond-num-column-add}, the condition number of $\mabar$ is less than $8 \cdot(\kappa(A)+\log(n))^7$.
\end{proof}

For the IPMs that use inverse maintenance (Sections \ref{sec:robust_ipm} and \ref{sec:sparse_ipm}), the bit complexities are analyzed in interaction with the inverse, and any rounding required to prevent the bit complexity of the resulting vectors from growing is done when we apply the inverse to a vector. For the IPM that works with shifted numbers to solve the linear systems (\cref{sec:inverse_free_ipm}), given an integer matrix and vector, the exact solution to the linear system is returned as a rational vector. To be sure that the bit complexities (of rational or real vectors) do not blow up, we need to switch between rational and real (fixed-point) vectors. For this purpose, we define the following functions that can be computed in $\Otil(n \ell)$, where $\ell$ is the bit complexity of the input vector and $q$ or $1/\epsilon$.

\begin{definition}
For a vector $\vx\in\R^n$, and a number $q\in\Q$, we define $\qround(\vx,q)$ to be a vector $\vu \in \Q^n$, where $\vu_i$ is the closest power of $q$ (or the negative of a power of $q$) to $\vx_i$. For a vector $\vx\in\R^n$ or $\vx\in\Q^n$, and a number $\veps\in\R_{>0}$, we define $\round(\vx,\veps)$ to be a fixed-point vector $\vu\in\R^n$, where $\abs{\vu_i - \vx_i} \leq \veps$.
\end{definition}

\noindent
We are now equipped to present our IPMs and analyze their running times in the next sections. Before doing so, we present an example in which the running time of the IPM with shifted numbers (\cref{thm:inverse_free_ipm}) is better than the IPM based on inverse maintenance (\cref{thm:robust_ipm}).
Let
\begin{align*}
\ma = \begin{bmatrix}
1 & 0 & 0 & 0 & \cdots & 0 & 0 & 0 \\
2 & 1 & 0 & 0 & \cdots & 0 & 0 & 0 \\
0 & 2 & 1 & 0 & \cdots & 0 & 0 & 0 \\
0 & 0 & 2 & 1 & \cdots & 0 & 0 & 0 \\
\vdots & \vdots & \vdots & \vdots & \ddots & \vdots & \vdots & \vdots \\
0 & 0 & 0 & 0 & \cdots & 1 & 0 & 0 \\
0 & 0 & 0 & 0 & \cdots & 2 & 1 & 0 \\
0 & 0 & 0 & 0 & \cdots & 0 & 2 & 1 \\
1 & 0 & 0 & 0 & \cdots & 0 & 0 & 0 \\
\end{bmatrix} \in \R^{n\times (n-1)}, \vb = \begin{bmatrix}
4/3 \\ 1 \\ 1 \\ 1 \\ \vdots \\ 1 \\ 1
\end{bmatrix} \in \R^{n-1}, \vc = \begin{bmatrix}
1 \\ 1 \\ 1 \\ 1 \\ \vdots \\ 1 \\ 1
\end{bmatrix} \in \R^n.
\end{align*}

\noindent
Then for the following linear program
\[
\min_{\vx \in \R^{n-1}: \ma^\top \vx = \vb, \vx \geq 0} \vc^\top \vx,
\]
$R \leq 2\sqrt{n}$ because for $\vx\geq 0$ with $\norm{\vx}_2 > 2 \sqrt{n}$, there exists $i\in[n]$ such that $\vx_i > 4$. Then one can see if $i\neq n$, $(\ma^\top \vx)_i > 4 > \vb_i$, and if $i=n$, then $(\ma^\top \vx)_{1} > 4 > \vb_{1}$. Moreover note that for $\vx = \frac{1}{3} \cdot \vc$, we have $\vx \geq 0$ and $\ma^\top \vx = \vb$. Therefore $r>\frac{1}{3}$. Hence $R/r \leq \frac{\sqrt{n}}{3}$. However $\kappa(\ma)$ as discussed in \cref{sec:discussion} is at least $2^{n-2}$ (check the vectors $\begin{bmatrix}
    1 & 0 & 0 & \cdots & 0
\end{bmatrix}$ and $\begin{bmatrix}
    (-1/2)^{n-1} & (-1/2)^{n-2} & \cdots & -1/2 & 1
\end{bmatrix}$ in $\R^{n-1}$ for the largest and smallest singular value, respectively). In this case the running time of \cref{thm:inverse_free_ipm} is $\Otil(n^{\omega+0.5} \log^2(1/\epsilon))$ and the running time of \cref{thm:robust_ipm} is $\Otil(n^{\omega+1} \log(1/\epsilon))$.

\subsection{Robust Interior Point Method For Solving Linear Programs}
\label{sec:robust_ipm}

The main result of this section is the following theorem that is achieved by Algorithm \ref{alg:ipm-robust}.

\RobustIPM*

\noindent
For this result we work with the potential function of the form $\phil(\frac{\vx \vs}{t}-1)$, where 
\[
\phil(\vu) = \sum_{i=1}^n \cosh(\lambda \vu_i) = \sum_{i=1}^n \frac{\exp(\lambda \vu_i)+\exp(-\lambda \vu_i)}{2}.
\]

\noindent
We use the following data structure to maintain the projection matrix $\ma (\ma^\top \mxbar \msbar^{-1}) \ma^\top$ and compute the changes $\vdeltil_{\vs}$ and $\vdeltil_{\vx}$.

\begin{theorem}[Projection maintenance data structure --- $\pds$]
\label{thm:proj-ds}
Let $n\geq d$.
There exists a data structure with the following operations that maintain an explicit matrix $\mz^{-1}$.
\begin{itemize}
    \item $\initialize(\ma\in\R^{n\times d}, \vx\in\R^{n},\vs\in\R^{n},\vr\in\R^{n}, f:\R\rightarrow\R, \omegadualcap, \veps)$: Sets $\vxbar = \vx$, $\vsbar = \vs$, $\vrbar = \vr$, and computes a linear operation $\mzhat$ such that 
    \[
    \norm{\mzhat - \ma(\ma^\top \mxbar \msbar^{-1} \ma)^{-1} \ma^\top}_{\fro} \leq \kappa^2 \cdot \veps.
    \]
    Moreover, sets $\vrtil=\vxtil=\vstil=\vec{0}\in\R^n$, $T=\emptyset$, and sets $\vwtil=\mzhat \msbar^{-1} f(\vrbar)$. 
    
    \item $\update(S\subseteq[n], \vx^{(u)}\in\R^{\abs{S}},\vs^{(u)}\in\R^{\abs{S}},\vr^{(u)}\in\R^{\abs{S}})$: Sets $\vxtil_S=\vx^{(u)} - \vxbar_{S}$, $\vstil_S=\vs^{(u)} - \vsbar_{S}$, $\vrtil_S=\vr^{(u)} - \vrbar_{S}$, and $T=T\cup S$. If $\abs{T}>n^{\omegadualcap}$, 
    sets $\vxbar = \vxbar + \vxtil$, $\vsbar = \vsbar + \vstil$, $\vrbar = \vrbar + \vrtil$, $\vrtil=\vxtil=\vstil=\vec{0}\in\R^n$, $T=\emptyset$, and
    updates $\mzhat$ such that after $k$ calls to $\update$, we have
\[
\norm{\mzhat - \ma(\ma^\top \mxbar \msbar^{-1} \ma) \ma^\top}_{\fro} \leq 50 \kappa^{12} \cdot (512 \cdot k \cdot (5\kappa^5)^{26} + 2 \cdot k \cdot (6\kappa^5)^2  + 1) \cdot \veps,
\]
where $\kappa>n+d$ is a number such that
\[
\norm{\ma}_{\fro},\norm{\mxbar \msbar^{-1}}_{\fro},\norm{\mxbar^{-1} \msbar}_{\fro},\norm{\mxtil \mstil^{-1}}_{\fro},\norm{\mxtil^{-1} \mstil}_{\fro}, \norm{\ma^\top \mxbar \msbar^{-1} \ma}_{\fro}, \norm{(\ma^\top \mxbar \msbar^{-1} \ma)^{-1}}_{\fro} \leq \kappa,
\]
during all the updates and 
\[
\veps < \frac{1}{10\kappa^5 \cdot (512 \cdot (k+1) \cdot (5\kappa^5)^{26} + 2 \cdot k \cdot (6\kappa^5)^2  + 1)}.
\]
After updating $\mzhat$, it sets $\vwtil=\mzhat \msbar^{-1} f(\vrbar)$.

    \item $\query()$: Compute
        $\md^{-1} \in \R^{\abs{T}\times\abs{T}}$ such that  \begin{align*}
        \norm{\md^{-1}-\left(\mxtil_{T,T}^{-1}\mstil_{T,T} + \mzhat_{T,T} \right)^{-1} }_{\fro} \leq \veps.
        \end{align*}
        Then it computes and returns 
    \[
    \vwtil + \mzhat (\msbar^{-1})_T^\top f(\vrtil_T) - \mzhat_{:T} \md^{-1} (\mzhat_{:T})^\top \msbar^{-1} f(\vrbar + \vrtil).
    \]
\end{itemize}
The running time of initialize, update, and query are $n^{\omega} \cdot \Otil(\log(\kappa/\veps))$, $\mtime(n,n,\abs{T}) \cdot \Otil(\log(\kappa/\veps))$, and $(n^{1+\omegadualcap} + n^{\omegadualcap \cdot \omega})\cdot \Otil(\log(\kappa/\veps))$, respectively.
\end{theorem}
\begin{proof}
We show that the data structure in Algorithm \ref{alg:proj-ds} achieves the desired result.
First note that $\mm$ is invertible since $\msbar \mxbar^{-1}$, and $\ma^\top \mxbar \msbar^{-1} \ma$ are invertible and
\[
\mm^{-1} = \begin{bmatrix}
(\ma^\top \mxbar \msbar^{-1} \ma)^{-1} & -(\ma^\top \mxbar \msbar^{-1} \ma)^{-1} \ma^\top \mxbar \msbar^{-1} \\
- \mxbar \msbar^{-1} \ma (\ma^\top \mxbar \msbar^{-1} \ma)^{-1} & \mxbar \msbar^{-1} + \mxbar \msbar^{-1} \ma (\ma^\top \mxbar \msbar^{-1} \ma)^{-1} \ma^\top \mxbar \msbar^{-1}
\end{bmatrix}.
\]
Note that by triangle inequality $\norm{\mm}_{\fro} \leq 3\kappa$, and $\norm{\mm^{-1}}_{\fro} \leq 2\kappa + 2 \kappa^3 + \kappa^5$.
Since $\mz$ is a matrix with $\norm{\mz^{-1} - \mm}_{\fro} \leq \veps$, taking $\mztil$ to be the $n$-by-$n$ bottom right block of $\mz$, then $\mzhat = \msbar \mxbar^{-1} (\mztil - \mxbar \msbar^{-1}) \mxbar^{-1} \msbar$ is a linear operator for $\ma (\ma^\top \mxbar \msbar^{-1} \ma)^{-1} \ma^\top$. 

We now bound the running times. The initialization requires computing the inverse of an $(n+d)\times (n+d)$ matrix with $n\geq d$. The required error bound and the condition number bounds give a running time of $n^{\omega}\cdot \Otil(\log (\kappa/\veps) )$. Then computing $\mzhat$ and $\vwtil$ according to Lines \ref{alg-step:proj-ds:mzhat-init} and \ref{alg-step:proj-ds:vwtil-init} of Algorithm \ref{alg:proj-ds} is done in $\Otil(n^2 \log(\kappa/\veps))$ time since $\mxbar$ and $\msbar$ are diagonal.

We now bound the running time of the update.
If $\abs{T} \leq n^{\hat{\alpha}}$, then the cost is bounded by $\Otil(n \log(\kappa/\veps))$ since we only set new values for entries of vectors according to the input. Otherwise, we update the inverse.
Computing $\md^{-1}$ takes $\abs{T}^{\omega} \cdot \Otil(\log(\kappa/\veps))$ because of the error bound and condition number bounds and the fact that $\matu \mz \matu$ only selects a submatrix of $\mz$. Then computing $\mz \matu^\top \md^{-1} \matu \mz$ takes $\mtime(n,n,\abs{T})\cdot \Otil(\log(\kappa/\veps))$ and having this matrix, we can update $\mz$ in time $ \Otil(n^2\log(\kappa/\veps))$. Finally, for the update, we need to recompute $\vwtil$, which can be done in $\Otil(n^2 \log(\kappa/\veps))$ time, similar to the initialization step.

Since the update procedure ensures that $\abs{T}\leq n^{\omegadualcap}$, computing $\md^{-1}$ in the query procedure takes at most $n^{\omegadualcap \cdot \omega}\cdot \Otil(\log(\kappa/\veps))$. Then computing $\vwtil + \mzhat (\msbar^{-1})_T^\top f(\vrtil_T) - \mzhat_{:T} \md^{-1} (\mzhat_{:T})^\top \msbar^{-1} f(\vrbar + \vrtil)$ according to the query step of Algorithm \ref{alg:proj-ds} takes only $n^{1+\omegadualcap }\cdot \Otil(\log(\kappa/\veps))$ time.
Note that we do not form the matrix $\mzhat$ for this procedure because forming this matrix would impose a cost of $\Omega(n^2)$.

We now need to bound the error of our matrix after $k$ updates.  We prove this by induction. For the base case, note that $\norm{\mz^{-1} - \mm}_{\fro} \leq \veps$.  Therefore since $\norm{\mm}_{\fro},\norm{\mm^{-1}}_{\fro} \leq 5 \kappa^5$, we have
\[
\norm{\mztil - (\mxbar \msbar^{-1} + \mxbar \msbar^{-1} \ma (\ma^\top \mxbar \msbar^{-1} \ma)^{-1} \ma^\top \mxbar \msbar^{-1})}_{\fro} \leq 5 \kappa^5 \cdot \veps.
\]
Therefore
\begin{align*}
\norm{\mzhat - \ma (\ma^\top \mxbar \msbar^{-1} \ma)^{-1} \ma^\top}_{\fro} 
& = 
\norm{\mxbar^{-1} \msbar(\mztil - \mxbar \msbar^{-1}) \mxbar^{-1} \msbar - \ma (\ma^\top \mxbar \msbar^{-1} \ma)^{-1} \ma^\top}_{\fro}
\\ & \leq 
\norm{\mxbar^{-1} \msbar}_{\fro}\norm{\mztil - (\mxbar \msbar^{-1} + \mxbar \msbar^{-1} \ma (\ma^\top \mxbar \msbar^{-1} \ma)^{-1} \ma^\top \mxbar \msbar^{-1})}_{\fro}\norm{\mxbar^{-1} \msbar}_{\fro} 
\\ & \leq 5\kappa^7 \cdot \veps.
\end{align*}
Now suppose after $k$ updates 
\[
\norm{\mz^{-1} - \mm}_{\fro} \leq (512 \cdot k \cdot (5\kappa^5)^{26} + 2 \cdot k \cdot (6\kappa^5)^2 + 1) \cdot \veps,
\]
and $\mxtil \mstil^{-1}$ is the diagonal matrix corresponding to the $k+1$'st update. Let 
\[
\my = \mz - \mz \matu^{\top} \md^{-1} \matu \mz ~~ , \text{with} ~~  \norm{\md^{-1}-\left(\mxtil_{T,T}^{-1}\mstil_{T,T} + \matu \mz \matu^{\top} \right)^{-1} }_{\fro} \leq \veps,
\]
and $\myhat$ is the matrix obtained by rounding the entries of $\my$ to numbers with $\ceil{\log(100 (n+d)^2  \cdot \kappa^5/\veps)}$ bits. \cref{lemma:smw-stability-asymm-fro} directly gives 
\[
\norm{\my^{-1} - \mmhat}_{\fro} \leq (512 \cdot (k+1) \cdot (5\kappa^5)^{26} + 2 \cdot k \cdot (6\kappa^5)^2  + 1) \cdot \veps,
\]
where $\mmhat$ is the matrix $\mm$ after the update, i.e.,
\[
\mmhat = \mm + \begin{bmatrix}
    \textbf{0}_{d\times d} & \textbf{0}_{d\times n} \\
    \textbf{0}_{n\times d} & \mstil \mxtil^{-1}
\end{bmatrix}.
\]
Moreover the norm bound assumptions imply $\norm{\mmhat}_{\fro},\norm{\mmhat^{-1}}_{\fro} \leq 5\kappa^5$, and 
\[
\norm{\my^{-1} - \mmhat}_{\fro} \leq (512 \cdot (k+1) \cdot (5\kappa^5)^{26} + 2 \cdot k \cdot (6\kappa^5)^2  + 1) \cdot \veps < \frac{1}{10\kappa^5},
\]
by \cref{lemma:forward-backward-equiv}, we have $\norm{\my}_{\fro} \leq 10\kappa^5$. Therefore the number of bits before decimal for the entries $\my$ is bounded by $\log_2(10\kappa^5) + 1$. Therefore the rounding only introduces error in the bits after the decimal point. Therefore
\[
\norm{\my - \myhat} \leq \veps.
\]
Invoking \cref{lemma:forward-backward-equiv} and because $\veps < \frac{1}{10\kappa^5}$ (by assumption), $\norm{\my}_{\fro} \leq 10\kappa^5$, and 
\[
\norm{\my^{-1}}_{\fro} \leq \norm{\my^{-1} - \mmhat}_{\fro} + \norm{\mmhat}_{\fro} \leq \frac{1}{10\kappa^5} + 5\kappa^5 \leq 6 \kappa^5,
\]
we have $\norm{\myhat^{-1}-\my^{-1}}_{\fro} \leq 2\cdot(6\kappa^5)^2 \cdot \veps$. Therefore by triangle inequality,
\[
\norm{\myhat^{-1} - \mmhat}_{\fro} \leq \norm{\myhat^{-1} - \my^{-1}}_{\fro} +
\norm{\my^{-1} - \mmhat}_{\fro}
\leq (512 \cdot (k+1) \cdot (5\kappa^5)^{26} + 2 \cdot (k+1) \cdot (6\kappa^5)^2  + 1) \cdot \veps.
\]
Therefore
\[
\norm{\myhat - \mmhat^{-1}}_{\fro} \leq 2\cdot (5\kappa^5)^2 \cdot (512 \cdot (k+1) \cdot (5\kappa^5)^{26} + 2 \cdot (k+1) \cdot (6\kappa^5)^2  + 1) \cdot \veps.
\]
Therefore after the update, we have, 
\[
\norm{\mzhat - \ma(\ma^\top \mxbar \msbar^{-1} \ma) \ma^\top}_{\fro} \leq 2 \kappa^2 \cdot (5\kappa^5)^2 \cdot (512 \cdot (k+1) \cdot (5\kappa^5)^{26} + 2 \cdot (k+1) \cdot (6\kappa^5)^2  + 1) \cdot \veps.
\]
\end{proof}

\RestyleAlgo{algoruled}
\IncMargin{0.15cm}
\begin{algorithm}[!th]

\textbf{Variables:} $n,d\in\N$ with $n>d$, $\ma\in\R^{n\times d}$, $\mz^{-1}\in\R^{d\times d}$, $\vxbar,\vsbar,\vrbar,\vxtil,\vstil,\vrtil\in \R^{n}$, $1>\veps>0$.

 \SetKwProg{pdsInit}{Procedure}{$(\ma\in\R^{n\times d}, \vx\in\R^{n},\vs\in\R^{n},\vr\in\R^{n}, f:\R\rightarrow\R, \omegadualcap, \veps)$}{}
  \pdsInit{\initialize}{
  Set $\ma$, $\veps$, $n$, and $d$ according to the input\\
  Set $\vxbar = \vx$, $\vsbar = \vs$, $\vrbar = \vr$, $\vrtil=\vxtil=\vstil=\vec{0}\in\R^n$, and $T=\emptyset$\\

  Set  
  \begin{align*}
      \mm = \begin{bmatrix}
      \textbf{0}_{d\times d} & \ma \\
      \ma^\top & \msbar \mxbar^{-1}
      \end{bmatrix}
  \end{align*}
  \\
  Compute the matrix $\mz$ such that $\norm{\mz^{-1} - \mm}_{\fro} \leq \veps$ \DontPrintSemicolon\tcp*{\textcolor{blue}{for example using the approach of \cite{DemmelDH07,DemmelDHK07}}}

    Set $\mzhat = \msbar \mxbar^{-1} (\mztil - \mxbar \msbar^{-1}) \mxbar^{-1} \msbar$, where $\mztil$ is the $n\times n$ bottom right block of $\mz$ \label{alg-step:proj-ds:mzhat-init} \\
  Set $\vwtil = \mzhat \msbar^{-1} f(\vrbar)$
  \label{alg-step:proj-ds:vwtil-init}
  }

\SetKwProg{pdsUpdate}{Procedure}{$(S\subseteq[n], \vx^{(u)}\in\R^{\abs{S}},\vs^{(u)}\in\R^{\abs{S}},\vr^{(u)}\in\R^{\abs{S}})$}{}
  \pdsUpdate{\update}{
    Set $\vxtil_S=\vx^{(u)} - \vxbar_{S}$, $\vstil_S=\vs^{(u)} - \vsbar_{S}$, $\vrtil_S=\vr^{(u)} - \vrbar_{S}$\\
    Set $T=T\cup S$ 
    
    \If{$\abs{T}>n^{\omegadualcap}$}{
    Let $\matu = \begin{bmatrix}
        \textbf{0}_{d\times d} & \mi_{T}
    \end{bmatrix} \in \R^{d\times (n+d)}$\\
    
    Update $\mz$ using Woodbury identity
    \begin{align}
    \label{eq:proj-ds-query}
    \mz \leftarrow \mz - \mz \matu^{\top} \md^{-1} \matu \mz,
    \end{align}
    where $\md^{-1} \in \R^{\abs{T}\times\abs{T}}$ is a matrix such that $\norm{\md^{-1}-\left(\mxtil_{T,T}^{-1}\mstil_{T,T} + \matu \mz \matu^{\top} \right)^{-1} }_{\fro} \leq \veps$, \\ Set $\vxbar = \vxbar + \vxtil$, $\vsbar = \vsbar + \vstil$, $\vrbar = \vrbar + \vrtil$, and $\vrtil=\vxtil=\vstil=\vec{0}\in\R^n$, $T=\emptyset$\\
    Round entries of $\mz$ to numbers with $\ceil{\log(100 (n+d)^2  \cdot \kappa^5/\veps)}$ bits\\
    Set $\mzhat = \msbar \mxbar^{-1} (\mztil - \mxbar \msbar^{-1}) \mxbar^{-1} \msbar$, where $\mztil$ is the $n\times n$ bottom right block of $\mz$  \\
  Set $\vwtil = \mzhat \msbar^{-1} f(\vrbar)$
    }
    }

\SetKwProg{pdsQuery}{Procedure}{$()$}{}
  \pdsQuery{\query}{
    Let $\mztil$ be the $n\times n$ bottom right block of $\mz$ \\

    Compute
        $\md^{-1} \in \R^{\abs{T}\times\abs{T}}$ such that  \begin{align*}\norm{\md^{-1}-\left(\mxtil_{T,T}^{-1}\mstil_{T,T} + (\msbar \mxbar^{-1})_{T,T}(\mztil_{T,T} - (\mxbar \msbar^{-1})_{T,T}) (\msbar \mxbar^{-1})_{T,T} \right)^{-1} }_{\fro} \leq \veps\end{align*}\\

    Compute 
    \begin{align*}
        \vh^{(1)} & = \msbar \mxbar^{-1}\left((\mztil (\mxbar^{-1}_T)^\top )  f(\vrtil_T) - (\msbar^{-1}_T)^\top  f(\vrtil_T) \right), \text{ and }
        \\
        \vh^{(2)} & = \md^{-1}\left((\msbar \mxbar^{-1})_{T,T}\left(\mztil_T \mxbar^{-1} f(\vrbar + \vrtil) - (\msbar^{-1})_T f(\vrbar + \vrtil)\right) \right), \text{and} 
        \\
        \vh^{(3)} & = \msbar \mxbar^{-1}\left( (\mztil_T)^\top \left( \left(\msbar \mxbar^{-1} \right)_{T,T}\vh^{(2)} \right) - (\mi_T)^\top\vh^{(2)} \right)
    \end{align*}
    
  Compute and return $\vwtil + \vh^{(1)} - \vh^{(3)}$ 
  }
  
\caption{Prjoection maintenance data structure $(\pds)$}
\label{alg:proj-ds}
\end{algorithm}

\RestyleAlgo{algoruled}
\begin{algorithm}[t]

\textbf{Input:} Full column rank matrix $\ma \in \R^{n\times d}$, initial feasible point $\vx^{(0)}$, slack $\vs^{(0)}$, centrality parameter $t^{(0)}$, final centrality parameter $t^{(\final)}$ all with bit complexity $\ell$, and condition number of $\ma$ less than $\kappa$. Error parameter $0<\epsilon_2<1$ \\

\textbf{Output:} $\vxhat\in\Q^{n}_{\geq 0}$ such that $\ma^\top \vxhat = \vb$ and $\vc^\top \vxhat \leq \min_{\vx:\ma^\top \vx = \vb,\vx\geq 0} \vc^\top \vx + 1.1\sqrt{n} \cdot t^{(\final)}$. \\

Set $\vxbar^{(1)} = \vx^{(0)}$, $\vsbar^{(1)} = \vs^{(0)}$, $\vrbar^{(1)} = \vr^{(0)} = \round( \frac{\vx^{(0)} \odot \vs^{(0)} - \vec{t}^{(0)}}{t^{(0)}}, \frac{1}{n})$, $\lambda = 16 \log 40 n$, $h=1/(128\lambda\sqrt{n})$, and $k=1$\\

$\pds.\initialize(\ma, \vxbar^{(1)},\vsbar^{(1)},\vrbar^{(1)},\nabla\phil, \min\{\omegadual,2/3\}, \frac{\epsilon_2}{10^6 \cdot (\kappa\cdot n)^{30}\cdot \log(t^{(0)}/t^{(\final)})})$

\While{$t^{(k-1)}\geq t^{(\final)}$}{
Set $\vdeltil_{\vs}^{(k)} = - \frac{t}{32\lambda\cdot \norm{\nabla\phil(\vrbar)}_2}\cdot\pds.\query()$ \DontPrintSemicolon\tcp*{$\ma (\ma^\top \mxbar^{(k)} (\msbar^{(k)})^{-1} \ma)^{-1} \ma^\top (\msbar^{(k)})^{-1} \vdeltil_{\vmu}^{(k)}$}
Compute $\vdeltil_{\vx}^{(k)} = - \frac{t}{32\lambda\cdot \norm{\nabla\phil(\vrbar)}_2}\cdot (\msbar^{(k)})^{-1} \nabla\phil(\vrbar) - \mxbar (\msbar^{(k)})^{-1} \vdeltil_{\vs}^{(k)}$\\
Update $\vx^{(k)} = \vx^{(k-1)} + \vdeltil_{\vx}^{(k)}$, $\vs^{(k)} = \vs^{(k-1)} + \vdeltil_{\vs}^{(k)}$, and $t^{(k)}=t^{(k-1)}/(1+h)$\\
Update $\vr^{(k)} = \frac{\vx^{(k)}\odot \vx^{(k)} - \vec{t}^{(k)}}{t^{(k)}}$\\
Let $S=\{i\in[n]: \abs{\log\vxbar_i^{(k)}-\log\vx_i^{(k)}}>\frac{1}{48} \text{ or } \abs{\log\vsbar_i^{(k)}-\log\vs_i^{(k)}}>\frac{1}{48} \text{ or } \abs{\log\vrbar_i^{(k)}-\log\vr_i^{(k)}}>\frac{1}{48 \lambda}\}$\\
$\pds.\update(S,\vx^{(k)}_S,\vs^{(k)}_S,\vr^{(k)}_S)$ \\
Set $k=k+1$\\
Set $\vxbar^{(k)}_S = \vx^{(k-1)}_S$, $\vsbar^{(k)}_S = \vs^{(k-1)}_S$, and $\vrbar^{(k)}_S = \vr^{(k-1)}_S$\\
Set $\vxbar^{(k)}_{[n]\setminus S} = \vxbar^{(k-1)}_{[n]\setminus S}$, $\vsbar^{(k)}_{[n]\setminus S} = \vsbar^{(k-1)}_{[n]\setminus S}$, and $\vrbar^{(k)}_{[n]\setminus S} = \vrbar^{(k-1)}_{[n]\setminus S}$
}
\Return $(\vx^{(k-1)},\vs^{(k-1)})$

\caption{Robust interior point method (IPM)}
\label{alg:ipm-robust}
\end{algorithm}

The robust interior point method converges if $\norm{(\vx \vs - \vt)/\vt}_{\infty}$, $\norm{\vdeltil_{\vx}/\vx}_2$, and $\norm{\vdeltil_{\vs}/\vs}_2$ are small throughout the algorithm. Here we argue that if we set the error parameters for solving the linear systems corresponding to iterations of robust IPM, then these quantities stay small. We first provide bounds for these for exact solves.

\begin{lemma}[\cite{lee2021tutorial}]
\label{lemma:robust-ipm-norm-bounds}
Let $\vxbar,\vsbar$ be vectors with $\norm{\log\vxbar-\log\vx}_{\infty} \leq \frac{1}{48}$, $\norm{\log\vsbar-\log\vs}_{\infty} \leq \frac{1}{48}$, $\mw = \mxbar \msbar^{-1}$, and $\matp = \sqrt{\mw} (\ma^\top \mw \ma)^{-1} \sqrt{\mw}$. Moreover let
\[
\vdel_{\vx}:= \frac{\mxbar}{\sqrt{\mxbar \msbar}} (\mi - \matp) \frac{\mi}{\sqrt{\mxbar \msbar}} \vdeltil_{\vmu}, ~~ \text{and} ~~\vdel_{\vs}:= \frac{\msbar}{\sqrt{\mxbar \msbar}}  \matp \frac{\mi}{\sqrt{\mxbar \msbar}} \vdeltil_{\vmu},
\]
where $\vdeltil_{\vmu} = - \frac{t \cdot \nabla\phil(\vrbar)}{32\lambda\cdot \norm{\nabla\phil(\vrbar)}_2}$, $\vrbar$ is a vector with $\norm{\vrbar - \frac{\vx \odot \vs - t}{t}}_{\infty} \leq \frac{1}{48 \lambda}$, and $\lambda = 16 \log 40 n$. Then under the invariant $\phil(\frac{\vx \odot \vs - t}{t}) \leq 16n$,
\[
\norm{\frac{\vx \odot \vs - t}{t}}_{\infty} \leq \frac{1}{16}, ~~ \text{and} ~~ \norm{\vdel_{\vx}/\vxbar}_2 \leq \frac{1}{20 \lambda}, ~~ \text{and} ~~ \norm{\vdel_{\vs}/\vsbar}_2 \leq \frac{1}{20 \lambda}.
\]
\end{lemma}

The bound $\norm{\frac{\vx \odot \vs - t}{t}}_{\infty} \leq \frac{1}{16}$ on the above lemma directly follows from $\phil(\frac{\vx \odot \vs - t}{t}) \leq 16n$ and does not depend on the computation of $\vdel_{\vx}$ and $\vdel_{\vs}$. 

\begin{remark}
\label{remark:robust-ipm-change-bound}
The upper bounds stated in \cite{lee2021tutorial} for $\norm{\vdel_{\vx}/\vx}_2$ and $\norm{\vdel_{\vs}/\vs}_2$ is $1/(16\lambda)$, but it can easily be strengthened to the bounds we stated above with the same argument.
Now note that instead of $\vdel_{\vx}$ and $\vdel_{\vs}$ we compute $\vdeltil_{\vx}$ and $\vdeltil_{\vs}$ by \cref{cor:lp-projection-regression}. Note that we use the matrix itself as the preconditioner and therefore, we only take one step by \cref{cor:lp-projection-regression}. This gives $\vutil$ such that $\norm{\vutil - \vu}_2 \leq \veps \norm{\vu}_2$, where $\vu = \matp \frac{\mi}{\sqrt{\mxbar \msbar}} \vdeltil_{\vmu}$. Therefore
\begin{align*}
\norm{\frac{\vdeltil_{\vx}}{\vxbar}}_2 
& \leq 
\frac{1}{20 \lambda} + \norm{\frac{\vdeltil_{\vx} - \vdel_{\vx}}{\vxbar}}_2 = \frac{1}{20 \lambda} + \norm{\frac{\mi}{\sqrt{\mxbar \msbar}} (\vutil - \vu)}_2 \leq \frac{1}{20 \lambda} + \frac{\veps}{\min_{i\in[n]} \sqrt{\vxbar_i \vsbar_i}}\norm{\vu}_2
\\ & \leq 
\frac{1}{20 \lambda} +\frac{\veps \cdot \max_{i\in[n]} \sqrt{\vxbar_i \vsbar_i} }{\min_{i\in[n]} \sqrt{\vxbar_i \vsbar_i}}\norm{\frac{\mi}{\sqrt{\mxbar \msbar}}\vu}_2 = \frac{1}{20 \lambda} +\frac{\veps \cdot \max_{i\in[n]} \sqrt{\vxbar_i \vsbar_i} }{\min_{i\in[n]} \sqrt{\vxbar_i \vsbar_i}} \norm{\frac{\vdel_{\vs}}{\vsbar}}_2
\\ & \leq
\frac{1}{20 \lambda} +\frac{\veps \cdot \max_{i\in[n]} \sqrt{\vxbar_i \vsbar_i} }{20\lambda\cdot \min_{i\in[n]} \sqrt{\vxbar_i \vsbar_i}}.
\end{align*}
A similar argument gives the same bound for $\norm{\vdeltil_{\vs}/\vsbar}_2$. Note that since entries of $\vx \odot \vs$ are close to $t$ and $\vxbar$ and $\vsbar$ are close to $\vx$ and $\vs$, respectively, we can take $\veps=\Omega(t^{(0)}/t^{(\final)})$, so that $\norm{\vdeltil_{\vs}/\vsbar}_2 \leq \frac{1}{16\lambda}$ and $\norm{\vdeltil_{\vx}/\vxbar}_2 \leq \frac{1}{16\lambda}$.
\end{remark}

\begin{remark}
\label{remark:robust-ipm-error-mu-correct}
Note that although we compute a vector $\vutil$ using \cref{cor:lp-projection-regression} and use that to compute $\vdeltil_{\vx}$ and $\vdeltil_{\vs}$ instead of using $\vu = \matp \frac{\mi}{\sqrt{\mxbar \msbar}} \vdeltil_{\vmu}$ to compute them, we still have $\msbar \vdeltil_{\vx} + \mxbar \vdeltil_{\vs} = \vdeltil_{\vmu}$ because 
\[
\msbar \vdeltil_{\vx} + \mxbar \vdeltil_{\vs} = \vdeltil_{\vmu} - \sqrt{\mxbar \msbar} \vutil + \sqrt{\mxbar \msbar} \vutil = \vdeltil_{\vmu}
\]
\end{remark}

The following is a combination of Lemma 16 and 18 of \cite{lee2021tutorial} that essentially follows from \cref{remark:robust-ipm-change-bound,remark:robust-ipm-error-mu-correct} by the same proof.

\begin{lemma}[\cite{lee2021tutorial}]
\label{lemma:robust-ipm-convergence}
Let $\lambda=16\log 40 n$, $\vt^{(0)} \in \R_{> 0}$, and $\vx^{(0)},\vs^{(0)}\in\R^n$ such that $\phil(\frac{\vx^{(0)} \odot \vs^{(0)} - t^{(0)}}{t^{(0)}}) \leq 16 n$. Moreover for $k\in\N$, let $\vx^{(k)} = \vx^{(k-1)} + \vdeltil_{\vx}^{(k)}$ and $\vs^{(k)} = \vs^{(k-1)} + \vdeltil_{\vs}^{(k)}$ be computed by an iteration of robust IPM (Algorithm \ref{alg:ipm-robust}) such that $\norm{\vdeltil_{\vs}^{(k)}/\vsbar^{(k-1)}}_2 \leq \frac{1}{16\lambda}$ and $\norm{\vdeltil_{\vx}^{(k)}/\vxbar^{(k-1)}}_2 \leq \frac{1}{16\lambda}$ where $\vxbar^{(k)}$, $\vsbar^{(k)}$ are vectors that satisfy $\norm{\log\vxbar^{(k)}-\log\vx^{(k)}}_{\infty} \leq \frac{1}{48}$, $\norm{\log\vsbar^{(k)}-\log\vs^{(k)}}_{\infty} \leq \frac{1}{48}$. Then for $\vr^{(k)}:= \frac{\vx^{(0)} \odot \vs^{(0)} - t^{(0)}}{t^{(0)}}$, $\norm{\vr^{(k+1)}-\vr^{(k)}}_2\leq \frac{1}{16\lambda}$. Moreover
$\norm{\log\vx^{(k+1)} - \log\vx^{(k)}}_2,\norm{\log\vs^{(k+1)} - \log\vs^{(k)}}_2 \leq \frac{1}{8\lambda}$.
In addition $\phil(\vr^{(k+1)}) \leq 12 n$ if $\phil(\vr^{(k)}) \leq 8 n$, and $\phil(\vr^{(k+1)}) \leq \phil(\vr^{(k)})$, otherwise.
\end{lemma}

The next lemma is useful for bounding the running time of inverse maintenance in Algorithm \ref{alg:ipm-robust}.

\begin{lemma}[\cite{lee2021tutorial}]
\label{lemma:robust-ipm-num-change}
Let $\vecv^{(0)},\vecv^{(1)},\vecv^{(2)},\ldots$ be vectors in $\R^{n}$ arriving in a stream with $\norm{\vecv^{(k+1)}-\vecv^{(k)}}_2 \leq \beta$ for all $k$. Then for $0<C<0.5$, we can pick $\vecvbar^{(0)},\vecvbar^{(1)},\vecvbar^{(2)},\ldots$, so that (see Algorithm 4 on \cite{lee2021tutorial})
\begin{itemize}
    \item $\norm{\vecvbar^{(k)}-\vecv^{(k)}}_{\infty} \leq C$ for all $k$.
    \item $\norm{\vecvbar^{(k)}-\vecvbar^{(k-1)}}_0 \leq O(2^{2q_k}(\beta/C)^2 \log^2(n))$ where $q_k$ is the largest integer with $k = 0 \mod 2^{q_k}$.
\end{itemize}
\end{lemma}

We are now equipped to prove our main result regarding the bit complexity of solving LPs.

\begin{proof}[Proof of \cref{thm:robust_ipm}]
We prove that Algorithm \ref{alg:ipm-robust} converges and outputs a near feasible solution, and we analyze the running time and bit complexity of this algorithm. Then this is combined with Algorithm \ref{alg:ipm-init} and \cref{lemma:init-feasible} (for finding the initial feasible solution) to give the desired result.

Note that by \cref{lemma:robust-ipm-norm-bounds,lemma:robust-ipm-convergence}, we have 
\[
\norm{\frac{\vxhat \odot \vshat - \widehat{t}}{\widehat{t}}}_{\infty} \leq \frac{1}{16},
\]
where $\vxhat,\vshat$ are the output of Algorithm \ref{alg:ipm-robust} and $\widehat{t}>0$ is a number smaller than $t^{(\final)}$. Therefore
\[
\vc^\top \vxhat \leq \min_{\vx:\ma^\top \vx = \vb, \vx \geq 0} \vc^\top \vx + \frac{n \cdot \widehat{t}}{16}.
\]
Taking $t^{(\final)}$ to be small enough, we can guarantee an upper bound on the error. We now discuss the feasibility of the returned solution. First, note that $\vxhat \geq 0$ by induction through the guarantee of \cref{remark:robust-ipm-change-bound}. Moreover in each iteration of Algorithm \ref{alg:ipm-robust}, we compute $\vdeltil_{\vx}$ as 
\[
\vdeltil_{\vx} =  \frac{\mi}{\msbar} \vdeltil_{\vmu} -\frac{\mxbar}{\sqrt{\mxbar \msbar}} \vutil,
\]
where $\vutil$ is a vector with $\norm{\vutil-\vu}_2 \leq \veps \norm{\vu}_2$ and $\vu=\matp \frac{\mi}{\sqrt{\mxbar \msbar}} \vdeltil_{\vmu}$. Therefore by \cref{cor:lp-projection-regression}, and the bounds on the condition number of $\mabar/\msbar$ and $\ma$,
\[
\norm{\ma^\top \vdeltil_{\vx}}_2 \leq \veps \cdot \kappa \cdot \frac{R}{r} \norm{\vu}_2 \leq \veps \cdot \kappa \cdot \frac{R}{r} \norm{\frac{\mi}{\sqrt{\mxbar \msbar}} \vdeltil_{\vmu}}_2,
\]
where the last inequality follows because $\matp$ is a projection matrix. Now since $\vdeltil_{\vmu} = - \frac{t \cdot \nabla\phil(\vrbar)}{32\lambda\cdot \norm{\nabla\phil(\vrbar)}_2}$ and by \cref{lemma:robust-ipm-norm-bounds} $\norm{\vrbar - \frac{\vx \odot \vs - t}{t}}_{\infty} \leq \frac{1}{48 \lambda}$, and $\norm{\frac{\vx \odot \vs - t}{t}}_{\infty}  \leq \frac{1}{16}$, $\norm{\frac{\mi}{\sqrt{\mxbar \msbar}} \vdeltil_{\vmu}}_2$ is bounded by $\poly(n)$. Therefore setting $\veps = \frac{\eps}{T \cdot \kappa \cdot \frac{R}{r} \norm{\frac{\mi}{\sqrt{\mxbar \msbar}} \vdeltil_{\vmu}}_2}$, where $T$ is the number of iterations of the algorithm, by triangle inequality we have the guarantee that $\norm{\ma^\top \vxhat - \vb}_2 \leq \epsilon$.

We now bound the running time of the algorithm. Based on the errors we discussed above and \cref{cor:lp-projection-regression} and \cref{remark:richardson-bit-complexity}, we need to take the bit complexity of $\Otil(\log(\frac{\kappa \cdot R}{\epsilon \cdot r}))$ for our inverses. By picking the right constants according to \cref{lemma:smw-stability-asymm-fro}, we can guarantee the stability of inverse maintenance and the data structure used in Algorithm \ref{alg:ipm-robust}.
By \cref{thm:proj-ds}, the data structure is initialized in time $\Otil(n^{\omega} \log(\frac{\kappa\cdot R}{\epsilon\cdot r}))$.

By construction of Algorithms \ref{alg:ipm-init} and \ref{alg:ipm-robust}, the number of iterations of our IPM is $\Otil(n^{0.5} \log(\frac{R}{\epsilon\cdot r}))$. By \cref{lemma:robust-ipm-convergence,lemma:robust-ipm-num-change}, for each $q_k$, there are $\Otil(\frac{n^{0.5}}{2^{q_k}} \log(\frac{R}{\epsilon\cdot r}))$ iterations where $\Otil(2^{2q_k})$ entries of our vectors $\vxbar,\vsbar,\vrbar$ change. Then by \cref{thm:proj-ds}, the running time of inverse maintenance is bounded by 
\[
\sum_{q_k : 2^{2q_k} > n^{\widehat{\alpha}}} \Otil \left( \frac{n^{0.5}}{2^{q_k}} \log(\frac{R}{\epsilon\cdot r}) \cdot  \mtime(n,n,2^{2q_k}) \log(\frac{\kappa\cdot R}{\epsilon\cdot r})\right),
\]
where $\widehat{\alpha}=\min\{\alpha,2/3\}$ and $\alpha$ is the dual of the matrix multiplication exponent. By \cref{lemma:rectangular-matrix-mult}, this is
\[
\sum_{q_k : 2^{2q_k} > n^{\widehat{\alpha}}} \Otil \left( \frac{n^{0.5}}{2^{q_k}} \log(\frac{R}{\epsilon\cdot r}) \cdot  \left( n^2 + (2^{q_k})^{2(\omega-2)/(1-\alpha)} n^{2-\alpha\cdot (\omega-2)/(1-\alpha)} \right)\log(\frac{\kappa\cdot R}{\epsilon\cdot r})\right),
\]
Since $2^{2q_k} > n^{\widehat{\alpha}}$, $\frac{n^{2.5}}{2^{q_k}}\leq \max\{n^{2.5-\alpha/2}, n^{2+1/6}\}$. Moreover the term 
\[
n^{2.5-\alpha\cdot (\omega-2)/(1-\alpha)} \cdot (2^{q_k})^{2(\omega-2)/(1-\alpha) - 1}
\]
is either increasing or decreasing in $q_k$ depending on whether $2(\omega-2)/(1-\alpha) - 1$ is positive or negative. If $2(\omega-2)/(1-\alpha) - 1>0$, the maximum happens for $2^{q_k}=n^{0.5}$ in which case 
\[
n^{2.5-\alpha\cdot (\omega-2)/(1-\alpha)} \cdot (2^{q_k})^{2(\omega-2)/(1-\alpha) - 1}=n^{\omega}.
\]
If $2(\omega-2)/(1-\alpha) - 1\leq 0$, the maximum happens for $2^{q_k}=n^{\widehat{\alpha}/2}$ in which case
\[
n^{2.5-\alpha\cdot (\omega-2)/(1-\alpha)} \cdot (2^{q_k})^{2(\omega-2)/(1-\alpha) - 1}=n^{2.5-\frac{\alpha\cdot(\omega-2)}{1-\alpha} + \frac{\widehat{\alpha}\cdot(\omega-2)}{1-\alpha} - \frac{\widehat{\alpha}}{2}}.
\]
If $\widehat{\alpha}=\alpha$, then this is bounded by $n^{2.5-\alpha/2}$. Otherwise, $\widehat{\alpha}<\alpha$ and this is bounded by $n^{2+1/6}$. We finally need to bound the running time of queries to the data structure which by \cref{thm:proj-ds} is 
\[
\Otil(n^{0.5} \log(\frac{R}{\epsilon\cdot r}) (n^{1+\widehat{\alpha}} + n^{\widehat{\alpha}\cdot \omega}) \cdot \log(\frac{\kappa \cdot R}{\epsilon\cdot r})),
\]
which since $\widehat{\alpha}\leq \frac{2}{3}$ is bounded by
\[
\Otil((n^{2+1/6} + n^{0.5+2 \omega/3})  \cdot \log(\frac{R}{\epsilon\cdot r}) \cdot \log(\frac{\kappa \cdot R}{\epsilon\cdot r})).
\]
Now since $\omega\geq 2$, $0.5 \leq \omega/4$ and therefore $0.5+2\omega/3 \leq 11\omega/12 < \omega$. Combining the above running times gives the desired result.
\end{proof}
\subsection{Inverse Free Path Following IPM}
\label{sec:inverse_free_ipm}

The main result of this section is the following theorem that is achieved by substituting Algorithm \ref{alg:ipm-simple-no-maintenance} in Steps \ref{alg-step:ipm-init-modified-initial} and \ref{alg-step:ipm-init-original-initial} of Algorithm \ref{alg:ipm-init}. For this we use a version of IPM that works with a $2$-norm potential function, uses $\vxbar$, and $\vsbar$ that have a relative error of $0.01$ compared to $\vx$ and $\vs$, and uses a $\vdeltil_{\vmu}$ that has a relative error of $0.001$ compared to $\vdel_{\vmu}$.

\thmInverseFreeIPM*

To solve the linear system corresponding to each iteration of the IPM, we use the following result that solves an integer linear system in the true matrix multiplication time times the bit complexity of the input (without paying $\log(\kappa)$ in the running time).

\begin{theorem}[\cite{storjohann2005shifted}]
\label{thm:storjohann-lin-sys}
Let $\mm\in\Z^{n\times n}$ be an invertible matrix and $\vb\in\Z^{n}$. There exists a Las Vegas algorithm that returns $\mm^{-1}\vb\in\Q^{n}$ with a expected cost of $O(n^{\omega}\cdot (\log n)\cdot (\log \vertiii{\mm}_{\infty}+\frac{\log\norm{\vb}_{\infty}}{n}+\log n)\cdot C^2)$, where $C = \log((\log \vertiii{\mm}_{\infty}+\frac{\log\norm{\vb}_{\infty}}{n}+\log n))$.
\end{theorem}

\noindent
The above theorem uses Cramer's rule to compute the solution of a linear system.

\begin{fact}[Cramer's rule]
Let $\mm$ be a nonsingular $n$-by-$n$ matrix, $\vb$ be an $n$-vector, and $\vx$ be a solution to the linear system $\mm \vx = \vb$. Then $\vx_i = \frac{\det(\mm_i)}{\det(\mm)}$, where $\mm_i$ is the matrix obtained by replacing column $i$ of $\mm$ with $\vb$.
\end{fact}

Therefore given an integer linear system, the algorithm of \cref{thm:storjohann-lin-sys} returns the solution of $\mm \vx = \vb$ as a rational number where the bit complexity of the denominators is $O(n \log \vertiii{\mm}_{\infty})$, and the bit complexity of the numerators is $O(n \log \vertiii{\mm}_{\infty}+\log\norm{\vb}_{\infty})$. Note that since the entries of $\vx$ have a common denominator of $\det(\mm)$, the vector $\det(\mm) \vx$ is an integer vector and its bit complexity (up to constant factors) is the same as the bit complexity of $\vx$, i.e., $O(n \log \vertiii{\mm}_{\infty}+\log\norm{\vb}_{\infty})$. Therefore the solution of the $\mm \vx = \vb$ can be used for the right-hand side of another linear system without increasing the running time solving the linear system --- note that the bit complexity of $\vb$ in \cref{thm:storjohann-lin-sys} appears with a factor of $1/n$ in the running time. 

If the right-hand side of the linear system does not have common denominators, then turning the right hand side to an integer vector can increase the bit complexity by $n$ times the bit complexity of the denominators. Thus if the bit complexity of the denominators is $n$, this leads to a bit complexity of $n^2$ which leads to increasing the running time of solving the linear system to $n^{\omega+1}$. To avoid this, we make sure that the right-hand side of the linear systems we solve have common denominators. This is achieved by rounding the vectors $\vx,\vs,\vdel_{\vmu}$ to rational vectors $\vxbar, \vsbar, \vdeltil_{\vmu}$ close to them. 

To guarantee that the right-hand side of linear systems we solve have common denominators, in each iteration,
instead of using $\vx,\vs$, and $\vdel_{\vmu}$ in the steps of IPM, we use $\vxbar,\vsbar,\vdeltil_{\vmu}$, where entries of $\vxbar,\vsbar$ are the closest powers of $\frac{101}{100}$ to entries of $\vx,\vs$, and entries of $\vdeltil_{\vmu}$ are the closest powers of $\frac{1001}{1000}$ to entries of $\vdel_{\vmu}$. In this section we show the IPM still converges with these vectors. Therefore, we make updates by solving the following linear system
\begin{align}
\label{eq:lin-sys-change}
\mxbar \vdeltil_{\vs} + \msbar \vdeltil_{\vx} & = \vdeltil_{\vmu}, \\
\ma^\top \vdeltil_{\vx} & = 0, \nonumber \\
\ma \vdeltil_{\vy} + \vdeltil_{\vs} & = 0. \nonumber
\end{align}
\begin{lemma}
The solution of \eqref{eq:lin-sys-change} is
\[
\vdeltil_{\vs} = \ma (\ma^\top \mxbar \msbar^{-1} \ma)^{-1} \ma^\top \msbar^{-1} \vdeltil_{\vmu} ~~ \text{, and } ~~
\vdeltil_{\vx} = \msbar^{-1} \vdeltil_{\vmu} - \mxbar \msbar^{-1} \vdeltil_{\vs}.\]
\end{lemma}
\begin{proof}
First note that since $\mxbar \vdeltil_{\vs} + \msbar \vdeltil_{\vx} = \vdeltil_{\vmu}$, $\msbar \vdeltil_{\vx} = \vdeltil_{\vmu} - \mxbar \vdeltil_{\vs}$. Therefore multiplying $\msbar^{-1}$ from left and using the fact that diagonal matrices commute with each other,
\[
\vdeltil_{\vx} = \msbar^{-1} \vdeltil_{\vmu} - \mxbar \msbar^{-1} \vdeltil_{\vs}.
\]
Therefore since $\ma^\top \vdeltil_{\vx} = 0$,
\[
\ma^\top \mxbar \msbar^{-1} \vdeltil_{\vs} = \ma^\top \msbar^{-1} \vdeltil_{\vmu}.
\]
Now since $\ma \vdeltil_{\vy} + \vdeltil_{\vs} = 0$,
\[
-\ma^\top \mxbar \msbar^{-1} \ma \vdeltil_{\vy} = \ma^\top \msbar^{-1} \vdeltil_{\vmu}.
\]
Now multiplying $\ma(\ma^\top \mxbar \msbar^{-1} \ma)^{-1}$ from left, we have
\[
-\ma \vdeltil_{\vy} =\ma(\ma^\top \mxbar \msbar^{-1} \ma)^{-1} \ma^\top \msbar^{-1} \vdeltil_{\vmu}.
\]
Finally the result follows by noting that $\vdeltil_{\vs} = -\ma \vdeltil_{\vy}$.
\end{proof}

\RestyleAlgo{algoruled}
\IncMargin{0.15cm}
\begin{algorithm}[t]

\textbf{Input:} Full column rank matrix $\ma \in \R^{n\times d}$, initial feasible point $\vx^{(0)}$, slack $\vs^{(0)}$, centrality parameter $t^{(0)}$, final centrality parameter $t^{(\final)}$ all with bit complexity $\ell$. \\

\textbf{Output:} $\vxhat\in\Q^{n}_{\geq 0}$ such that $\ma^\top \vxhat = \vb$ and $\vc^\top \vxhat \leq \min_{\vx:\ma^\top \vx = \vb,\vx\geq 0} \vc^\top \vx + 1.1\sqrt{n} \cdot t^{(\final)}$. \\

Set $\mu^{(0)} = \mx^{(0)} \vs^{(0)}$, $\beta=10^4$, and $k=1$\\

\While{$t^{(k-1)}\geq t^{(\final)}$}{
Set $\vdel_{\vmu}^{(k)} = - 0.5 (\vmu^{(k-1)}-\vec{t}^{(k-1)})$\\
Set $\vxbar^{(k)}=\qround(\vx^{(k)},\frac{101}{100}),\vsbar^{(k)}=\qround(\vs^{(k)},\frac{101}{100}),\vdeltil_{\vmu}^{(k)}=\qround(\vdel_{\vmu}^{(k)},\frac{1001}{1000})$\\
Compute $\vdeltil_{\vs}^{(k)} = \ma (\ma^\top \mxbar^{(k)} (\msbar^{(k)})^{-1} \ma)^{-1} \ma^\top (\msbar^{(k)})^{-1} \vdeltil_{\vmu}^{(k)}$ using \cref{thm:storjohann-lin-sys}\\
Compute $\vdeltil_{\vx}^{(k)} = (\msbar^{(k)})^{-1} \vdeltil_{\vmu} - \mxbar^{(k)} (\msbar^{(k)})^{-1} \vdeltil_{\vs}^{(k)}$\\
Update $\vx^{(k)} = \vx^{(k-1)} + \round(\vdeltil_{\vx}^{(k)},\exp{(-4n)})$, $\vs^{(k)} = \vs^{(k-1)} + \round(\vdeltil_{\vs}^{(k)},\exp{(-4n)})$, and $t^{(k)}=\round((1-\frac{1}{\beta\sqrt{n}}) t^{(k-1)},\exp{(-4n)})$\\
For all $i \in [n]$ such that $\vx^{(k)}_i \leq \exp(-3n)$, set $\vx^{(k)}_i = 0$, remove its corresponding entry in $\vc$ and corresponding row in $\ma$\\
Set $k=k+1$\\
}
\Return $(\vx^{(k-1)},\vs^{(k-1)})$

\caption{$\ifltwoipm$ - Inverse Free path following interior point method (IPM)}
\label{alg:ipm-simple-no-maintenance}
\end{algorithm}

\begin{remark}
For $i\in[n]$, we define $\vxbar_i$ and $\vsbar_i$ as the closest power of $\frac{101}{100}$ to $\vx_i$ and $\vs_i$, respectively. Then $\norm{\frac{\vxbar-\vx}{\vx}}_{\infty},\norm{\frac{\vsbar-\vs}{\vs}}_{\infty}\leq 0.01$.
\end{remark}

For the rest of the section, we prove the convergence of IPM when we use the approximate values for taking a step. This is achieved by showing that: 1) $\vx$ and $\vs$ stay positive through the whole course of the algorithm (see \cref{lemma:xs-stay-pos-storjohann}); and 2) $\norm{(\vx \odot \vs - \vec{t})/t}$ stays less than or equal to $0.01$ (see \cref{lemma:xs-stay-close-t-storjohann}). The latter implies that when $t$ is small enough, we are close to the optimal objective value.

\begin{lemma}
\label{lemma:xs-stay-pos-storjohann}
Let $t>0$, $\vmu = \vx \odot \vs$, $\vdel_{\vmu} = -h(\mu - t)$, $\norm{\frac{\vxbar-\vx}{\vx}}_{\infty}\leq 0.01,\norm{\frac{\vsbar-\vs}{\vs}}_{\infty}\leq 0.01$, $\norm{\frac{\vdeltil_{\vmu}-\vdel_{\mu}}{\vdel_{\mu}}}_{\infty} \leq 0.01$ and
\[
\vdeltil_{\vs} = \ma (\ma^\top \mxbar \msbar^{-1} \ma)^{-1} \ma^\top \msbar^{-1} \vdeltil_{\vmu} ~~ \text{, and } ~~
\vdeltil_{\vx} = \msbar^{-1} \vdeltil_{\vmu} - \mxbar \msbar^{-1} \vdeltil_{\vs}.\]
Moreover suppose $\norm{\frac{\vmu-\vec{t}}{t}}_2 \leq 0.01$.
Then
$\norm{\mx^{-1} \vdeltil_{\vx}}_2,\norm{\ms^{-1} \vdeltil_{\vs}}_2 \leq 0.15 h$.
\end{lemma}
\begin{proof}
First note that
\[
\norm{(\vx \odot \vs - \vec{t})/t}_{\infty} \leq \norm{(\vx \odot \vs - \vec{t})/t}_{2} \leq 0.01. 
\]
Therefore $0.99 \cdot t\leq \norm{\vx \odot \vs}_{\infty}\leq 1.01 \cdot t$.
Moreover since $\norm{\frac{\vxbar-\vx}{\vx}}_{\infty},\norm{\frac{\vsbar-\vs}{\vs}}_{\infty}\leq 0.01$, for all $i\in[n]$, $0.99 \vx_i\leq \vxbar_{i}\leq 1.01 \vx_i$, and $0.99 \vs_{i}\leq \vsbar_{i}\leq 1.01 \vs_{i}$. Thus
\[
\frac{99^3}{100^3} t \leq \norm{\vxbar \odot \vsbar}_{\infty} \leq \frac{101^3}{100^3} t.
\]
Denoting $\mpbar:=\mxbar^{1/2}\msbar^{-1/2} \ma (\ma^\top \mxbar \msbar^{-1} \ma)^{-1} \ma^\top \msbar^{-1/2} \mxbar^{1/2}$, we have
\[
\msbar^{-1} \vdeltil_{\vs} = \mxbar^{-1/2}\msbar^{-1/2} \mpbar \mxbar^{-1/2}\msbar^{-1/2} \vdeltil_{\vmu}.
\]
Now since $\mpbar$ is an orthogonal projection matrix, we have
\begin{align*}
\norm{\msbar^{-1} \vdeltil_{\vs}}_2 & = \norm{\mxbar^{-1/2}\msbar^{-1/2} \mpbar \mxbar^{-1/2}\msbar^{-1/2} \vdeltil_{\vmu}}_2
\\ & \leq
\norm{\vxbar^{-1/2} \odot \vsbar^{-1/2}}_{\infty} \norm{\mpbar \mxbar^{-1/2}\msbar^{-1/2} \vdeltil_{\vmu}}_2
\\ & \leq \frac{100^{1.5}}{99^{1.5}} \cdot \frac{1}{\sqrt{t}} \norm{\mpbar \mxbar^{-1/2}\msbar^{-1/2} \vdeltil_{\vmu}}_2
\\ & \leq \frac{100^{1.5}}{99^{1.5}} \cdot \frac{1}{\sqrt{t}} \norm{ \mxbar^{-1/2}\msbar^{-1/2} \vdeltil_{\vmu}}_2
\\ & \leq \frac{100^3}{99^3} \cdot \frac{1}{t} \norm{ \vdeltil_{\vmu}}_2.
\end{align*}
Now we have
\[
\norm{\ms^{-1} \vdeltil_{\vs}}_2 \leq \frac{100}{99} \norm{\msbar^{-1} \vdeltil_{\vs}}_2 \leq \frac{100^4}{99^4} \cdot \frac{1}{t} \norm{ \vdeltil_{\vmu}}_2 \leq \frac{100^4}{99^4} \cdot  \frac{101}{100} \cdot \frac{1}{t} \norm{ \vdel_{\vmu}}_2
\]
Now since $\vdeltil_{\vmu} = -h(\vmu-\vec{t})$ and $\norm{(\vmu-\vec{t})/t}_2\leq 0.1$, we have $\norm{\ms^{-1} \vdeltil_{\vs}}_2 \leq 0.05 h$.
For $\mx^{-1} \vdel_{\vx}$, we similarly have
\[
\mxbar^{-1} \vdeltil_{\vx} = \mxbar^{-1/2}\msbar^{-1/2} (\mi - \mpbar) \mxbar^{-1/2}\msbar^{-1/2} \vdeltil_{\vmu}
\]
Since $(\mi - \mpbar)$ is also a projection matrix, by a similar argument as above, we have
\begin{align*}
\norm{\mx^{-1} \vdeltil_{\vx}}_2 \leq 0.05 h.
\end{align*}
\end{proof}
\begin{remark}
Let $h\leq 1$.
Then since $\norm{\mx^{-1} \vdeltil_{\vx}}_{\infty} \leq \norm{\mx^{-1} \vdeltil_{\vx}}_{2} \leq 0.15$, we have for any entry $i\in[n]$, $| \vdeltil_{\vx}(i)/\vx(i)|\leq 0.05$. Therefore , $\vx(i)+\vdeltil_{\vx}(i) \geq 0.95 \vx(i) > 0$. A similar argument gives $\vs(i)+\vdeltil_{\vs}(i) > 0$. Therefore the entries of $\vx$ and $\vs$ stay positive after an update.
\end{remark}

\begin{lemma}
\label{lemma:xs-stay-close-t-storjohann}
Let $\ma$ be an integer matrix, $\norm{\frac{\vxbar-\vx}{\vx}}_{\infty},\norm{\frac{\vsbar-\vs}{\vs}}_{\infty}\leq 0.01$, $\vdel_{\vmu} = -h(\vmu_1-\vec{t}_1)$, and $\vmu_1=\vx \odot \vs$, such that $h= 0.1$. Moreover let $\vdeltil_{\vmu}\in\R^{n}$ such that and  $\norm{\frac{\vdeltil_{\vmu}-\vdel_{\mu}}{\vdel_{\mu}}}_{\infty} \leq 0.001$. Let
\[
\vdeltil_{\vs} = \ma (\ma^\top \mxbar \msbar^{-1} \ma)^{-1} \ma^\top \msbar^{-1} \vdeltil_{\vmu} ~~ \text{, and } ~~
\vdeltil_{\vx} = \msbar^{-1} \vdeltil_{\vmu} - \mxbar \msbar^{-1} \vdeltil_{\vs}.\]
Moreover suppose $\norm{\frac{\vmu_1-\vec{t_1}}{t_1}}_2 \leq 0.01$.
Let $\vmu_2 = (\vx+\vdeltil_{\vx})\odot (\vs+\vdeltil_{\vs})$ and $t_2 = (1-1/(\beta\sqrt{n})) t_1$, where $\beta=10^4$. Then $\norm{\frac{\vmu_2-\vec{t_2}}{t_2}}_2 \leq 0.01$.
\end{lemma}
\begin{proof}
By triangle inequality,
\begin{align*}
\norm{\frac{\vmu_2-\vec{t_2}}{t_2}}_2 
& = 
\norm{\frac{(\vx+\vdeltil_{\vx})\odot (\vs+\vdeltil_{\vs})-\vec{t_2}}{t_2}}_2
\\ & \leq 
\norm{\frac{\vx\odot \vs-\vec{t_2} + \vxbar \odot \vdeltil_{\vs} +\vsbar \odot \vdeltil_{\vx}}{t_2}}_2 + \norm{\frac{(\vx - \vxbar)\odot \vdeltil_{\vs}}{t_2}}_2 + \norm{\frac{(\vs - \vsbar)\odot \vdeltil_{\vx}}{t_2}}_2 +  \norm{\frac{\vdeltil_{\vx} \odot \vdeltil_{\vs}}{t_2}}_2.
\end{align*}
Now by \cref{lemma:xs-stay-pos},
\begin{align*}
\norm{\frac{\vdeltil_{\vx} \odot \vdeltil_{\vs}}{t_2}}_2 
& = 
\norm{\mx \ms\frac{\mx^{-1}\vdeltil_{\vx} \odot \ms^{-1} \vdeltil_{\vs}}{t_2}}_2 
\\ &
\leq 
\norm{\frac{\vx \odot \vs}{t_2}}_{\infty} \norm{\mx^{-1}\vdeltil_{\vx}}_{\infty} \norm{\ms^{-1}\vdeltil_{\vs}}_{2}
\\ & \leq 
\frac{1.01}{1-1/(\beta\sqrt{n})} (0.05h)^2 
\\ & \leq 0.00003.
\end{align*}
Now denoting $\mpbar:=\mxbar^{1/2}\msbar^{-1/2} \ma (\ma^\top \mxbar \msbar^{-1} \ma)^{-1} \ma^\top \msbar^{-1/2} \mxbar^{1/2}$, we have
\[
\vxbar \odot \vdeltil_{\vs} +\vsbar \odot \vdeltil_{\vx} = \mxbar^{1/2}\msbar^{1/2} (\mi-\mpbar)\mxbar^{-1/2}\msbar^{-1/2} \vdeltil_{\vmu}+\mxbar^{1/2}\msbar^{1/2} \mpbar\mxbar^{-1/2}\msbar^{-1/2} \vdeltil_{\vmu} = \vdeltil_{\vmu} = -h(\vmu_1-\vec{t}_1).
\]
Therefore
\begin{align*}
\norm{\frac{\vx\odot \vs-\vec{t_2} + \vxbar \odot \vdeltil_{\vs} +\vsbar \odot \vdeltil_{\vx}}{t_2}}_2 & \leq
\frac{1}{1-1/(\beta\sqrt{n})}\norm{\frac{\vx\odot \vs-\vec{t_1} + \vxbar \odot \vdeltil_{\vs} +\vsbar \odot \vdeltil_{\vx}}{t_1}}_2 + \frac{1}{1-1/(\beta\sqrt{n})}\norm{\frac{1}{\beta\sqrt{n}}}_2
\\ & \leq
\frac{1}{1-1/(\beta\sqrt{n})}\norm{\frac{\vx\odot \vs-\vec{t_1} +\vdel_{\vmu}}{t_1}}_2 + \frac{1}{1-1/(\beta\sqrt{n})}\norm{\frac{\vdeltil_{\vmu}-\vdel_{\vmu}}{t_1}}_2 + 0.0002
\\ & \leq
\frac{1}{1-1/(\beta\sqrt{n})}\norm{\frac{\vx\odot \vs-\vec{t_1} +\vdel_{\vmu}}{t_1}}_2 + \frac{0.001}{1-1/(\beta\sqrt{n})}\norm{\frac{\vdel_{\vmu}}{t_1}}_2 + 0.0002
\\& \leq \frac{1}{1-1/(\beta\sqrt{n})}\cdot\norm{\frac{(1-h)(\vmu_1 - \vec{t}_1)}{\vec{t}_1}}_2 + 0.0004
\\ & \leq
0.9001 \cdot \norm{\frac{\vmu_1 - \vec{t}_1}{\vec{t}_1}}_2 + 0.0004.
\end{align*}
Moreover we have 
\begin{align*}
 \norm{\frac{(\vx - \vxbar)\odot \vdeltil_{\vs}}{t_2}}_2 & \leq \frac{0.01}{1-1/(\beta\sqrt{n})}\cdot \norm{\frac{\vx \odot \vdeltil_{\vs}}{t_1}}_2 
 \\ & = 
 \frac{0.01}{1-1/(\beta\sqrt{n})}\cdot \norm{\frac{\mx \ms \ms^{-1} \vdeltil_{\vs}}{t_1}}_2
 \\ & \leq \frac{0.01 \cdot 1.01}{1-1/(\beta\sqrt{n})}\cdot \norm{\ms^{-1} \vdeltil_{\vs}}_2
 \leq 
 0.0002,
\end{align*}
where the last inequality follows from \cref{lemma:xs-stay-pos-storjohann}. Similarly we can show $\norm{\frac{(\vs - \vsbar)\odot \vdeltil_{\vx}}{t_2}}_2 \leq 0.0002$.
Combining all of the above and since $\norm{\frac{\vmu_1 - \vec{t}_1}{\vec{t}_1}}_2 \leq 0.01$, we have
\[
\norm{\frac{\vmu_2 - \vec{t}_2}{\vec{t}_2}}_2 \leq 0.01.
\]
\end{proof}

We are now equipped to prove the main result of this section.

\begin{proof}[Proof of \cref{thm:inverse_free_ipm}]
For ease of notation, we drop the superscripts denoting iteration number for $\mxbar,\msbar,\vdeltil_{\vmu}$ in this proof.
First note that in each iteration of the IPM (Algorithm \ref{alg:ipm-simple-no-maintenance}), we first compute  \[\vdeltil_{\vs} = \ma (\ma^\top \mxbar \msbar^{-1} \ma)^{-1} \ma^\top \msbar^{-1} \vdeltil_{\vmu},\] and then use this to compute $\vdeltil_{\vx}^{(k)} = \msbar^{-1} \vdeltil_{\vmu} - \mxbar^{(k)} \msbar^{-1} \vdeltil_{\vs}$. Note that by construction $\mxbar,\msbar,\vdeltil_{\vmu}$ are rational matrices and vector.

Now without loss of generality, we assume $\ma$ is an integer matrix. Otherwise if the bit complexity of $\ma$ is $\ell$, we multiply $\ma$ by $2^{\ell}$. This converts the matrix to an integer matrix. Note that this does not change the bit complexity of the matrix, and we later divide the solution of the linear system $(\ma^\top \mxbar \msbar^{-1} \ma)\vz =  \ma^\top \msbar^{-1} \vdeltil_{\vmu}$ by $2^{2\ell}$. To find the value of $(\ma^\top \mxbar \msbar^{-1} \ma)^{-1} \ma^\top \msbar^{-1} \vdeltil_{\vmu}$, in each iteration, we find $\vz\in\Q^{d}$ such that $101^{s_1}\cdot 100^{s_2}(\ma^\top \mxbar \msbar^{-1} \ma)\vz = 101^{s_1} \cdot 1000^{s_3} \ma^\top \msbar^{-1}\cdot\vdeltil_{\vmu}$, where $101^{s_1}$ is the denominator of the smallest entry of $1/\vsbar$, $100^{s_2}$ is the denominator of the smallest entry of $\vxbar$, and $1000^{s_3}$ is the denominator of the smallest entry of $\vdeltil_{\vmu}$. 
We then use this to find $\vdeltil_x,\vdeltil_s$, round these to a relative error of $\exp{(-O(n))}$ and add it to $\vx$ and $\vs$. This rounding introduces an error of size $\exp{(-O(n))} \kappa(\ma)$ in $\ma^\top \vx = \vb$. Picking a large enough constant, we can make sure the total aggregate error over the whole course of the algorithm caused by this rounding is exponentially small.

Note that $s_1,s_2,s_3$ determine the bit complexity of the matrix and the vector of the corresponding integer linear systems we solve, and these numbers depend on the magnitude of smallest and largest entries of $\vx,\vs,\vdel_{\vmu}$.

First note that by definition $\vdel_{\vmu} = -0.5(\vmu - \vec{t})$. Therefore since by \cref{lemma:xs-stay-close-t-storjohann}, it is maintained that over the course of the algorithm $\norm{(\vmu-\vec{t})/t}_2 \leq 0.01$, for all $i\in[n]$,
\[
\frac{0.99}{2} t \leq \abs{(\vdel_{\vmu})_i} \leq \frac{1.01}{2} t. 
\]
Therefore over the whole course of the algorithm 
\[
\frac{0.99}{2} t^{(\final)} \leq \abs{(\vdel_{\vmu})_i} \leq \frac{1.01}{2} t^{(0)}. 
\]
Now note that since for all $\vx$ such that $\ma^\top \vx = \vb$ and $\vx\geq 0$, $\norm{\vx}_2 \leq R$, for all $\vx$ such that $\norm{\ma^\top \vx - \vb} \leq \exp(-Cn)$ and $\vx\geq 0$, $\norm{\vx}_2 \leq R + \exp(-(C-2)n) \leq 2R$. Moreover since $\mu=\vx\odot \vs$ and $\norm{(\vmu-\vec{t})/t}_2 \leq 0.01$, $\vs_i \geq 0.99\frac{t^{(\final)}}{2R}$. 
Now note that by construction $\vx_i$ is always either greater than or equal to $\exp(-3n)$ or is zero. Therefore $s_i \leq \exp(3n) \cdot 1.01 \cdot t^{(0)}$.

Now we need to bound these values for Steps \ref{alg-step:ipm-init-modified-initial} and \ref{alg-step:ipm-init-original-initial} of Algorithm \ref{alg:ipm-init}. For Step \ref{alg-step:ipm-init-modified-initial}, we have $t^{(0)} = 2^{16} \veps^{-3} n^2 \frac{R}{r} \cdot \norm{\vc}_{\infty}R$ and $t^{(\final)} = \norm{\vc}_{\infty}R$. Therefore noting that $\veps = 1/(100\sqrt{n})$, the number of iterations is $O(n^{0.5}\cdot \log(\frac{n \cdot R}{r}))$. Moreover the outer radius of the modified LP is less than $200(n+1) \sqrt{n} \cdot R$. Therefore by above arguments the bit complexity of right-hand side of linear systems we solve for Step \ref{alg-step:ipm-init-modified-initial} of Algorithm \ref{alg:ipm-init} is bounded by 
\[
O(n + \log(n \norm{\vc}_{\infty} R/r)).
\]
Combining this with number of iterations and \cref{thm:storjohann-lin-sys}, bounds the running time of Step \ref{alg-step:ipm-init-modified-initial} of Algorithm \ref{alg:ipm-init}.

For Step \ref{alg-step:ipm-init-original-initial} of Algorithm \ref{alg:ipm-init}, $t^{(0)}\leq \frac{7}{6} \norm{\vc}_{\infty}R$, and $t^{(\final)}$ is $\epsilon/2n$. Therefore the number of iterations is $O(\sqrt{n} \log(n\norm{\vc}_{\infty}R/\epsilon))$.
This gives a bit complexity of $O(n+\log(n\norm{\vc}_{\infty}R/\epsilon))$ for the right-hand side of the linear systems solved for Step \ref{alg-step:ipm-init-original-initial} of Algorithm \ref{alg:ipm-init}, and combining this with iteration number and \cref{thm:storjohann-lin-sys} bounds the running time for this step.
\end{proof}

\subsection{Solving Sparse LP faster than Matrix Multiplication for \texorpdfstring{$\omega>2.5$}{omega>2.5}}
\label{sec:sparse_ipm}

The main result of this section is the following theorem achieved by Algorithm \ref{alg:ipm-simple}.

\SparseIPM*

We define the vector $\vmu^{(k)}:=\vx^{(k)} \odot \vs^{(k)}$. We want to maintain $\norm{\vmu^{(k)}-\vec{t}^{(k)}}_2=O(t^{(k)})$. The update is by taking the gradient of $\norm{\vmu^{(k)}-\vec{t}^{(k)}}_2$, i.e., $\vmu^{(k+1)} =  \vmu^{(k)}+\vdel_{\vmu}^{(k)}$ where $\vdel_\vmu^{(k)} = -h(\vmu^{(k)}-\vec{t}^{(k)})$. We later show that we can set $t^{(k+1)} = \left(1 - \frac{1}{\sqrt{n}}\right)t^{(k)}$. Let $\vxbar,\vsbar\in\R^{n}$ such that $\norm{\frac{\vxbar-\vx}{\vx}}_{2},\norm{\frac{\vsbar-\vs}{\vs}}_{2} \leq \frac{1}{\sqrt{n}}$.

We need to prove the following.
\begin{enumerate}
    \item $\vx+\vdel_{\vx}$ and $\vs+\vdel_{\vs}$ stay nonnegative.
    \item Given $\norm{\vmu^{(k)}-\vec{t}^{(k)}}_2=O(t^{(k)})$, we have $\norm{\vmu^{(k+1)}-\vec{t}^{(k+1)}}_2=O(t^{(k+1)})$.
\end{enumerate}

\RestyleAlgo{algoruled}
\IncMargin{0.15cm}
\begin{algorithm}[t]

\textbf{Input:} Full column rank matrix $\ma\in \R^{n\times d}$ ($n > d$), $\vb\in \R^{d}$, $\vc \in\R^n$, $0<\epsilon_1,\epsilon_2<1$ such that $\norm{\ma^\top \ma}_{\fro}, \norm{(\ma^\top \ma)^{-1}}_{\fro}, \norm{\vb}_2, \norm{\vc}_2 \leq \kappa$, for $\kappa>1$. \\

\textbf{Output:} $\vxhat\in\R^{n}_{\geq 0}$ such that $\norm{\ma \vxhat - \vb}_2 \leq \epsilon_1$ and $\vc^\top \vxhat \leq \min_{\vx:\ma^\top \vx = \vb,\vx\geq 0} \vc^\top \vx + \epsilon_2$. \\

Compute the initial feasible primal solution $\vx^{(0)}$ and slack $\vs^{(0)}$ and change the matrices. \\

Set $t^{(0)} = 1$, $\mu^{(0)} = \mx^{(0)} \vs^{(0)}$, and $\vwhat^{(0)}=\vw^{(0)} = (\ms^{(0)})^{-1} \vx^{(0)}$\\

Set $T = \sqrt{n} \cdot \log(\kappa/\epsilon_2)$ and $k=1$\\

Set $\ds$ to be the inverse maintenance data structure \DontPrintSemicolon \tcp*{\textcolor{blue}{either the dense data structure $\dds$ or the sparse data structure $\sds$}}

$\ds.\initialize(\ma,\mwhat,\frac{\epsilon_1}{10^6 (\kappa\cdot n\cdot  \log(1/\epsilon_2))^{30}})$

\While{$k < T$}{
Set $\vdel_{\vmu}^{(k)} = - 0.5 (\vmu^{(k-1)}-\vec{t}^{(k-1)})$\\
Compute $\vu^{(k)} = \mx^{1/2} \ms^{-1/2} \ma (\ma^\top \mx \ms^{-1} \ma)^{-1} \ma^\top \ms^{-1} \vdel_{\vmu}$ using $\ds.\query$ and Richardson's iteration\\
Compute $\vdel_{\vs}^{(k)} = \mx^{-1/2} \ms^{1/2} \vu$ and $\vdel_{\vs}^{(k)} = \ms^{-1} \vdel_{\vmu} - \mx^{1/2} \ms^{-1/2} \vu$\\
Update $\vx^{(k)} = \vx^{(k-1)} + \vdel_{\vx}^{k}$, $\vs^{(k)} = \vs^{(k-1)} + \vdel_{\vs}^{k}$, $\vw^{(k)} = (\ms^{(k)})^{-1} \vx^{(k)}$, and $t^{(k)}=(1-\frac{1}{1000\sqrt{n}}) t^{(k-1)}$\\
Let $Q = \{i: \frac{\abs{\vwhat^{(k-1)}_i - \vw^{(k)}_i}}{\abs{\vw^{(k)}_i}}>0.5\}$\\
Set $\vwhat_{Q}^{(k)} = \vw_{Q}^{(k)}$, $\vwhat_{[n]\setminus Q}^{(k)} = \vwhat_{[n]\setminus Q}^{(k-1)}$, and call $\ds.\update(Q,\vw^{(k)}_Q)$\\
Set $k=k+1$\\
}
\Return $\vx^{(T)}$

\caption{Path following interior point method (IPM)}
\label{alg:ipm-simple}
\end{algorithm}

\begin{lemma}
\label{lemma:xs-stay-pos}
Let
\begin{align}
\label{eq:xs-stay-pos-norm-bound}
\norm{\mx}_{\fro},\norm{\mx^{-1}}_{\fro},\norm{\ms}_{\fro},\norm{\ms^{-1}}_{\fro},\norm{\ma}_{\fro} \leq \kappa.
\end{align}
Let 
\[
\vdel_{\vx} = \mx^{1/2}\ms^{-1/2} (\mi-\mptil)\mx^{-1/2}\ms^{-1/2} \vdel_{\vmu} ~~ \text{, and } ~~
\vdel_{\vs} = \mx^{-1/2}\ms^{1/2} \mptil \mx^{-1/2}\ms^{-1/2} \vdel_{\vmu},
\]
where $\mptil := \mx^{1/2}\ms^{-1/2} \ma \md^{-1} \ma^\top \ms^{-1/2} \mx^{1/2}$, $\vdel_{\vmu} = -h(\vmu-\vec{t})$, and $\vmu=\vx \odot \vs$ such that $\norm{(\ma^\top \mx \ms^{-1} \ma)^{-1} - \md^{-1}}_{\fro} \leq \veps$ such that $\kappa^4 \veps \leq 0.1$. Moreover suppose $\norm{\frac{\vmu-\vec{t}}{t}}_2 \leq 0.1$.
Then
$\norm{\mx^{-1} \vdel_{\vx}}_2,\norm{\ms^{-1} \vdel_{\vs}}_2 \leq 0.15 h$.
\end{lemma}
\begin{proof}
First note that
\[
\norm{(\vx \odot \vs - \vec{t})/t}_{\infty} \leq \norm{(\vx \odot \vs - \vec{t})/t}_{2} \leq 0.1. 
\]
Therefore $0.9 \cdot t\leq \norm{\vx \odot \vs}_{\infty}\leq 1.1 \cdot t$.
We have 
\begin{align*}
\norm{\ms^{-1} \vdel_{\vs}}_2 & = \norm{\mx^{-1/2}\ms^{-1/2} \mptil \mx^{-1/2}\ms^{-1/2} \vdel_{\vmu}}_2
\\ & \leq
\norm{\mx^{-1/2}\ms^{-1/2}}_{\infty} \norm{\mptil \mx^{-1/2}\ms^{-1/2} \vdel_{\vmu}}_2
\\ & \leq \frac{10}{9} \cdot \frac{1}{\sqrt{t}} \norm{\mptil \mx^{-1/2}\ms^{-1/2} \vdel_{\vmu}}_2.
\end{align*}
Now denoting $\matp:=\mx^{1/2}\ms^{-1/2} \ma (\ma^\top \mx \ms^{-1} \ma)^{-1} \ma^\top \ms^{-1/2} \mx^{1/2}$, since $\norm{(\ma^\top \mx \ms^{-1} \ma)^{-1} - \md^{-1}}_{\fro} \leq \veps$, by \eqref{eq:xs-stay-pos-norm-bound},
$
\norm{\mptil - \matp}_{\fro} \leq \kappa^4 \veps
$.
Now by triangle inequality and since $\matp$ is an orthogonal projection matrix,
\begin{align*}
\norm{\mptil \mx^{-1/2}\ms^{-1/2} \vdel_{\vmu}}_2 
& \leq 
\norm{\matp \mx^{-1/2}\ms^{-1/2} \vdel_{\vmu}}_2 + \norm{(\mptil - \matp) \mx^{-1/2}\ms^{-1/2} \vdel_{\vmu}}_2
\\ & \leq
(1 + \kappa^4 \veps) \norm{ \mx^{-1/2}\ms^{-1/2} \vdel_{\vmu}}_2.
\end{align*}
Therefore
\begin{align*}
\norm{\ms^{-1} \vdel_{\vs}}_2 
& \leq 
\frac{11}{9} \cdot \frac{1}{\sqrt{t}} \norm{ \mx^{-1/2}\ms^{-1/2} \vdel_{\vmu}}_2
\\ & \leq 
\frac{11}{9} \cdot \frac{1}{\sqrt{t}} \norm{ \mx^{-1/2}\ms^{-1/2}}_{\infty} \norm{ \vdel_{\vmu}}_2
\\ & \leq
\frac{110}{81} \cdot \frac{1}{t} \norm{ \vdel_{\vmu}}_2
\end{align*}
Now since $\vdel_{\vmu} = -h(\vmu-\vec{t})$ and $\norm{(\vmu-\vec{t})/t}_2\leq 0.1$, we have $\norm{\ms^{-1} \vdel_{\vs}}_2 \leq 0.15 h$.
For $\mx^{-1} \vdel_{\vx}$, we similarly have
\begin{align*}
\norm{\mx^{-1} \vdel_{\vx}}_2 & = \norm{\mx^{-1/2}\ms^{-1/2} (\mi - \mptil) \mx^{-1/2}\ms^{-1/2} \vdel_{\vmu}}_2
\\ & \leq
\norm{\mx^{-1/2}\ms^{-1/2}}_{\infty} \norm{(\mi - \mptil) \mx^{-1/2}\ms^{-1/2} \vdel_{\vmu}}_2
\\ & \leq \frac{10}{9} \cdot \frac{1}{\sqrt{t}} \norm{(\mi - \mptil) \mx^{-1/2}\ms^{-1/2} \vdel_{\vmu}}_2.
\end{align*}
Then a similar triangle inequality, and bound on $\norm{(\vmu-\vec{t})/t}_2\leq 0.1$, gives
\begin{align*}
\norm{\mx^{-1} \vdel_{\vx}}_2 & \leq \frac{10}{9} \cdot \frac{1}{\sqrt{t}} (1+\kappa^4 \veps) \norm{ \mx^{-1/2}\ms^{-1/2}}_{\infty} \norm{ \vdel_{\vmu}}_2
\leq
0.15 h.
\end{align*}
\end{proof}
\begin{note}
Let $h\leq 1$.
Then since $\norm{\mx^{-1} \vdel_{\vx}}_{\infty} \leq \norm{\mx^{-1} \vdel_{\vx}}_{2} \leq 0.15$, we have for any entry $i\in[n]$, $| \vdel_{\vx}(i)/\vx(i)|\leq 0.15$. Therefore , $\vx(i)+\vdel_{\vx}(i) \geq 0.85 \vx(i) > 0$. A similar argument gives $\vs(i)+\vdel_{\vs}(i) > 0$. Therefore the entries of $\vx$ and $\vs$ stay nonnegative after an update.
\end{note}

\begin{lemma}
\label{lemma:xs-stay-close-t}
Let
\begin{align}
\label{eq:xs-stay-close-to-t}
\norm{\mx}_{\fro},\norm{\mx^{-1}}_{\fro},\norm{\ms}_{\fro},\norm{\ms^{-1}}_{\fro},\norm{\ma}_{\fro} \leq \kappa.
\end{align}
Let 
\[
\vdel_{\vx} = \mx^{1/2}\ms^{-1/2} (\mi-\mptil)\mx^{-1/2}\ms^{-1/2} \vdel_{\vmu} ~~ \text{, and } ~~
\vdel_{\vs} = \mx^{-1/2}\ms^{1/2} \mptil \mx^{-1/2}\ms^{-1/2} \vdel_{\vmu},
\]
where $\mptil := \mx^{1/2}\ms^{-1/2} \ma \md^{-1} \ma^\top \ms^{-1/2} \mx^{1/2}$, $\vdel_{\vmu} = -h(\vmu_1-\vec{t}_1)$, and $\vmu_1=\vx \odot \vs$ such that $h\leq 0.5$, $\norm{(\ma^\top \mx \ms^{-1} \ma)^{-1} - \md^{-1}}_{\fro} \leq \veps$, and $\kappa^4 \veps \leq 0.1$. Moreover suppose $\norm{\frac{\vmu_1-\vec{t_1}}{t_1}}_2 \leq 0.1$.
Let $\vmu_2 = (\vx+\vdel_{\vx})\odot (\vs+\vdel_{\vs})$ and $t_2 = (1-1/(1000\sqrt{n})) t_1$, and $n\geq 10$. Then $\norm{\frac{\vmu_2-\vec{t_2}}{t_2}}_2 \leq 0.1$.
\end{lemma}
\begin{proof}
By triangle inequality,
\begin{align*}
\norm{\frac{\vmu_2-\vec{t_2}}{t_2}}_2 
& = 
\norm{\frac{(\vx+\vdel_{\vx})\odot (\vs+\vdel_{\vs})-\vec{t_2}}{t_2}}_2
\\ & \leq 
\norm{\frac{\vx\odot \vs-\vec{t_2}}{t_2} + \frac{\vx \odot \vdel_{\vs} +\vs \odot \vdel_{\vx}}{t_2} }_2 + \norm{\frac{\vdel_{\vx} \odot \vdel_{\vs}}{t_2}}_2.
\end{align*}
Now by \cref{lemma:xs-stay-pos},
\begin{align*}
\norm{\frac{\vdel_{\vx} \odot \vdel_{\vs}}{t_2}}_2 
& = 
\norm{\mx \ms\frac{\mx^{-1}\vdel_{\vx} \odot \ms^{-1} \vdel_{\vs}}{t_2}}_2 
\\ &
\leq 
\norm{\frac{\vx \odot \vs}{t_2}}_{\infty} \norm{\mx^{-1}\vdel_{\vx}}_{\infty} \norm{\ms^{-1}\vdel_{\vs}}_{2}
\\ & \leq 
\frac{1.1}{1-1/\sqrt{n}} (0.15h)^2 
\\ & \leq 0.0125.
\end{align*}
Now denoting $\matp:=\mx^{1/2}\ms^{-1/2} \ma (\ma^\top \mx \ms^{-1} \ma)^{-1} \ma^\top \ms^{-1/2} \mx^{1/2}$, 
\[
\vx \odot \vdel_{\vs} +\vs \odot \vdel_{\vx} = \vdel_{\vmu} + \mx^{1/2}\ms^{1/2} (\matp-\mptil)\mx^{-1/2}\ms^{-1/2} \vdel_{\vmu}+\mx^{1/2}\ms^{1/2} (\mptil-\matp)\mx^{-1/2}\ms^{-1/2} \vdel_{\vmu} = \vdel_{\vmu} = -h(\vmu_1-\vec{t}_1).
\]
Therefore
\begin{align*}
\norm{\frac{\vx\odot \vs-\vec{t_2}}{t_2} + \frac{\vx \odot \vdel_{\vs} +\vs \odot \vdel_{\vx}}{t_2} }_2 & = \frac{1}{1-1/\sqrt{n}}\cdot\norm{\frac{(1-h)(\vmu_1 - \vec{t}_1)}{\vec{t}_1}}_2 + \frac{1}{1-1/\sqrt{n}}\norm{\frac{1}{1000\sqrt{n}}}_2
\\ & \leq
0.75 \cdot \norm{\frac{\vmu_1 - \vec{t}_1}{\vec{t}_1}}_2 + 0.002.
\end{align*}
Combining the above and using $\norm{\frac{\vmu_1 - \vec{t}_1}{\vec{t}_1}}_2\leq 0.1$, we have $\norm{\frac{\vmu_2 - \vec{t}_2}{\vec{t}_2}}_2\leq 0.1$.
\end{proof}

\begin{lemma}
\label{lemma:how-much-infeasible}
Let
\begin{align}
\norm{\mx}_{\fro},\norm{\mx^{-1}}_{\fro},\norm{\ms}_{\fro},\norm{\ms^{-1}}_{\fro},\norm{\ma}_{\fro} \leq \kappa.
\end{align}
Let 
\[
\vdel_{\vx} = \mx^{1/2}\ms^{-1/2} (\mi-\mptil)\mx^{-1/2}\ms^{-1/2} \vdel_{\vmu} ~~ \text{, and } ~~
\vdel_{\vs} = \mx^{-1/2}\ms^{1/2} \mptil \mx^{-1/2}\ms^{-1/2} \vdel_{\vmu},
\]
where $\mptil := \mx^{1/2}\ms^{-1/2} \ma \md^{-1} \ma^\top \ms^{-1/2} \mx^{1/2}$, $\vdel_{\vmu} = -h(\vmu-\vec{t})$, and $\vmu=\vx \odot \vs$ such that $\norm{(\ma^\top \mx \ms^{-1} \ma)^{-1} - \md^{-1}}_{\fro} \leq \veps_1$. Moreover suppose the initial feasible solution satisfies $\norm{\ma^\top \vx^{(0)} - \vb}_2 \leq \veps_2$. Then after $k$ updates, we have $\norm{\ma^\top \vx^{(k)} - \vb}_2 \leq k \kappa^7 \veps_1 + \veps_2$.
\end{lemma}
\begin{proof}
By triangle inequality, we have
\begin{align*}
\norm{\ma^\top \vx^{(k)} - \vb}_2 \leq \norm{\ma^\top \vx^{(0)} - \vb}_2 + \sum_{j=1}^k \norm{\ma^\top \vdel_{\vx}^{(k)}}_2 \leq \veps_2 + k \kappa^7 \veps_1.
\end{align*}

\end{proof}

\begin{lemma}
\label{lemma:num-changes-LP-simple}
Over the span of $k$ iterations, Algorithm \ref{alg:ipm-simple} makes at most $O(k^2\log{n})$ changes to $\vwtil$.
\end{lemma}
\begin{proof}
\end{proof}

\begin{lemma}
\label{lemma:LP-sparse-time}
Suppose an algorithm uses the sparse data structure of \cref{thm:sparse-ds}. Moreover assume that the algorithm runs for $n^{1/2}$ iterations and after every $k$ iterations, at most $k^2$ many entries are updated. Then for $\omega>2.5$, the total running time of the algorithm is
\[
\Otil\left(\left(\nnz(\ma)\cdot m^2\cdot n + \frac{n^{\omega}}{m^{\omega-2.5}} \right)\cdot \log^2((\kappa+\norm{\vb}_2)/\veps))\right).
\]
\end{lemma}
\begin{proof}
First note that the total contribution of terms of the form $\nnz(\ma)\cdot m^2\cdot \abs{S}$ in the updates is $\nnz(\ma)\cdot m^2\cdot n$ since there are at most $n$ updates over the course of the algorithm. So in the following, we omit the contribution of these terms. Moreover the total running time of query over the course of the algorithm is
\[
\Otil(\nnz(\ma)\cdot m^2\cdot n^{0.5}+n^{2.5}\cdot\log^2(\kappa/\veps))
\]
Now note that the total cost of initialization and updates of rank more than $n/m$ is
\[
\Otil\left(\left(d\cdot \nnz\left(\ma\right)\cdot m^{1.5} + \left(\frac{d}{m}\right)^\omega m^{2.5}\right)\log^2\left(\kappa/\veps\right)\right)
\]
because the number of such updates is at most $m^{0.5}$. Now note that the cost of an update of rank less than $n^{\omegadual}$, is $O(n^2)$. Therefore the total cost of such updates over the course of the algorithm is $O(n^{2.5})$.
For $k=2^j$, the total cost of updates of rank at most $k^2$ (modulo $\log^2(\kappa/\veps)$) is
\begin{align*}
\left(n^2\cdot (k^2)^{\omega-2} + \mtime(n,k^2,k^2) + k^{2\omega} \right) \cdot \frac{n^{0.5}}{k} 
& = 
n^{2.5} \cdot k^{2\omega - 5} + \mtime(n,k^2,k^2) \cdot \frac{n^{0.5}}{k} + n^{0.5}\cdot k^{2\omega-1}
\end{align*}
Moreover since $\omega> 2.5$ and $k^2 \leq n/m$, we have
\[
n^{2.5} \cdot k^{2\omega-5} \leq n^{2.5} \left( \frac{n}{m} \right)^{(2\omega-5)/2} \leq \frac{n^{\omega}}{m^{\omega-2.5}}.
\]
Now since $3\omega-1 > 0$, and $k^2\leq n/m$,
\[
n^{0.5}\cdot k^{2\omega-1} \leq n^{0.5} \cdot \left( \frac{n}{m} \right)^{(2\omega-1)/2} = \frac{n^{\omega}}{m^{\omega-0.5}}.
\]
We also have
\[
\mtime(n,k^2,k^2) \cdot \frac{n^{0.5}}{k} \leq n^{\omega-2+0.5}\cdot k^3 \leq \frac{n^{\omega}}{m^{1.5}}.
\]
Moreover since $3\geq \omega> 2.5$,
\[
\frac{n^{\omega}}{m^{\omega-0.5}} \leq \frac{n^{\omega}}{m^{1.5}} \leq \frac{n^{\omega}}{m^{\omega-2.5}}.
\]
We can finally bound the cost of all updates by considering all powers of two between $1$ and $(n/m)^{1/2}$ for $k$.
\end{proof}
\section{\texorpdfstring{$p$}{p}-Norm Regression}
\label{sec:OuterLoop}

In this section, we consider the problems of the following form.
\begin{align}
\label{eq:p-norm-problem}
\min_{\vx: \ma^\top \vx = \vb} \norm{\vx}_p^p,
\end{align}
where $\ma\in\R^{n\times d},\vb\in\R^d$, and $p>1$.
We follow the approach of \cite{AdilKPS19}. We first discuss a residual problem for \eqref{eq:p-norm-problem} in \cref{subsec:res-problem}. We show that \eqref{eq:p-norm-problem} can be solved by solving $O_p(\alpha \log(n/\epsilon))$ instances of the residual problem to $\alpha$-approximation. 
We discuss how this residual problem can be turned into a mixed $(2,p)$-norm minimization problem with an extra linear constraint in \cref{subsec:solve-res-problem}.
We can either directly optimize over this mixed $(2,p)$-norm problem (\cref{sec:two-p-norm}) or use a mixed $(2,\infty)$-norm problem as a proxy (\cref{subsec:two-inf-norm}). The latter approach introduces an extra factor of $n^{2/p}$ in the running time. However, since this approach is simpler, we first focus on this. Moreover, computing a constant factor approximation to mixed $(2,\infty)$-norm is of independent interest. Both approaches require solutions to a series of weighted linear regression problems. Therefore before diving into either, we discuss the bit complexity of this weighted linear regression problem in \cref{subsec:weighted-lin-reg}. We extensively use the following inequality in this section.

\begin{fact}[Holder's inequality]\label{fact:holders}
Let $\vx,\vy \in \R^{n}$ and $p,q\in[1,\infty]$ such that $\frac{1}{p}+\frac{1}{q} =1$. Then \[\norm{\vx \odot \vy}_1 \leq \norm{\vx}_{p} \norm{\vy}_{q}.\]
\end{fact}

\subsection{Residual Problem}
\label{subsec:res-problem}

We start this section by defining the smoothed $p$-norm function which was first introduced in \cite{bubeck2018homotopy} and has been used extensively in the $p$-norm minimization literature \cite{AdilKPS19,AdilPS19,AdilS20} since. We also refer to this function as a mixed $(2,p)$-norm function because, under a certain threshold, it is a quadratic function and above the threshold, it is a power $p$ function.

\begin{definition}\label{def:quadratic-smooth}
For $p\geq 1$ and a threshold $t \in \R_{\geq 0}$, we define the (quadratically) smoothed $p$-norm function $\gamma_{p}(t,\cdot): \R \to \R$ as
\begin{align}
\gamma_p(t, x) := \begin{cases}
\frac{p}{2} t^{p-2} x^2 & \text{ if } |x|\leq t,\\
|x|^p + (\frac{p}{2} - 1) t^p & \text{ otherwise.}
\end{cases}
\end{align}
Overloading the notation, for a threshold vector $\vt\in\R^{n}_{\geq 0}$, we define $\gamma_{p}(\vt,\cdot): \R^n \to \R$ as
\begin{align}
\gamma_p(\vt, \vx) := \sum_{i=1}^n \gamma_p(\vt_i, \vx_i).
\end{align}
\end{definition}

The smoothed $p$-norm function gives a decent approximation for the Bregman divergence of the $p$-norm function. An important observation is that the smoothed $p$-norm function is symmetric (i.e., $\gamma_p(t,x)=\gamma_p(t,-x)$), while the Bregman divergence is not necessarily symmetric.

\begin{lemma}[\cite{AdilKPS19}]
\label{lemma:taylor}
Let $p>1$. Then for any $\vx,\vdelta\in\R^n$,
\[
\norm{\vx}_p^p - \vg^\top \vdelta + \frac{p-1}{p \cdot 2^p} \gamma_p(\abs{\vx},\vdelta) \leq \norm{\vx-\vdelta}_p^p \leq \norm{\vx}_p^p - \vg^\top \vdelta + 2^p \cdot \gamma_p(\abs{\vx},\vdelta),
\]
where $\vg$ is the gradient of $p$-norm at $\vx$, i.e., $\vg=p\cdot \abs{\vx}^{p-2} \odot \vx$.
\end{lemma}

Equipped with the above lemma, a natural approach is to take second-order Newton steps according to the smoothed $p$-norm function. In other words, we take steps according to the following residual problem.

\begin{definition}[Residual problem]
\label{def:res-problem}
Given $\vx \in \R^{n}$ and $p>1$, we define the mixed $(2,p)$-norm residual problem at $\vx$ as
\[
\argmax_{\ma^\top \vdelta = 0} ~ \vg^\top \vdelta - \frac{p-1}{p\cdot 2^p} \cdot \gamma_p(\abs{\vx}, \vdelta),
\]
where $\vg = p \cdot \abs{\vx}^{p-2} \odot \vx$ is the gradient of $\norm{\vx}_p^p$.
\end{definition}

To perform this second-order Newton approach, we require an initial point that is fairly close to the optimal. The next lemma states that the optimal solution to the quadratic problem is close to the optimal solution of the $p$-norm problem. This is similar to Lemma 4.8 of \cite{AdilKPS19}, but they only consider the case of $p\geq 2$ and the exact solution to the quadratic problem.

\begin{lemma}
\label{lemma:p-norm-initial-solution}
Let $p>1$, $\epsilon>0$, $\vx^{(0)}\in \R^n$ such that  $\norm{\vx^{(0)}}_2 \leq (1+\epsilon)\cdot \min_{\vx: \ma^\top \vx = \vb} \norm{\vx}_2$, and $\vx^{*}=\argmin_{\vx: \ma^\top \vx = \vb} \norm{\vx}_p$. Then
\[
\norm{\vx^{(0)}}_p^p \leq (1+\epsilon)^p \cdot  n^{\abs{p-2}/2} \norm{\vx^*}_p^p.
\]
\end{lemma}
\begin{proof}
We have two cases. For $1 < p < 2$, we have $\frac{\abs{p-2}}{2} = \frac{2-p}{2}$. Moreover $\norm{\vx}_2 \leq \norm{\vx}_p$. By taking $r=2/p$ and $s=2/(2-p)$ for Holder's inequality over vectors $\vecv = [\vx_i^p]_i, \vu=[1]_i \in \R^n$, respectively, we have $r,s\geq 1$, and
\[
\norm{\vx}_p^p = \sum_{i=1}^n \vx_i^p \leq \norm{\vecv}_{2/p} \norm{\vu}_{2/(2-p)} = \left(\sum_{i=1}^n \vx_i^2\right)^{p/2} n^{(2-p)/2} = n^{(2-p)/2} \norm{\vx}_2^p.
\]
Therefore since by construction $\norm{\vx^{(0)}}_2 \leq \norm{\vxstar}_2$,
\[
\norm{\vx^{(0)}}_p^p \leq 
n^{(2-p)/2} \norm{\vx^{(0)}}_2^p \leq n^{(2-p)/2} \cdot (1+\epsilon)^p \norm{\vxstar}_2^p \leq n^{(2-p)/2} \cdot (1+\epsilon)^p \norm{\vxstar}_p^p.
\]
For $p\geq 2$, $\frac{\abs{p-2}}{2} = \frac{p-2}{2}$, and $\norm{\vx}_p \leq \norm{\vx}_2$. Taking $r=p/2$, $s=p/p-2$ for Holder's inequality over vectors $\vecv=[\vx_i^p]_i$, $\vu=[1]_i\in\R^n$, respectively, we have $r,s\geq 1$, and
\[
\norm{\vx}_2^2 = \sum_{i=1}^n \vx_i^2 \leq \norm{\vecv}_{p/2} \norm{\vu}_{p/(p-2)} = \left(\sum_{i=1}^n \vx_i^p\right)^{2/p} n^{(p-2)/p} = n^{(p-2)/p} \norm{\vx}_p^2.
\]
Thus,
\[
\norm{\vx^{(0)}}_p^p \leq 
\norm{\vx^{(0)}}_2^p \leq (1+\epsilon)^p \norm{\vxstar}_2^p \leq n^{(p-2)/2} \cdot (1+\epsilon)^p \norm{\vxstar}_p^p.
\]
\end{proof}

We now show that by finding an approximate solution to the residual problem, we can move closer to the optimal. The following lemma, which is derived by \cref{lemma:taylor} is useful for this purpose.

\begin{lemma}[\cite{AdilKPS19}]
\label{lemma:p-norm-lambda-shift}
Let $p>1$ and $\lambda \leq \left(\frac{p-1}{p\cdot 4^p}\right)^{1/\min\{1, p-1\}}$. Then for any $\vx,\vdelta\in\R^n$,
\[
\norm{\vx}_p^p - f(\lambda \vdelta) \leq \norm{\vx - \lambda \vdelta}_p^p \leq \norm{\vx}_p^p - \lambda f(\vdelta),
\]
where $f(\vdelta) = \vg^\top \vdelta - \frac{p-1}{p\cdot 2^p} \cdot \gamma_p(\abs{\vx}, \vdelta)$.
\end{lemma}

\noindent
Using this, we can prove the following lemma.

\begin{lemma}[\cite{AdilKPS19}]
\label{lemma:res-improvement}
Let $\alpha, p>1$, $\vx,\widehat{\vdelta}\in\R^n$ such that $\widehat{\vdelta}$ is an $\alpha$-approximate solution to the mixed $(2,p)$-norm residual problem at $\vx$ , i.e., 
\[
\vg^\top \widehat{\vdelta} - \frac{p-1}{p\cdot 2^p} \cdot \gamma_p(\abs{\vx}, \widehat{\vdelta})
\geq \frac{1}{\alpha} \cdot \max_{\ma^\top \vdelta = 0} ~ \vg^\top \vdelta - \frac{p-1}{p\cdot 2^p} \cdot \gamma_p(\abs{\vx}, \vdelta).
\]
Then with $\lambda = \left(\frac{p-1}{p\cdot 4^p}\right)^{1/\min\{1, p-1\}}$ and $\opt = \min_{\vx: \ma^\top \vx = \vb} \norm{\vx}_p^p$,
\[
\norm{\vx - \lambda \widehat{\vdelta}}_p^p - \opt \leq \left( 1- \frac{\lambda}{\alpha} \right) \cdot \left( \norm{\vx}_p^p - \opt \right).
\]
\end{lemma}
\begin{proof}
We define $f(\vdelta) = \vg^\top \vdelta - \frac{p-1}{p\cdot 2^p} \cdot \gamma_p(\abs{\vx}, \vdelta)$ and $\vdeltastar = \argmax_{\ma^\top \vdelta = 0}  f(\vdelta)$. Then by \cref{lemma:taylor}
\[
f(\vdeltahat) \geq \frac{1}{\alpha} f(\vdeltastar) \geq \frac{1}{\alpha} f(\vx - \vxstar) \geq \frac{1}{\alpha} \left( \norm{\vx}_p^p - \norm{\vxstar}_p^p \right) = \frac{1}{\alpha} \left( \norm{\vx}_p^p - \opt \right).
\]
Moreover by \cref{lemma:p-norm-lambda-shift},
\begin{align*}
\norm{\vx - \lambda \widehat{\vdelta}}_p^p - \opt 
\leq 
\norm{\vx}_p^p - \lambda f(\vdeltahat) - \opt
\leq 
- \frac{\lambda}{\alpha}(\norm{\vx}_p^p - \opt) + \norm{\vx}_p^p - \opt \leq (1-\frac{\lambda}{\alpha}) \cdot (\norm{\vx}_p^p - \opt).
\end{align*}
\end{proof}

Then \cref{lemma:p-norm-initial-solution,lemma:res-improvement} imply that Algorithm \ref{alg:p-norm-outer-loop} finds an approximate solution. Note that our algorithm considers the possible errors in solving the subproblems, e.g., the fact that the solution of $\ma^\top \vx = \vb$ might not have a finite representation in fixed-point arithmetic and we have to have some error in our output.

\RestyleAlgo{algoruled}
\IncMargin{0.15cm}
\begin{algorithm}[t]

\textbf{Input:} Full column rank matrix $\ma\in \R^{n\times d}$ ($n>d$), $\vb\in \R^n$, $p > 1$,  $\epsilon>0$, $\alpha>1$, where the residual problem can be solved to $\alpha$-approximation. \\
\textbf{Output:} $\vxhat\in\R^{n}$ such that $\norm{\ma \vxhat - \vb}_2 \leq \epsilon$ and $\norm{\vxhat}_p^p \leq (1+\epsilon) \min_{\vx:\ma^\top \vx = \vb} \norm{\vx}_p^p$.\\

Set $\lambda = \left(\frac{p-1}{p\cdot 4^p}\right)^{1/\min\{1,p-1\}}$ and $T = \ceil{\frac{\alpha}{\lambda} \cdot \log(\frac{1.1^p \cdot n^{\abs{p-2}/2}}{\epsilon} )}$ \label{alg-step:iter-refine-iter-count}\\

Compute $\vx^{(0)} \in \R^n$ such that $\norm{\vx^{(0)}}_2 \leq 1.1 \cdot \norm{\vxstar}_2$  and $\norm{\pi_{\ma} \vx^{(0)} - \vxstar}_2 \leq \frac{\epsilon}{2T}\cdot \norm{\vxstar}_2$, where $\vxstar = \argmin_{\vx:\ma^\top \vx = \vb} \norm{\vx}_2$
\\

\For{t = 1,\ldots, T}{
Compute $\vdelta^{(t)}\in \R^n$ such that $\norm{\pi_{\ma}\vdelta^{(t)}}_2 \leq \frac{\epsilon}{2T \cdot \lambda} \cdot \norm{\vx^*}_2$ and $\vg^\top \vdelta^{(t)} - \frac{p-1}{p \cdot 2^p} \gamma_p(\abs{\vx^{(t-1)}},\vdelta^{(t)}) \leq \alpha \cdot \max_{\ma^\top \vdelta = 0} \vg^\top \vdelta - \frac{p-1}{p \cdot 2^p} \gamma_p(\abs{\vx^{(t-1)}},\vdelta)$ \label{alg-step:iter-refine-res-problem}\\

Set $\vx^{(t)} = \vx^{(t-1)} - \lambda \vdelta^{(t)}$
}
\Return $\vxhat:=\vx^{(T)}$
\caption{$p$-Norm Minimization by Approximately Solving a Series of Residual Problems}
\label{alg:p-norm-outer-loop}
\end{algorithm}

\begin{theorem}[Iterative refinement for $p$-norm minimization]
\label{thm:iter-refine}
Algorithm \ref{alg:p-norm-outer-loop} computes $\vxhat \in \R^n$ such that
\[
\norm{\vxhat}_p^p \leq (1+\epsilon) \norm{\vxstar}_p^p, ~~ \text{ and } ~~ \norm{\pi_{\ma} \vxhat - \vxstar}_2 \leq \epsilon \cdot \norm{\vxstar}_2,
\]
in $O_p(\alpha \cdot \log(n/\epsilon))$ iterations, where $\vxstar = \argmin_{\vx: \ma^\top \vx = \vb} \norm{\vx}_p^p$ and $\alpha$ is the approximation factor for solving the residual problem.
\end{theorem}

\begin{proof}
The number of iterations easily follows by Line \ref{alg-step:iter-refine-iter-count} of the algorithm and noting that $\lambda$ is a function of only $p$. Now since $\vxhat = \vx^{(0)} -\lambda \sum_{t=1}^T \vdelta^{(t)}$, by triangle inequality,
\begin{align*}
\norm{\pi_{\ma} \vxhat - \vxstar}_2 & \leq \norm{\pi_{\ma} \vx^{(0)} - \vxstar}_2 + \lambda\sum_{t=1}^T \norm{\pi_{\ma} \vdelta^{(t)}}_2 \leq \frac{\epsilon}{2T} \cdot \norm{\vxstar}_2 + \lambda \cdot T \cdot \frac{\epsilon}{2T\cdot \lambda} \cdot \norm{\vxstar}_2 \leq \epsilon \cdot \norm{\vxstar}_2,
\end{align*}
where the second inequality follows by construction of $\vx^{(0)}$ and $\vdelta^{(t)}$ (see Algorithm \ref{alg:p-norm-outer-loop}). Finally, since $1-\frac{\lambda}{\alpha} \leq \exp(-\frac{\lambda}{\alpha})$, by \cref{lemma:res-improvement},
\begin{align*}
\norm{\vxhat}_p^p - \norm{\vxstar}_p^p \leq \exp(-\frac{T\cdot \lambda}{\alpha})\norm{\vx^{(0)}}_p^p.
\end{align*}
Therefore since by \cref{lemma:p-norm-initial-solution}, $\norm{\vx^{(0)}}_p^p \leq 1.1^p \cdot  n^{\abs{p-2}/2} \norm{\vx^*}_p^p$, $\norm{\vxhat}_p^p - \norm{\vxstar}_p^p \leq \epsilon \cdot \norm{\vxstar}_p^p$, and the result follows.
\end{proof}

The only remaining part of solving the $p$-norm minimization problem is to devise an algorithm for solving the residual problem. We focus on this for the rest of the section.

\subsection{Solving The Residual Problem}
\label{subsec:solve-res-problem}

The objective of the residual problem (\cref{def:res-problem}) is a linear combination of a linear function and the smoothed $p$-norm function. We first discuss how the linear function can be removed from the objective and added as one of the constraints. This is essentially done by ``guessing'' the value of this linear term for the optimal solution.

\begin{lemma}[\cite{AdilKPS19}]
\label{lemma:binary-search}
Let $p>1$, $\vx\in\R^n$, $\vb\in\R^d$, and $f:\R^n \to \R$ with $f(\vdelta) = \vg^\top \vdelta - \frac{p-1}{p \cdot 2^p} \gamma_p(\abs{\vx},\vdelta)$, where $\vg=p \abs{\vx}^{p-2} \odot \vx$ is the gradient of $\norm{\vx}_p^p$. Moreover suppose $\vdelta^* = \argmax_{\vx:\ma^\top\vx=\vec{0}} f(\vdelta)$ and $f(\vdelta^*) \in \left[ 2^{j-1},2^{j} \right)$ for some $j \in \Z$. Let 
\begin{align}
    \label{eq:res-prob-with-grad-constraint}
   \vdeltahat = \argmin_{\vdelta} ~~ & ~~ \gamma_p(\abs{\vx},\vdelta) \\ \nonumber \st ~~ & ~~
    \vg^\top \vdelta = 2^{j-1}, \\ \nonumber
    ~~ & ~~ \ma \vdelta = 0.
\end{align}
For $\beta>1$, let $\vdeltatil\in\R^n$ such that $\gamma_p(\abs{\vx},\vdeltatil) \leq \beta \cdot \frac{p}{p-1} \cdot 2^{j+p}$ and $\vg^\top \vdeltatil \geq 2^{j-2}$. Then
\begin{enumerate}
    \item $\gamma_p(\abs{\vx}, \vdeltahat) \leq \frac{p}{p-1} \cdot 2^{j+p}$.
    
    \item $f(\mu \vdeltatil) \geq \frac{1}{8 \cdot (4 \beta p)^{1/(\min\{p,2\}-1)}} \cdot \frac{p - 1}{p} \cdot f(\vdeltastar)$ for $\mu=\left(\frac{1}{4\beta p}\right)^{1/(p-1)}$, if $1< p\leq 2$, and $\mu = \frac{1}{8\beta}$, otherwise.
\end{enumerate}
\end{lemma}

The counterpart of \cref{lemma:binary-search} in \cite{AdilKPS19} assumes we have $\vdeltatil$ such that $\vg^\top \vdeltatil = 2^{j-1}$ but since we cannot guarantee the existence of such a vector in fixed-point arithmetic, we replace this with the assumption that $\vg^\top \vdeltatil \geq 2^{j-2}$. However, the proof is similar to that of \cite{AdilKPS19} and only requires adjusting the constants.
\cref{lemma:binary-search} implies that instead of approximately solving the residual problem, we can guess the interval $[2^{j-1},2^{j})$ that contains the optimal objective value of the residual problem and approximately solve a problem of the form \eqref{eq:res-prob-with-grad-constraint}. Therefore to solve the residual problem, we need to iterate over such intervals, compute an approximate solution of \eqref{eq:res-prob-with-grad-constraint} for each, and take the one that achieves the maximum value for the function $f$. Now, the question is how many intervals we need to iterate over.
The next lemma asserts that we only need to try a logarithmic number of intervals.

\begin{lemma}[\cite{AdilKPS19}]
\label{lemma:binary-search-range}
Let $p>1$ and  $\vx^{(0)}=\argmin_{\ma^\top \vx=\vb} \norm{\vx}_2$. Moreover let $\vxbar\in\R^{n}$ such that 
\[
\norm{\vxbar}_p^p > (1+\epsilon) \min_{\ma^\top \vx=\vb} \norm{\vx}_p^p.
\]
Let $f(\vdelta) = \vg^\top \vdelta - \frac{p-1}{p \cdot 2^p} \gamma_p(\abs{\vxbar},\vdelta)$ and $\lambda = \left(\frac{p-1}{p\cdot 4^p}\right)^{1/\min\{1,p-1\}}$. Then
\[
\min_{\ma^\top \vdelta = 0}f(\vdelta) \in \left[\frac{\epsilon \norm{\vx^{(0)}}_p^p}{n^{\abs{p-2}/2}}, \frac{ \norm{\vx^{(0)}}_p^p}{\lambda}\right].
\]
\end{lemma}

\cref{lemma:binary-search-range} asserts that if our current solution is not a $(1+\epsilon)$-approximation, we only need to iterate over $\ceil{\log(\lambda n^{\abs{p-2}/2}/\epsilon)} +1$ intervals for \eqref{eq:res-prob-with-grad-constraint} in order to approximately solve the residual problem. Moreover by substituting $\alpha = 16 \beta \cdot \frac{p}{p-1}$ from \cref{lemma:binary-search}, in Algorithm \ref{alg:p-norm-outer-loop}, we have 
\[
T = \ceil{\frac{16\beta \cdot p}{\lambda \cdot (p-1)} \cdot \log(\frac{1.1^p \cdot n^{\abs{p-2}/2}}{\epsilon} )}.
\]
Then Line \ref{alg-step:iter-refine-res-problem} of Algorithm \ref{alg:p-norm-outer-loop} can be performed by finding a $\beta$-approximation for $\ceil{\log(\lambda n^{\abs{p-2}/2}/\epsilon)} +1$ instances of problem \eqref{eq:res-prob-with-grad-constraint} and taking the maximum. However, note that searching over such instances only improves the solution if the current solution is not a $(1+\epsilon)$-approximation (see \cref{lemma:binary-search-range}). Therefore in this approach, we need to add a conditional statement to the loop of Algorithm \ref{alg:p-norm-outer-loop} to break and return $\vx^{(t-1)}$ if $\norm{\vx^{(t)}}_p^p > \norm{\vx^{(t-1)}}_p^p$. 

Now, we need to approximately solve mixed $(2,p)$-norm minimization problems of the form \eqref{eq:res-prob-with-grad-constraint}. For the rest of the section, we focus on the case of $p\geq 2$. Our first approach is to solve such problems by approximately solving instances of a mixed $(2,\infty)$-norm minimization problem.

The next lemma connects the smoothed $p$-norm function to a mixed $(2,\infty)$-norm function, which in turn allows us to approximately minimize the smoothed $p$-norm function, by approximately minimizing the mixed $(2,\infty)$-norm function.

\begin{lemma}
\label{lemma:mixed-smooth}
Let $\mabar\in\R^{n\times d}$, $\vbbar\in\R^d$, $\vt\in\R_{\geq 0}^n$, $p\geq 2$, $\widehat{j}\in\Z$,
\[
\vdeltahat = \argmin_{\vdelta:\mabar^\top \vdelta = \vbbar} \gamma_p(\vt, \vdelta),
\]
and $\gamma_p(\vt, \vdeltahat) \in [2^{\widehat{j}-1},2^{\widehat{j}})$. Let $\vr,\vs\in\R^n$, and $q\in\Z$ with $q\leq -2$, $\vr_i = \frac{\vt_i^{p-2}}{2^{\max\{\widehat{j},q\}+2}}$, and $\vs_i=\left(\frac{1}{2^{\max\{\widehat{j},q+1\}+p+1}}\right)^{1/p}$. Then
\[
\min_{\vdelta:\mabar^\top \vdelta = \vbbar}\norm{\vdelta}_{\vr}^2 + \norm{\vs \odot \vdelta}_{\infty} \leq 1,
\]
and if $\vdeltabar$ such that $\norm{\vdeltabar}_{\vr}^2 + \norm{\vs \odot \vdeltabar}_{\infty} \leq \theta$, then
\[
\gamma_{p}(\vt,n^{-1/p}\cdot \vdeltabar) \leq p \cdot \left(4\theta +2^{p+1}\theta^p \right) \cdot \max\{\gamma_p(\vt,\vdeltahat),2^q\}.
\]
\end{lemma}
\begin{proof}
First, note that since $p\geq 2$, for $x,t\in \R$ and $t\geq 0$, $\gamma_{p}(t,x) \geq \max\{t^{p-2} x^2, \abs{x}^p\}$. Therefore,
\[
\sum_{i=1}^n \vt_i^{p-2} \vdeltahat_i^2 + \abs{\vdeltahat_i}^p \leq 2\gamma_p(\vt,\vdeltahat) < 2^{\widehat{j}+1}.
\]
Then by the construction of $\vr,\vs$ and since $\vt_i^{p-2} \vdeltahat_i^2$ and $\abs{\vdeltahat_i}^p$ are nonnegative, 
\[
\norm{\vdeltahat}_{\vr}^2 \leq \frac{1}{2} ~~ \text{, and } ~~ \norm{\vs \odot \vdeltahat}_{\infty} \leq \norm{\vs \odot \vdeltahat}_{p} \leq \frac{1}{2}.
\]
Therefore 
\[
\min_{\vdelta:\mabar^\top \vdelta = \vbbar}\norm{\vdelta}_{\vr}^2 + \norm{\vs \odot \vdelta}_{\infty}  \leq \norm{\vdeltahat}_{\vr}^2 + \norm{\vs \odot \vdeltahat}_{\infty} \leq 1.
\]
Now since for $x,t\in\R$ and $\abs{x} > t\geq 0$, $\frac{p}{2} \abs{\vx}^p > \abs{\vx}^p + (\frac{p}{2} - 1) t^p$, $\gamma_p(t,x) \leq \frac{p}{2} (t^{p-2} x^2 + \abs{x}^p)$. Therefore,
\[
\gamma_{p}(\vt,n^{-1/p} \cdot\vdeltabar) \leq \frac{p}{2} \sum_{i=1}^n  \vt_i^{p-2} (n^{-1/p} \cdot\vdeltabar_i)^2 + \frac{p}{2} \sum_{i=1}^n \abs{n^{-1/p} \cdot\vdeltabar_i}^p.
\]
We now bound the terms on the right-hand side. Since $\norm{\vdeltabar}_{\vr}^2 + \norm{\vs \odot \vdeltabar}_{\infty} \leq \theta$,
\[
\sum_{i=1}^n \frac{t_i^{p-2}}{2^{\max\{\widehat{j},q\}+2}} \cdot (n^{-1/p} \cdot\vdeltabar_i)^2 \leq \sum_{i=1}^n \frac{t_i^{p-2}}{2^{\max\{\widehat{j},q\}+2}} \cdot \vdeltabar_i^2 = \norm{\vdeltabar}_{\vr}^2 \leq \theta.
\]
Moreover
\begin{align*}
\norm{n^{-1/p}\vdeltabar}_p & = 2^{(\max\{\widehat{j},q+1\}+p+1)/p} \norm{n^{-1/p} \cdot \vs \odot \vdeltabar}_{p}\leq 2^{(\max\{\widehat{j},q+1\}+p+1)/p} \cdot n^{1/p}\norm{n^{-1/p}\cdot \vs \odot \vdeltabar}_{\infty} 
\\ & \leq 
2^{(\max\{\widehat{j},q+1\}+p+1)/p} \cdot \theta.
\end{align*}
Thus since $\max\{\gamma_p(t,\vdeltahat), 2^q\} \geq 2^{\max\{\widehat{j},q+1\}-1}$,
\[
\gamma_{p}(\vt,n^{-1/p}\cdot \vdeltabar) \leq \frac{p}{2} \left( 2^{\max\{\widehat{j},q\}+2} \theta + 2^{\max\{\widehat{j},q+1\}+p+1} \theta^p \right) \leq p \cdot \left( 4 \theta + 2^{p+1} \theta^p \right)  \cdot \max\{\gamma_p(\vt,\vdeltahat),2^q\}.
\]

\end{proof}

\cref{lemma:mixed-smooth} limits the values of $\vr$ and $\vs$ if $\widehat{j}$ is small. This combined with \cref{lemma:binary-search} implies that if $\argmax_{\ma^\top \vdelta = 0} ~ \vg^\top \vdelta - \frac{p-1}{p\cdot 2^p} \cdot \gamma_p(\abs{\vx}, \vdelta) < 2^j$, with $\mabar = \begin{bmatrix}
    \ma | \vg
\end{bmatrix}$ and $\vbbar=\begin{bmatrix}
    \vec{0} \\ 2^{j-1}
\end{bmatrix}$, we only need to try $j+p-q+1+ \log(\frac{p}{p-1})$ values for $\vr$ and $\vs$ to find a vector $\vdeltabar$ with small $\gamma_p$ value.

Although \cref{lemma:mixed-smooth} implies that optimizing over the mixed $(2,\infty)$-norm function gives a vector $\vdeltabar$ with a small value for the $\gamma_p$ function, note that after multiplying $\vdeltabar$ by $n^{-1/p}$, the value of $\vg^\top \vdelta$ decreases. Therefore we cannot use \cref{lemma:binary-search} to bound the value of the residual function for $n^{-1/p} \vdeltabar$. To obtain such a bound, we use the following lemma.

\begin{lemma}[\cite{AdilKPS19}]
\label{lemma:gamma-scale}
Let $p>1$, $\lambda\geq 0$, and $\vt,\vdelta\in\R^n$ with $\vt \geq 0$. Then
\[
\min\{\lambda^2, \lambda^p\} \gamma_p(\vt,\vdelta) \leq \gamma_p(\vt,\lambda \vdelta) \leq \max\{\lambda^2, \lambda^p\} \gamma_p(\vt,\vdelta).
\]
\end{lemma}

Now by further scaling of $\vdeltabar$, we obtain a vector that gives a constant factor approximation for the residual problem. Note that by \cref{lemma:binary-search} and picking $q=\min\{j+p, -2\}$ in \cref{lemma:mixed-smooth}, we have  $\gamma_{p}(\vt,n^{-1/p}\cdot \vdeltabar) \leq p \cdot \left( 4 \theta + 2^{p+1} \theta^p \right) \cdot \frac{p}{p-1} \cdot 2^{j+p}$ assuming that the optimal value of the residual function is in $[2^{j-1},2^j)$. Moreover, the optimal value of the residual function is bounded by \cref{lemma:binary-search-range}.

\begin{lemma}
Let $p \geq 2$, $\vx\in\R^n$, $\vb\in\R^d$, and $f:\R^n \to \R$ with $f(\vdelta) = \vg^\top \vdelta - \frac{p-1}{p \cdot 2^p} \gamma_p(\abs{\vx},\vdelta)$, where $\vg=p \abs{\vx}^{p-2} \odot \vx$ is the gradient of $\norm{\vx}_p^p$. Moreover suppose $\vdelta^* = \argmax_{\vx:\ma^\top\vx=\vec{0}} f(\vdelta)$ and $f(\vdelta^*) \in \left[ 2^{j-1},2^{j} \right)$ for some $j \in \Z$. Moreover let $\vdeltabar\in\R^n$ and $\theta \geq 1$ such that $\vg^\top \vdelta \geq 2^{j-2}$ and $\gamma_{p}(\vt,n^{-1/p}\cdot \vdeltabar) \leq p \cdot \left( 4 \theta + 2^{p+1} \theta^p \right) \cdot \frac{p}{p-1} \cdot 2^{j+p}$.
Then for $\lambda = \frac{1}{8 p \cdot \left(4\theta +2^{p+1}\theta^p \right)}$,
\[
f(n^{-2/p} \lambda\vdeltabar) \geq \frac{n^{-2/p}}{64 p \cdot \left(4\theta +2^{p+1}\theta^p \right)} f(\vdeltastar).
\]
\end{lemma}
\begin{proof}
By \cref{lemma:gamma-scale} since $p\geq 2$ and $\lambda<1$, 
\begin{align*}
 \vg^\top (n^{-2/p} \lambda\vdeltabar) - \frac{p-1}{p \cdot 2^p} \gamma_p(\abs{\vx},n^{-2/p} \lambda \vdeltabar) 
 & \geq
n^{-2/p} \lambda \cdot 2^{j-2}  - n^{-2/p} \lambda^2 \cdot \frac{p-1}{p \cdot 2^p} \cdot \gamma_p(\vt, n^{-1/p}\vdeltabar)
\\ & \geq
n^{-2/p} \lambda \cdot 2^{j-2}  - n^{-2/p} \lambda^2 \cdot \frac{p-1}{p \cdot 2^p} p \cdot \left(4\theta +2^{p+1}\theta^p \right) \frac{p}{p-1} 2^{j+p}
\\ & = n^{-2/p} \lambda \cdot 2^{j-2}  - n^{-2/p}\lambda^2 \cdot  p \cdot \left(4\theta +2^{p+1}\theta^p \right) 2^{j}
\\ & \geq
\frac{n^{-2/p}}{8 p \cdot \left(4\theta +2^{p+1}\theta^p \right)} (2^{j-2} - 2^{j-3})
\\ & \geq
\frac{n^{-2/p}}{64 p \cdot \left(4\theta +2^{p+1}\theta^p \right)} \left( \vg^\top \vdeltastar - \frac{p-1}{p \cdot 2^p} \gamma_p(\abs{\vx},\vdeltastar) \right).
\end{align*}
\end{proof}

Now that we established we can find a constant factor approximation for the residual problem by guessing the value of the linear term in the residual function and approximately solving the mixed $(2,p)$-norm problem directly or by approximately solving the mixed $(2,\infty)$-norm problem, we discuss how adding the linear constraint affects the condition number of our matrix. This is important since the bit complexity of inversion and inverse maintenance depends on the condition number of the matrix.

We show that the gradient term (arising from Taylor's expansion of the $p$-norm --- see \cref{lemma:taylor}) can be incorporated to the inverse because if the current solution is not close to the optimum of $p$-norm, the projection of the gradient vector into the kernel of matrix $\ma$ is large, and therefore the matrix $\mabar:=[\ma|\vg]$ does not have a large condition number. We first show that the projection is large. 

\begin{lemma}
Let $\vb\in\R^{d}$, and $\ma\in\R^{n\times d}$ be a matrix with full column rank.
Let $p\geq 1$, $\vg:= p \cdot \abs{\vxhat}^{p-2} \odot \vxhat$, be the gradient of $\norm{\vx}_p^p$ at $\vxhat$, and
\[
\vxstar = \argmin_{\vx:\ma^\top \vx = \vb} \norm{\vx}_p^p.
\]
Let
$0<\veps<1$,
and $\vxhat\in \R^{d}$ such that $\norm{\ma^\top \vxhat - \vb}_2 \leq \frac{\veps}{2\kappa^{3.5}}$ and $\norm{\vxhat}_p^p > (1+\veps) \norm{\vxstar}_p^p$.
Let $\kappa>1$, $\norm{\vxstar}_p^p \geq 1/\kappa$, and
\[\norm{\ma^\top \ma}_{\fro},\norm{(\ma^\top \ma)^{-1}}_{\fro},\norm{\vg}_2,\norm{\vxstar - \vxhat}_2 \leq \kappa.\]
Then $\norm{(\mi-\ma(\ma^{\top} \ma)^{-1} \ma^\top)\vg}_2 \geq \frac{\veps}{2\kappa^2}$.
\end{lemma}
\begin{proof}
Since $\norm{\vx}_p^p$ is a convex function, we have
\[
(1+\veps) \norm{\vxstar}_p^p + \vg^\top (\vxstar - \vxhat) <\norm{\vxhat}_p^p + \vg^\top (\vxstar - \vxhat) \leq \norm{\vxstar}_p^p.
\]
Therefore
\[
\veps \norm{\vxstar}_p^p \leq \vg^\top (\vxhat - \vxstar).
\]
We have
\[
(\vxhat - \vxstar) = (\mi-\ma(\ma^{\top} \ma)^{-1} \ma^\top)(\vxhat - \vxstar) + \ma(\ma^{\top} \ma)^{-1} \ma^\top(\vxhat - \vxstar)
\]
Since $\ma^\top \vxstar = \vb$, $\norm{\ma^\top (\vxhat - \vxstar)}_2 \leq \frac{\veps}{2\kappa^{3.5}}$. Therefore by Cauchy-Schwarz and triangle inequalities,
\begin{align*}
\veps \frac{1}{\kappa} 
& \leq 
\norm{\vg^\top (\mi-\ma(\ma^{\top} \ma)^{-1} \ma^\top)}_2 \norm{\vxhat - \vxstar}_2 + \norm{\vg^\top}_2 \norm{\ma(\ma^{\top} \ma)^{-1}}_2 \norm{\ma^\top (\vxhat - \vxstar) }_2 
\\ & \leq 
\norm{\vg^\top (\mi-\ma(\ma^{\top} \ma)^{-1} \ma^\top)}_2 \kappa + \kappa^{2.5} \cdot \frac{\veps}{2\kappa^{3.5}}
\end{align*}
Therefore
\[
\norm{\vg^\top (\mi-\ma(\ma^{\top} \ma)^{-1} \ma^\top)}_2 \geq \frac{\veps}{2\kappa^2}.
\]
\end{proof}

The next lemma states that if we add a new column $\vg$ to the matrix $\ma$ forming the matrix $\mabar=[\ma|\vg]$, given that the projection of $\vg$ into the kernel of $\ma$ is not small, the condition number of $\mabar$ is small.

\begin{lemma}
\label{lemma:cond-num-column-add}
Let $\ma \in \R^{n\times d}$, $n > d$, be a matrix with full column rank. Moreover let $\vg \in \R^n$.
Suppose $\kappa>1$, and
\[\norm{\ma^\top \ma}_{\fro},\norm{(\ma^{\top} \ma)^{-1}}_{\fro},\norm{\vg}_2, 1/\norm{(\mi-\ma(\ma^{\top} \ma)^{-1} \ma^\top)\vg}_2 \leq \kappa.\] 
Then
$\norm{\mabar^\top \mabar}_{\fro}, \norm{(\mabar^\top \mabar)^{-1}}_{\fro}\leq 8\kappa^7$,
where
$
\mabar = \begin{bmatrix}
\ma | \vg
\end{bmatrix}.
$
\end{lemma}
\begin{proof}
First note that $\vg$ is not in the range of $\ma$, since if $\vg=\ma\vy$, then
\[
\norm{(\mi-\ma(\ma^{\top} \ma)^{-1} \ma^\top)\vg}_2 = \norm{\ma\vy-\ma(\ma^{\top} \ma)^{-1} \ma^\top \ma\vy}_2 = 0,
\]
which is in contrast with the assumption. Therefore $\mabar$ has full column rank and $\mabar^\top \mabar$ is invertible. Now note that
\[
\mabar^\top \mabar = \begin{bmatrix}
\ma^\top \ma & \ma^\top \vg \\
\vg^{\top} \ma & \vg^\top \vg.
\end{bmatrix}
\]
Therefore by triangle inequality and consistency of the Frobenius norm.
\[
\norm{\mabar^\top \mabar}_{\fro} \leq \norm{\ma^\top \ma}_{\fro} + 2\norm{\ma^\top \vg} + \norm{\vg}_2^2 \leq 4 \kappa^2 \leq 8\kappa^7.
\]
Let $s:= \norm{\vg}_2^2 - \vg^\top \ma (\ma^\top \ma)^{-1} \ma^\top \vg$ be the Schur complement of $\mabar^\top \mabar$. By matrix inversion lemma, since $\ma^\top \ma$ and $\mabar^\top \mabar$ are invertible, $s$ is also invertible and
\[
(\mabar^\top \mabar)^{-1} = \begin{bmatrix}
(\ma^\top \ma)^{-1} + \frac{(\ma^\top \ma)^{-1} \ma^\top \vg \vg^\top \ma (\ma^\top \ma)^{-1}}{s} & -\frac{(\ma^\top \ma)^{-1} \ma^\top \vg }{s}
\\
-\frac{\vg^\top \ma (\ma^\top \ma)^{-1}}{s} & \frac{1}{s}
\end{bmatrix}.
\]
Now note that
\begin{align*}
\norm{(\mi-\ma(\ma^{\top} \ma)^{-1} \ma^\top)\vg}_2^2 
& = 
\vg^\top (\mi-\ma(\ma^{\top} \ma)^{-1} \ma^\top) (\mi-\ma(\ma^{\top} \ma)^{-1} \ma^\top) \vg
\\ & = \norm{\vg}_2^2 - 2 \cdot  \vg^\top \ma(\ma^{\top} \ma)^{-1} \ma^\top \vg + \vg^\top \ma(\ma^{\top} \ma)^{-1} \ma^\top \ma(\ma^{\top} \ma)^{-1} \ma^\top \vg
\\ & = 
\norm{\vg}_2^2 - \vg^\top \ma(\ma^{\top} \ma)^{-1} \ma^\top \vg
\\ & = s.
\end{align*}
Therefore by assumption $1/s \leq \kappa^2$. Now by the triangle inequality and the consistency of the Frobenius norm, we have
\begin{align*}
\norm{(\mabar^\top \mabar)^{-1}}_{\fro} 
& \leq
\norm{(\ma^\top \ma)^{-1} + \frac{(\ma^\top \ma)^{-1} \ma^\top \vg \vg^\top \ma (\ma^\top \ma)^{-1}}{s}}_{\fro} + 2\cdot \norm{\frac{(\ma^\top \ma)^{-1} \ma^\top \vg }{s}}_{\fro} + \frac{1}{s}
\\ & \leq 
\norm{(\ma^\top \ma)^{-1}}_{\fro} + \kappa^2(\norm{(\ma^\top \ma)^{-1} \ma^\top \vg }_{\fro}^2 + 2\cdot\norm{(\ma^\top \ma)^{-1} \ma^\top \vg }_{\fro} +1)
\\ & \leq 
\kappa + \kappa^2 (\norm{(\ma^\top \ma)^{-1} \ma^\top \vg }_{\fro} +1)^2
\leq \kappa + \kappa^2 (\kappa^{2.5}+1)^2 \leq 8\kappa^7.
\end{align*}
\end{proof}

\subsection{Weighted Linear Regression with Equality Constraints}
\label{subsec:weighted-lin-reg}

In this section, we examine computing a high-accuracy solution to a weighted constrained linear regression problem using an erroneous inverse of a preconditioner. The inverse has error because we are working under the fixed-point arithmetic. To approximately solve the mixed $(2,p)$-norm minimization problem, or the mixed $(2,\infty)$-norm minimization problem, we need to solve $\Otil_p(n^{1/3})$ such weighted constrained linear regression problem. We later discuss that for these problems, using inverse maintenance techniques, we can maintain an erroneous constant-factor spectral approximation of the inverse as the perconditioner.

Note that when solving the problem $\argmin_{\vx: \ma^\top \vx = \vb} \frac{1}{2} \norm{\vx}_{\mw}^2$, we require the error of the solution to be small in two different norms: the norms defined on matrices $\mw$ and $\pi_{\ma}$. Interestingly, as we see in the next lemma, $\norm{\vxstar}_2$ is within a factor $R$ of $\norm{\pi_{\ma} \vxstar}_2$. Therefore, we do not need $\log(\kappa(\ma))$ iterations of Richardson to achieve this.

\begin{lemma}
\label{lemma:2-proj-norm-connection-xstar}
Let $\ma\in \R^{n\times d}$ with full column rank, $\vb\in\R^d$, $\mw\in\R^{n\times n}$ be a diagonal matrix with $R \mi \succeq \mw \succeq \mi$, and
\[
\vxstar := \argmin_{\vx:\ma^\top \vx = \vb} \frac{1}{2} \norm{\vx}_{\mw}^2.
\]
Then 
\[
\norm{\vxstar}_2\leq R \cdot \norm{\pi_{\ma} \vxstar}_2.
\]
\end{lemma}
\begin{proof}
By \cref{lemma:constrained-richardson}, $\vxstar = \mw^{-1} \ma (\ma^\top \mw^{-1} \ma)^{-1} \vb$. Therefore since $\mw^{-1} \preceq \mi$,
\begin{align*}
\norm{\vxstar}_2 & \leq \vb^\top (\ma^\top \mw^{-1} \ma)^{-1} \ma^\top \mw^{-2} \ma (\ma^\top \mw^{-1} \ma)^{-1} \vb
\\ & \leq
\vb^\top (\ma^\top \mw^{-1} \ma)^{-1} \ma^\top \mw^{-1} \ma (\ma^\top \mw^{-1} \ma)^{-1} \vb
\\ & =
\vb^\top (\ma^\top \mw^{-1} \ma)^{-1} \vb.
\end{align*}
Moreover
\begin{align*}
\norm{\pi_{\ma} \vxstar}_2 & = {\vxstar}^\top \ma (\ma^\top \ma)^{-1} \ma^\top \vxstar
\\ & = 
\vb^\top (\ma^\top \mw^{-1} \ma)^{-1} \ma^\top \mw^{-1} \ma (\ma^\top \ma)^{-1} \ma^\top \mw^{-1} \ma (\ma^\top \mw^{-1} \ma)^{-1} \vb
\\ & =
\vb^{\top} (\ma^\top \ma)^{-1} \vb.
\end{align*}
Now note that since $\frac{1}{R} \mi \preceq \mw^{-1}$, we have $\frac{1}{R} \ma^\top \ma \preceq \ma^\top \mw^{-1} \ma$. Therefore $R(\ma^\top \ma)^{-1} \succeq (\ma^\top \mw^{-1} \ma)^{-1}$. Thus
\[
R \cdot \norm{\pi_{\ma} \vxstar}_2 = R \cdot \vb^{\top} (\ma^\top \ma)^{-1} \vb \geq \vb^\top (\ma^\top \mw^{-1} \ma)^{-1} \vb \geq \norm{\vxstar}_2.
\]
\end{proof}

We are now equipped to prove the main result of this subsection, which is the main subprocedure for both mixed $(2,\infty)$-norm minimization and mixed $(2,p)$-norm minimization.

\constrainedRichardson*

\begin{proof}
Note that the gradient of $\frac{1}{2} \norm{\vx}_{\mw}^2$ is $\mw \vx$ and for any $\vx$ in the kernel of $\ma^\top$, $\ma^\top(\vx^* + \vx) = \vb$. Therefore $\mw \vx^*$ should be orthogonal to the kernel of $\ma^\top$. Therefore there exists $\vy$ such that $\ma \vy = \mw \vxstar$. Therefore $\vx^* = \mw^{-1} \ma \vy$ and $\vb = \ma^\top \vxstar = \ma^\top \mw^{-1} \ma \vy$. Solving for $\vy$, we have $\vy = (\ma^\top \mw^{-1} \ma)^{-1}\vb$. Thus by $\ma \vy = \mw \vxstar$, we have
$\vxstar = \mw^{-1} \ma (\ma^\top \mw^{-1} \ma)^{-1}\vb$. 

Since $\ma^\top \mw^{-1} \ma$ is full-rank, $(\ma^\top \mw^{-1} \ma)^{-1}\vb$ corresponds to a linear system of the form $(\ma^\top \mw^{-1} \ma) \vz = \vb$.
Therefore by using Richardson's iteration (\cref{lemma:richardson}) as $\vz^{(k+1)} = \vz^{(k)} - \mmtil^{-1}(\ma^\top \mw^{-1} \ma \vz^{(k)} - \vb)$ with $\vz^{(0)}$ and $\norm{\mmtil^{-1} - \mm^{-1}}_{\fro} \leq \frac{\veps}{d \cdot \lambda\cdot \norm{\ma^\top \mw^{-1} \ma}_2}$, we can guarantee that
\[
\norm{\vz^{(k)} - \vz^*}_{\mm} \leq (1-\lambda^{-1} +\veps)^{k} \norm{\vz^*}_{\mm},
\]
where $\vz^* = (\ma^\top \mw^{-1} \ma)^{-1} \vb$. Since $\vxstar = \mw^{-1} \ma \vz^*$, we have
\[
\norm{\vz^{(k)} - \vz^*}_{\ma^\top \mw^{-1} \ma} \leq \lambda \cdot (1-\lambda^{-1} +\veps)^{k} \norm{\vz^*}_{\ma^\top \mw^{-1} \ma}.
\]
Setting $\vx^{(k)} = \mw^{-1} \ma \vz^{(k)}$, we have
\begin{align*}
\norm{\vz^{(k)} - \vz^*}_{\ma^\top \mw^{-1} \ma}^2 
& = 
(\vz^{(k)} - \vz^*)^\top \ma^\top \mw^{-1} \ma (\vz^{(k)} - \vz^*)
\\ & =
(\vz^{(k)} - \vz^*)^\top \ma^\top \mw^{-1} \mw \mw^{-1} \ma (\vz^{(k)} - \vz^*)
\\ & = \norm{\vx^{(k)} - \vxstar}_{\mw}^2.
\end{align*}
Similarly,
\begin{align*}
\norm{\vz^*}_{\ma^\top \mw^{-1} \ma}^2 = (\vz^*)^\top \ma^\top \mw^{-1} \ma \vz^* = (\vz^*)^\top \ma^\top \mw^{-1} \mw \mw^{-1} \ma \vz^* = \norm{\vxstar}_{\mw}^2.
\end{align*}
Thus
\begin{align}
\label{eq:weighted-norm-error}
\norm{\vx^{(k)} - \vxstar}_{\mw} \leq \lambda \cdot (1-\lambda^{-1} +\veps)^{k} \norm{\vxstar}_{\mw}.
\end{align}
Therefore, taking $k>\frac{1}{\lambda^{-1}-\veps}\log(R^2\cdot \lambda/\epsilon)$, since $R \mi \succeq \mw\succeq \mi$, we have
\[
\norm{\vx^{(k)} - \vxstar}_{2} \leq \norm{\vx^{(k)} - \vxstar}_{\mw} \leq \frac{\epsilon}{R^2} \cdot \norm{\vxstar}_{\mw} \leq \frac{\epsilon}{R} \cdot \norm{\vxstar}_{2}.
\]
Thus by triangle inequality,
\[
\norm{\vx^{(k)}}_{\mw} \leq (1+\epsilon) \norm{\vxstar}_{\mw}.
\]
Moreover, since projection only decreases the length of a vector and by \cref{lemma:2-proj-norm-connection-xstar}, we have
\[
\norm{\pi_{\ma}(\vx^{(k)} - \vxstar)}_{2} \leq \norm{\vx^{(k)} - \vxstar}_{2} \leq \frac{\epsilon}{R} \cdot \norm{\vxstar}_{2} \leq \epsilon \cdot \norm{\pi_{\ma}\vxstar}_2.
\]
\end{proof}

The multiplicative weights update algorithms that we employ in the next two sections are susceptible to error (as opposed to interior point methods). More specifically, they require high-accuracy solutions to the weighted linear regression problems in the sense that if we output $\vxhat\in\R^n$ for the problem $\vxstar = \argmin_{\vx:\ma^\top \vx=\vb} \frac{1}{2} \norm{\vx}_{\mw}^2$, we need $\norm{\vxhat - \vxstar}_2 \leq \frac{1}{\poly(n)}$. This is required to guarantee certain potential functions are increasing rapidly and is implied by our iterative method with preconditioning for solving the weighted linear regression problems by taking an appropriate error parameter (Richardson's iteration of \cref{lemma:constrained-richardson}).

\subsection{Mixed \texorpdfstring{$(2,\infty)$}{(2,∞)}-Norm Minimization}
\label{subsec:two-inf-norm}

In this section, we discuss a multiplicative weights update approach to find a constant-factor approximation to the weighted mixed $(2,\infty)$-norm problem. This problem is of independent interest, but as shown in \cref{subsec:solve-res-problem}, it can also be used to find a high-accuracy solution to the $p$-norm minimization problem.

\begin{theorem}[continues=thm:main-mixed-2-infty]

Let $\ma\in\R^{n\times d}$ and $\vb\in \R^{d}$, $\vr,\vs\in\R^{n}_{\geq 0}$, $n\geq d$, such that the condition number of $\ma$ is less than $\kappa$ and the bit complexity of $\vr,\vs$, and $\vb$ are bounded by $\log(\kappa)$.
For $0<\epsilon<1$, and $m\leq n^{(\omega-2)/(\omega-1)}$, there is an algorithm that outputs $\vxhat$ such that $\norm{\pi_{\ma} (\vxhat - \vxstar)}_2 \leq \epsilon \norm{\pi_{\ma}\vxstar}_2$ and
\begin{align}
    \norm{\vxhat}_{\vr}^2 + \norm{\vs \odot\vxhat}_{\infty} =O(1) \cdot(\norm{\vxstar}_{\vr}^2 + \norm{\vs \odot\vxstar}_{\infty}),
\end{align}
where $\vxstar = \argmin_{\vx:\ma^\top \vx = \vb} \norm{\vx}_{\vr}^2 + \norm{\vs \odot\vx}_{\infty}$,
in time
\[
\Otil_p( (n^{\omega}+n^{7/3} \cdot  \log^2(1/\epsilon)) \log(\alpha_2\kappa/\epsilon) \log(\alpha_1 \kappa) \log(\kappa/\epsilon)),
\]
where $\alpha_1=1/(\min_{i\in[n]} \vr_i+\vs_i^2)$ and $\alpha_2 = (\max_{i\in[n]} \vr_i + \max_{i\in[n]} \vs_i)/\min_{i\in[n]} \vr_i$.
Moreover, for sparse matrices, there is an algorithm that returns an output with the same guarantees 
with probability at least $1-n^{-10} \cdot \log(\alpha_2)$
in time 
\[
\Otil\left(\left(n^{\omega} m^{7/3-\omega} + \nnz(\ma)\cdot m^{4/3}\cdot n \cdot \log^2(1/\epsilon) + n^{7/3} \cdot \log^2(1/\epsilon) \right)\cdot \log(\alpha_2\kappa/\epsilon) \log(\alpha_1 \kappa)\log^2(\kappa/\epsilon))\right).
\]

\end{theorem}

In this section we present our  multiplicative weight update algorithm to solve a mixed $(2,\infty)$-norm problem --- see Algorithm \ref{alg:mwu-inf}.

We first show that a certain weighted linear regression problem is related to the mixed $(2,\infty)$-norm minimization, and bounds on the optimum value of the mixed norm gives several bounds on the optimum solution of the weighted linear regression problem. Such a weighted linear regression problem is solved in every iteration of our multiplicative weights update algorithm (Algorithm \ref{alg:mwu-inf})

\begin{lemma} 
\label{lemma:all-bounds-w}
Let $\vr,\vs \in \R^{n}_{\geq 0}$, $\ma \in \R^{n\times d}$, and $\vb\in\R^d$. Let $\vw \in \R_{\geq 0}^{n}$, and for all $i\in[n]$, define
$\vwtil_i = \norm{\vw}_1 \cdot \vr_i + \left( \vw_i + \frac{1}{n} \norm{\vw}_1 \right) \cdot \vs_i^2$ (similar to Step \ref{alg:mwu-inf-step-wtil} of Algorithm \ref{alg:mwu-inf}). Let
\begin{align*}
\vdeltabar = 
\argmin_{\vdelta: \ma^\top \vdelta = \vb} \norm{\vdelta}_{\vr}^2 + \norm{\vs \odot \vdelta}_{\infty}, ~~ \text{, and } ~~ \vdeltahat = \argmin_{\vdelta: \ma^\top \vdelta = \vb} \norm{\vdelta}_{\vwtil}^2,
\end{align*}
and suppose
\begin{align}
    \label{eq:deltabar-assumption}
    \norm{\vdeltabar}_{\vr}^2 + \norm{\vs \odot \vdeltabar}_{\infty} \leq 1.
\end{align}
Then
\begin{enumerate}
    \item $\norm{\vdeltahat}_{\vwtil}^2 \leq 6 \cdot \norm{ \vw}_1$. \label{lemma:all-bounds-w-part-1} 
    
    \item $\norm{\vdeltahat}_{\vr}^2 \leq 6$. \label{lemma:all-bounds-w-part-2}
    
    \item $\sum_{i=1}^n \vw_i \abs{\vs_i \vdeltahat_i} \leq \sqrt{6} \norm{\vw}_1$. \label{lemma:all-bounds-w-part-3}
    
    \item For all $i\in[n]$, $\abs{\vs_i \vdeltahat_i} \leq \sqrt{6 n}$. \label{lemma:all-bounds-w-part-4}
\end{enumerate}
\end{lemma}
\begin{proof}
By definition of $\vwtil$, we have
\begin{align}
\norm{\vdeltahat}_{\vwtil}^2
& =
\min_{\vdelta: \ma^\top\vdelta = \vb} \sum_{i=1}^n \left(\norm{\vw}_1 \cdot \vr_i+ (\vw_i+\frac{1}{n} \norm{\vw}_1) \vs_i^2 \right) \vdelta_i^2 
\\ & \leq 
\sum_{i=1}^n \left(\norm{\vw}_1 \cdot \vr_i + (\vw_i+\frac{1}{n} \norm{\vw}_1) \vs_i^2 \right) \vdeltabar_i^2
\label{eq:less-sum-w-1}
\end{align}
By assumption \eqref{eq:deltabar-assumption} and definition of $\vdeltabar$,
\begin{align}
    \label{eq:less-sum-w-2}
\sum_{i=1}^n \norm{\vw}_1 \cdot \vr_i \vdeltabar_i^2 \leq \norm{\vw}_1.
\end{align}
Moreover
\begin{align*}
\sum_{i=1}^n (\vw_i+\frac{1}{n} \norm{\vw}_1) \vs_i^2 \cdot  \vdeltabar_i^2 = \sum_{i=1}^n \vw_i \vs_i^2 \cdot \vdeltabar_i^2 + \frac{1}{n} \norm{\vw}_1 \sum_{i=1}^n \vs_i^2 \cdot \vdeltabar_i^2
\end{align*}
Since $\norm{\vs \odot \vdeltabar}_{\infty} \leq 1$, each $\vs_i^2 \cdot \vdeltabar_i^2 \leq 1$ and
\begin{align}
    \label{eq:less-sum-w-3}
\sum_{i=1}^n \vw_i \vs_i^2 \cdot \vdeltabar_i^2 \leq \norm{\vw}_1.
\end{align}
Moreover $\frac{1}{n}\norm{\vs \odot \vdeltabar}_2^2 \leq \norm{\vs \odot \vdeltabar}_{\infty}^2 \leq 1$. Therefore 
\begin{align}
\label{eq:less-sum-w-4}
\frac{1}{n} \norm{\vw}_1 \sum_{i=1}^n \vs_i^2 \cdot \vdeltabar_i^2 \leq \norm{\vw}_1.
\end{align}
Then the first part of the lemma follows by combining \eqref{eq:less-sum-w-1}, \eqref{eq:less-sum-w-2}, \eqref{eq:less-sum-w-3}, and \eqref{eq:less-sum-w-4}. Now since $\vr,\vs$ are nonnegative $\vwtil_i \geq \norm{\vw}_1 \cdot \vr_i$. Therefore $\norm{\vdeltahat}_{\vwtil}^2 \geq \norm{\vw}_1 \norm{\vdeltahat}_{\vr}^2$, and the second part of the lemma follows from the first part. By Cauchy-Schwarz on the vectors $[\sqrt{w_i}]_i$ and $[\sqrt{\vw_i} \abs{\vs_i \vdeltahat_i}]_i$, definition of $\vwtil$ and first part of the lemma, the third part follows as the following,
\[
\sum_{i=1}^n \vw_i \abs{\vs_i \vdeltahat_i} \leq \sqrt{ \left(\sum_{i=1}^n \vw_i \right) \cdot \left( \sum_{i=1}^n \vw_i \vs_i^2 \vdeltahat_i^2 \right)} \leq \sqrt{\norm{\vw}_1 \cdot \norm{\vdeltahat}_{\vwtil}^2} \leq \sqrt{6} \cdot \norm{ \vw}_1.
\]
Now by definition of $\vwtil$ and first part of the lemma as the following,
\[
\sum_{i=1}^n \frac{1}{n} \norm{\vw}_1 \cdot \vs_i^2 \cdot \vdeltahat _i^2 \leq \norm{\vdeltahat}_{\vwtil}^2 \leq 6 \cdot \norm{\vw}_1.
\]
Therefore the fourth part follows by
\[
\norm{\vs \odot \vdeltahat}_{\infty}^2 \leq \norm{\vs \odot \vdeltahat}_{2}^2 \leq 6n.
\]
\end{proof}

\RestyleAlgo{algoruled}
\IncMargin{0.15cm}
\begin{algorithm}[!ht]

\textbf{Input:} Full column rank matrix $\ma\in \R^{n\times (d+1)}$ ($n > d$), $\vb\in \R^{d+1}$, $\vr,\vs \in\R^n$,  $0<\veps<1$ such that $\norm{\ma^\top \ma}_{\fro}, \norm{(\ma^\top \ma)^{-1}}_{\fro}, \norm{\vb}_2 \leq \kappa$, for $\kappa>1$. \\

\textbf{Output:} $\vyhat\in\R^{n}$ such that $\norm{\ma \vyhat - \vb}_2 \leq \veps$ and $\norm{\vyhat}_{\vr}^2+\norm{\vs \odot \vyhat}_{\infty} \leq \alpha \min_{\vy:\ma^\top \vy = \vb} \norm{\vy}_{\vr}^2+\norm{\vs \odot \vy}_{\infty}$. \\

Set $\rho = 8 \cdot n^{1/3} \cdot \log\left(\frac{\left(18\sqrt{6}+18 \right) n^2}{\norm{\pi_{\ma} \vb}_2^2 \cdot  \min_{i\in[n]}\{\vr_i +\vs_i^2\}} \right)$ \label{alg:mwu-inf-step-rho} \\
Set $T = \ceil{2\rho\cdot \log n}$ \label{alg:mwu-inf-step-big-T}, $t = 0$, $k=0$, and 
$\vw^{(0,0)}\in\R^n$ to be a vector of all ones \label{alg:mwu-inf-step-w-init}\\

For $i\in[n]$, $\eta\in \{0,\ldots,\ceil{16\rho \cdot \log(n)}\}$, set $c_{i,\eta} = 0$\\

Let $\mu=\norm{\vw^{(t,k)}}_1$ and $\vp_i =\vw_i^{(t,k)}$ \\

Set $\vwtil^{(t,k)}_i = \vwhat_i^{(t,k)} = \mu\cdot \vr_i + (\vp_i+\frac{\mu}{n})\cdot \vs_i^2$ for all $i\in[n]$ \label{alg:mwu-p-step-what}\\

Set $\ds$ to be the inverse maintenance data structure \DontPrintSemicolon \tcp*{\textcolor{blue}{either the dense data structure $\dds$ or the sparse data structure $\sds$ with parameter $m$}}

$\ds.\initialize(\ma,\vwhat^{(t,k)},\frac{\epsilon}{10^6 (\kappa\cdot n)^{30}})$

\While{$t < T$}{
Set $\vwtil^{(t,k)}_i = \norm{\vw^{(t,k)}}_1 \cdot \vr_i + \left( \vw^{(t,k)}_i + \frac{1}{n} \norm{\vw^{(t,k)}}_1 \right) \cdot \vs_i^2$ \label{alg:mwu-inf-step-wtil}\\

\If{$\norm{\vw^{(t,k)}}_1 > 2 \mu$ or ($\ds$ is sparse and $\floor{(n/ m)^{1/3}}$ divides $t+k$)}{
Set $S=[n]$, $\mu=\norm{\vw^{(t,k)}}_1$, $\vp_i = \vw_i^{(t,k)}$, and $\vwhat_i^{(t,k)} = \mu\cdot \vr_i + (\vp_i+\frac{\mu}{n})\cdot \vs_i^2$ \label{alg:mwu-inf-step-change-p-whole}\\
For $i\in[n]$, $\eta\in \{0,\ldots,\ceil{16\rho \cdot \log(n)}\}$, set $c_{i,\eta} = 0$\\
} \Else{
Set $S \leftarrow \displaystyle \bigcup_{\eta: t+k+1 \mod{2^\eta}\equiv 0} \{i \in[n]: c_{i,\eta} \geq \frac{2^{\eta}}{\ceil{\log_2(16 \rho \cdot \log n)}\cdot \log(2)}\}$ \DontPrintSemicolon \tcp*{\textcolor{blue}{this guarantees that if $i\notin S$ then $\vw_i^{(t,k)}\leq 2\vp_i$}}
Set $\vp_i = \vw_i^{(t,k)}$ and $\vwhat_i^{(t,k)} = \mu\cdot \vr_i + (\vp_i+\frac{\mu}{n})\cdot \vs_i^2$, for all $i\in S$ \label{alg:mwu-inf-step-change-p-not-whole}\\
For $i\in S$, $\eta\in \{0,\ldots,\ceil{16\rho \cdot \log(n)}\}$, set $c_{i,\eta} = 0$\\
}
$\ds.\update(S, \vwhat_S^{(t,k)})$

Let $\vdelta^{(t,k)}\in\R^n$ such that $\norm{\pi_{\ma}(\vdelta^{(t,k)} - {\vdeltastar}^{(t,k)})}_{2} \leq \frac{\epsilon}{16\rho\cdot \log n} \norm{\pi_{\ma}{\vdeltastar}^{(t,k)}}_{2}$ and $\norm{\vdelta^{(t,k)} - {\vdeltastar}^{(t,k)}}_2 \leq \frac{\epsilon}{2\kappa^6} \cdot \norm{{\vdeltastar}^{(i,k)}}_{\mwtil^{(t,k)}}$ where
${\vdeltastar}^{(i,k)} = \argmin_{\Delta:\ma^\top \vdelta = \vb} \norm{\vdelta}_{\mwtil^{(t,k)}}^2$ \label{alg:mwu-inf-step-delta-approx} \DontPrintSemicolon \tcp*[h]{\textcolor{blue}{see \cref{lemma:constrained-richardson}}}

\If(\tcp*[h]{\textcolor{blue}{regular step}}){$\norm{\vs \odot \vdelta^{(t,k)}}_{3} \leq \rho$ }{ 
Set $\vw^{(t+1,k)}_i = \vw^{(t,k)}_i \cdot \left(1+\frac{\abs{\vs_i \cdot \vdelta_i^{(t,k)}}}{\rho}\right)$ for all $i\in[n]$ \label{alg:mwu-inf-step-w-update-regular} \\
Set $\vy^{(t+1)} = \vdelta^{(t,k)}$\\
Increase $t = t + 1$
}\Else(\tcp*[h]{\textcolor{blue}{width-reduction step}}){
Set vector $\vu^{(k)} = \frac{\rho}{\norm{\vs \odot \vdelta^{(t,k)}}_3} \cdot \vs \odot \vdelta^{(t,k)}$\\
Set $\vw^{(t,k+1)}_i = \vw^{(t,k)}_i \cdot \left(1+\frac{\abs{\vu^{(k)}_i}}{\rho}\right)$ for all $i\in[n]$ \label{alg:mwu-inf-step-w-update-width-reduct}\\
Set $k = k + 1$
}
For all $i\in[n]$, increase $c_{i,\eta}$ by one where $\eta\in\{0,\ldots,\ceil{16\rho\cdot\log n}\}$ such that $\frac{\vw_i^{(t,k)} - \vw_i^{(\old)}}{\vw_i^{(\old)}}\in (2^{-\eta-1},2^{-\eta}]$, where $\vw_i^{(\old)}$ is $\vw_i^{(t-1,k)}$ or $\vw_i^{(t,k-1)}$ depending on type of the step.
}
\Return $\frac{1}{T} \sum_{t=1}^T \vy^{(t)}$

\caption{Multiplicative weights update to solve mixed $(2,\infty)$-norm minimization}
\label{alg:mwu-inf}
\end{algorithm}

For the rest of the proof, we use two potential functions. The first one is $\norm{\vw^{(t,k)}}_1$, which we show only increases slowly over the course of the algorithm. The second potential function is $\min_{\vdelta:\ma^\top \vdelta = \vb} \norm{\vdelta}_{\vwtil^{(t,k)}}$ that we show increases significantly when a width reduction steps happen. Moreover, since \cref{lemma:all-bounds-w} guarantees that $\min_{\vdelta:\ma^\top \vdelta = \vb} \norm{\vdelta}_{\vwtil^{(t,k)}}=O(1)\cdot \norm{\vw^{(t,k)}}_1$, we get a bound on the maximum number of width reduction steps. 

\begin{lemma}
\label{lemma:first-pot-increase}
Let $t,k\geq 0$, and $\vw$ be as defined in Algorithm \ref{alg:mwu-inf} (initialized on Step \ref{alg:mwu-inf-step-w-init} and updated on Steps \ref{alg:mwu-inf-step-w-update-regular} and \ref{alg:mwu-inf-step-w-update-width-reduct}). Then we have
\[
\norm{\vw^{(t,k+1)}}_1 \leq \left( 1+\frac{\sqrt{6}+1}{\rho} \right) \norm{\vw^{(t,k)}}_1 ~~ \text{, and } \norm{\vw^{(t+1,k)}}_1 \leq \left( 1+\frac{\sqrt{6}+1}{\rho} \right) \norm{\vw^{(t,k)}}_1.
\]
Moreover for $\vyhat:=\frac{1}{T} \sum_{t=1}^T \vy^{(t)}$ and $K$ equal to the number of width reduction steps, we have
\[
\norm{\vyhat}_{\vr}^2 + \norm{\vs \odot \vyhat}_{\infty} \leq  7\cdot \frac{T+K}{T} + 13.
\]
\end{lemma}
\begin{proof}
First note that an update to $\vw^{(t+1,k)}$ happens in a regular step and an update to $\vw^{(t,k+1)}$ only happens in a width reduction step. However both updates are the same. Therefore we denote either of $\vw^{(t+1,k)}$ or $\vw^{(t,k+1)}$ with $\vw^{(\text{new})}$ in this proof, and prove the bound for $\vw^{(\text{new})}$.
Let 
\[
{\vdeltastar}^{(t,k)} = \argmin_{\vdelta:\ma^\top \vdelta =\vb}\norm{\vdelta}_{\vwtil^{(t,k)}}^2,
\]
and $\vdelta^{(t,k)}$ be as defined on Step \ref{alg:mwu-inf-step-delta-approx} of Algorithm \ref{alg:mwu-inf}.
By triangle inequality
\begin{align}
\norm{\vw^{(\text{new})}}_1 & = \sum_{i=1}^n \vw_i^{(t,k)} + \frac{1}{\rho} \sum_{i=1}^n \vw_i^{(t,k)} \abs{\vs_i \cdot \vdelta^{(t,k)}_i} 
\\ & \leq
\sum_{i=1}^n \vw_i^{(t,k)} + \frac{1}{\rho} \sum_{i=1}^n \vw_i^{(t,k)} \abs{\vs_i\cdot {\vdeltastar}^{(t,k)}_i} + \frac{1}{\rho} \sum_{i=1}^n \vw_i^{(t,k)}  \abs{\vs_i \cdot (\vdelta^{(t,k)}_i - {\vdeltastar}^{(t,k)}_i)}
\end{align}
Therefore by Part \ref{lemma:all-bounds-w-part-3} of \cref{lemma:all-bounds-w},
\begin{align*}
\norm{\vw^{(\text{new})}}_1 \leq (1+\frac{\sqrt{6}+\norm{\vs \odot (\vdelta^{(t,k)} - {\vdeltastar}^{(t,k)})}_{\infty}}{\rho})
\norm{\vw^{(t,k)}}_1.
\end{align*}
By \cref{lemma:constrained-richardson}, we have $\norm{\vs \odot (\vdelta^{(t,k)} - {\vdeltastar}^{(t,k)})}_{\infty} \leq \norm{\vs \odot (\vdelta^{(t,k)} - {\vdeltastar}^{(t,k)})}_{2} \leq 2\kappa^6 \veps$. Therefore 
\begin{align*}
\norm{\vw^{(\text{new})}}_1 \leq (1+\frac{\sqrt{6}+1}{\rho})
\norm{\vw^{(t,k)}}_1.
\end{align*}
Taking this over all iterations, denoting the number of width reduction steps with $K$, and noting that $\norm{\vw^{(0,0)}} = n$,
\begin{align}
\label{eq:bound-infty-norm-1}
\norm{\vw^{(T,K)}}_1 \leq \left(1+\frac{\sqrt{6}+1}{\rho}\right)^{T+K} n \leq \exp\left((T+K)\cdot \frac{\sqrt{6}+1}{\rho}\right) \cdot n.
\end{align}
Moreover denoting the set of all pairs $(t,k)$, for which $\vdelta^{(t,k)}$ is computed, by $S$, we have
\begin{align}
\vw_i^{(T,K)} = \prod_{(t,k)\in S} \left(1 + \frac{\abs{\vs_i \cdot \vdelta^{(t,k)}_i}}{\rho} \right) 
\geq
\prod_{t=1}^T \left(1+\frac{\abs{\vs_i \cdot \vy_i^{(t)}}}{\rho} \right).
\end{align}
Now note that by construction $\abs{\vs_i \cdot \vy_i^{(t)}} \leq \rho$. Therefore
\begin{align}
\label{eq:bound-infty-norm-2}
\vw_i^{(T,K)} \geq \prod_{t=1}^T \exp\left(\frac{\abs{\vs_i \cdot \vy_i^{(t)}}}{2\rho} \right)
\end{align}
Combining \eqref{eq:bound-infty-norm-1} and \eqref{eq:bound-infty-norm-2}, taking the logarithm, and using triangle inequality, we have
\[
(T+K)\cdot \left(\frac{\sqrt{6}+1}{\rho}\right) + \log n \geq \sum_{t=1}^T \left(\frac{\abs{\vs_i \cdot \vy_i^{(t)}}}{2\rho} \right) \geq T \cdot \left(\frac{\abs{\vs_i \cdot \frac{1}{T} \sum_{t=1}^T \vy_i^{(t)}}}{2\rho} \right) = T \cdot \left(\frac{\abs{\vs_i \cdot \vyhat_i}}{2\rho} \right)
\]
Therefore by definition of $T$ (Step \ref{alg:mwu-inf-step-big-T} of Algorithm \ref{alg:mwu-inf}), we have
\[
\abs{\vs_i \cdot \vyhat_i} \leq 7\cdot \frac{T+K}{T} + \frac{2\rho}{T}\log(n) \leq 7\cdot \frac{T+K}{T} + 1
\]
Finally by Cauchy-Schwarz inequality
\[
\norm{\vyhat}_{\vr}^2 =
\sum_{i=1}^n \vr_i \left( \frac{1}{T}\sum_{t=1}^T \vy^{(t)}_i \right)^2 \leq \frac{1}{T} \sum_{i=1}^n \vr_i \sum_{t=1}^T (\vy^{(t)}_i)^2 = \frac{1}{T} \sum_{t=1}^T \norm{\vy^{(t)}}_{\vr}^2
\]
This just comes from the convexity of $2$-norm and Part \ref{lemma:all-bounds-w-part-2} of \cref{lemma:all-bounds-w}. Let $(t,k_t)$ be the pair corresponding to $\vy^{(t)}$. Then by \cref{lemma:constrained-richardson} and \cref{lemma:all-bounds-w}, we have
\[
\norm{\vy^{(t)}}_{\vr} \leq 
\norm{\vy^{(t)} - {\vdeltastar}^{(t,k_t)}}_{\vr} + \norm{{\vdeltastar}^{(t,k_t)}}_{\vr} \leq 2\kappa^6 \veps + \norm{{\vdeltastar}^{(t,k_t)}}_{\vwtil^{(t,k_t)}} \leq 1+ \sqrt{6}
\]
Therefore $\norm{\vyhat}_{\vr}^2 \leq 12$.
\end{proof}

Now we use \cref{lemma:all-bounds-w}, to prove that the width reduction steps increase the second potential function significantly if the weights have a large increase.

\begin{lemma}
\label{lemma:QuadraticSolIncrease}
Let $\ma \in\R^{n\times d}$, $\vb \in\R^d$, and $\vw^{(1)},\vw^{(2)}\in\R^{n}_{\geq 0}$ such that $\vw^{(2)} - \vw^{(1)} \geq 0$. Moreover for $j=1,2$, let
\[
\vdelta^{(j)} = \argmin_{\vdelta:\ma^\top \vdelta =\vb} \frac{1}{2} \norm{\vdelta}_{\vw^{(j)}}^2.
\]
Then
\[
\frac{1}{2}\norm{\vdelta^{(2)}}_{\vw^{(2)}}^2 \geq \frac{1}{2}\norm{\vdelta^{(1)}}_{\vw^{(1)}}^2 + \frac{1}{4}\norm{\vdelta^{(1)}}_{\frac{\vw^{(2)} - \vw^{(1)}}{\vw^{(2)}}\cdot \vw^{(1)} }^2
\]
\end{lemma}
\begin{proof}
Throughout the proof let $j\in\{1,2\}$.
We look at the dual of 
\[
\min_{\ma^\top \vdelta = \vb} \frac{1}{2}\norm{\vdelta}_{\vw^{(j)}}^2,
\]
which is
\[
\max_{\vz} -\vb^{\top} \vz - \frac{1}{2}\norm{\ma \vz}_{(\vw^{(j)})^{-1}}^2.
\]
Let $\vz^{(j)}$ be the optimal solution of the dual problem for $\vw^{(j)}$. 
The gradient of $\norm{\vdelta}_{\vw^{(j)}}^2$ at $\vdelta^{(j)}$ is orthogonal to the kernel of $\ma^\top$. Therefore there exists $\vu^{(j)}\in\R^{d}$ such that
\[
\mw^{(j)} \vdelta^{(j)} = \ma \vu^{(j)}.
\]
Therefore $\vdelta^{(j)} =  (\mw^{(j)})^{-1}\ma \vu^{(j)}$, and since $\ma^\top \vdelta^{(j)} = \vb$, we have
\[
\vu^{(j)} = (\ma^\top (\mw^{(j)})^{-1}\ma)^{-1} \vb ~~ \text{, and } ~~ \vdelta^{(j)} = (\mw^{(j)})^{-1}\ma (\ma^\top (\mw^{(j)})^{-1}\ma)^{-1} \vb
\]
For the dual solution, we have that the gradient of $-\vb^{\top} \vz - \frac{1}{2}\norm{\ma \vz}_{(\vw^{(j)})^{-1}}^2$ at $\vz^{(j)}$ is zero. Therefore
\[
-\vb - \ma^\top (\mw^{(j)})^{-1} \ma \vz^{(j)} = 0.
\]
Therefore $\vz^{(j)} = - (\ma^\top (\mw^{(j)})^{-1} \ma)^{-1}\vb$, and
\[
(\ma \vz^{(j)})_i^2 = \left(\ma (\ma^\top (\mw^{(j)})^{-1} \ma)^{-1}\vb\right)_i^2 ~~ \text{, and } (\vdelta^{(j)}_i)^2 = (\vw^{(j)}_i)^{-2} \left(\ma (\ma^\top (\mw^{(j)})^{-1} \ma)^{-1}\vb\right)_i^2.
\]
Thus,
\begin{align}
\label{eq:optimal-weighted-lin-reg}
(\vw^{(j)}_i)^{-1} (\ma \vz^{(j)})_i^{2} = \vw^{(j)}_i (\vdelta_i^{(j)})^2 ~~ \text{, and } \norm{\vdelta^{(j)}}_{\mw^{(j)}}^2 = \norm{\ma \vz^{(j)}}_{\mw^{(j)}}^2 = \vb^\top (\ma^\top (\mw^{(j)})^{-1} \ma)^{-1}\vb,
\end{align}
where the last equality follows by substituting the value of $\vz^{(j)}$.
We have
\[
-\frac{1}{\vw^{(2)}_i} = -\frac{1}{\vw^{(1)}_i} +\frac{\vw^{(2)}_i - \vw^{(1)}_i}{\vw^{(2)}_i \cdot \vw^{(1)}_i}. 
\]
Therefore,
\begin{align*}
\frac{1}{2}\norm{{\vdelta}^{(2)}}_{\mw^{(2)}}^2 
& = 
-\vb^{\top} {\vz}^{(2)} - \frac{1}{2}\norm{\ma {\vz}^{(2)}}_{(\mw^{(2)})^{-1}}^2 
\\ & \geq
-\vb^{\top} {\vz}^{(1)} - \frac{1}{2}\norm{\ma {\vz}^{(1)}}_{(\mw^{(2)})^{-1}}^2 
\\ & =
-\vb^{\top} {\vz}^{(1)} - \frac{1}{2} \sum_{i\in[n]} (\vw^{(2)}_i)^{-1} (\ma {\vz}^{(1)})_i^2
\\ & =
-\vb^{\top} {\vz}^{(1)} - \frac{1}{2} \sum_{i\in[n]} (\vw^{(1)}_i)^{-1} (\ma {\vz}^{(1)})_i^2 + \frac{1}{2} \sum_{i\in[n]} \frac{(\vw^{(2)}_i-\vw^{(1)}_i)}{\vw^{(2)}_i \cdot \vw^{(1)}_i} (\vw^{(1)}_i)^2 (\vdelta_i^{(j)})^2 
\\ & = 
\frac{1}{2}\norm{{\vdelta}^{(1)}}_{\mw^{(1)}}^2 + \frac{1}{2} \sum_{i\in[n]} \frac{(\vw^{(2)}_i-\vw^{(1)}_i)}{\vw^{(2)}_i}\cdot (\vw_i^{(1)}) ({\vdelta}^{(1)}_i)^2.
\end{align*}
\end{proof}

We are now equipped to show that in a width reduction step, the value of the second potential function increases significantly. This combined with \cref{lemma:first-pot-increase} implies that the number of width reduction steps is at most $\Otil(\rho)$.

\begin{lemma}
\label{lemma:width-reduct-pot-increase}
Let 
\[
{\vdeltastar}^{(t,k)} = \argmin_{\vdelta:\ma^\top \vdelta =\vb} \frac{1}{2} \norm{\vdelta}_{\mwtil^{(t,k)}}^2,
\]
where $\mwtil^{(t,k)}$ is defined as Step \ref{alg:mwu-inf-step-wtil} of Algorithm \ref{alg:mwu-inf}. Then after a width-reduction step
\[
\frac{1}{2}\norm{{\vdeltastar}^{(t,k+1)}}_{\vwtil^{(t,k+1)}}^2 \geq \left(1+ \frac{\rho^2}{72 \cdot n}\right) \cdot \frac{1}{2}\norm{{\vdeltastar}^{(t,k)}}_{\vwtil^{(t,k)}}^2,
\]
where $\rho$ is defined as Step \ref{alg:mwu-inf-step-rho}
of Algorithm \ref{alg:mwu-inf}.
\end{lemma}
\begin{proof}
First note that for positive numbers $a,b,c$ and $r\geq 1$ with $b\geq a$, 
\begin{align}
\label{eq:num-denom-removal}
\frac{a+c}{b+rc} \geq \frac{1}{r}\cdot \frac{a}{b}.
\end{align}
This holds by dividing both sides of the following by $r$
\[
\frac{ra+rc}{b+rc} \geq \frac{ra}{b} \geq \frac{a}{b}.
\]
By \cref{lemma:QuadraticSolIncrease},
in the width reduction step,

\begin{align*}
\frac{1}{2}\norm{{\vdeltastar}^{(t,k+1)}}_{\vwtil^{(t,k+1)}}^2
& \geq 
\frac{1}{2}\norm{{\vdeltastar}^{(t,k)}}_{\vwtil^{(t,k)}}^2 + \frac{1}{2} \sum_{i\in[n]} \frac{\vwtil_i^{(t,k+1)} - \vwtil_i^{(t,k)}}{\vwtil_i^{(t,k+1)}} \cdot \vwtil_i^{(t,k)} ({\vdeltastar}_i^{(t,k)})^2
\end{align*}
Now by construction and \cref{lemma:first-pot-increase},  $\norm{\vw^{(t,k)}}_1\leq \norm{\vw^{(t,k+1)}}_1\leq (1+\frac{12}{\rho})\norm{\vw^{(t,k)}}_1 \leq 2 \norm{\vw^{(t,k)}}_1$, and since $\vw^{(t,k+1)}_i\geq 1$ and $\norm{\vw^{(t,k+1)}}_1 \leq 6n$ over the course of the algorithm,
\[
\vw^{(t,k+1)}_i + \frac{1}{n} \norm{\vw^{(t,k+1)}}_1 \leq \vw^{(t,k+1)}_i + 6 = (1+\frac{\vu_i^{(k)}}{\rho}) \cdot \vw^{(t,k)}_i + 6 \vw_i^{(t,k)} \leq 8 \cdot \vw^{(t,k)}_i,
\]
where the last inequality follows from the construction of $\vu^{(k)}$. Then by \eqref{eq:num-denom-removal},
\begin{align*}
\frac{\vwtil_i^{(t,k)}}{\vwtil_i^{(t,k+1)}}
& =
\frac{\norm{\vw^{(t,k)}}_1 \cdot \vr_i + \left( \vw^{(t,k)}_i + \frac{1}{n} \norm{\vw^{(t,k)}}_1 \right) \cdot \vs_i^2}{\norm{\vw^{(t,k+1)}}_1 \cdot \vr_i + \left( \vw^{(t,k+1)}_i + \frac{1}{n} \norm{\vw^{(t,k+1)}}_1 \right) \cdot \vs_i^2}
\\ & \geq 
\frac{\norm{\vw^{(t,k)}}_1 \cdot \vr_i + \vw^{(t,k)}_i \cdot \vs_i^2 + \frac{1}{n} \norm{\vw^{(t,k)}}_1  \cdot \vs_i^2}{2\cdot\norm{\vw^{(t,k)}}_1 \cdot \vr_i +  2\cdot \vw^{(t,k)}_i \cdot \vs_i^2 + 6 \cdot \vw^{(t,k)}_i \cdot \vs_i^2}
\\ & \geq
\frac{1}{12} \cdot \frac{\frac{1}{n} \norm{\vw^{(t,k)}}_1  \cdot \vs_i^2}{\vw^{(t,k)}_i \cdot \vs_i^2}.
\end{align*}
Moreover since $\norm{\vw^{(t,k)}}_1\leq \norm{\vw^{(t,k+1)}}_1$,
\begin{align*}
\vwtil_i^{(t,k+1)} - \vwtil_i^{(t,k)} \geq (\vw_i^{(t,k+1)} - \vw_i^{(t,k)}) \cdot \vs_i^2.
\end{align*}
Therefore
\begin{align*}
\frac{1}{2}\norm{{\vdeltastar}^{(t,k+1)}}_{\vwtil^{(t,k+1)}}^2
& \geq 
\frac{1}{2}\norm{{\vdeltastar}^{(t,k)}}_{\vwtil^{(t,k)}}^2 + \frac{1}{24} \sum_{i\in[n]} \frac{(\vw_i^{(t,k+1)} - \vw_i^{(t,k)}) \cdot \vs_i^2}{\vw^{(t,k)}_i \cdot \vs_i^2} \cdot \frac{1}{n} \norm{\vw^{(t,k)}}_1  \cdot \vs_i^2 ({\vdeltastar}_i^{(t,k)})^2
\end{align*}
Now by construction of $\vu^{(k)}$ (since $\norm{\vs \odot {\vdeltastar}^{(t,k)}}_3 > \rho$), and construction of $\vw^{(t,k+1)}$, we have
\begin{align*}
\frac{1}{2}\norm{{\vdeltastar}^{(t,k+1)}}_{\vwtil^{(t,k+1)}}^2
& \geq 
\frac{1}{2}\norm{{\vdeltastar}^{(t,k)}}_{\vwtil^{(t,k)}}^2 + \frac{\norm{\vw^{(t,k)}}_1}{24\cdot n} \sum_{i\in[n]} \frac{\vu_i^{(k)}}{\rho}  \cdot (\vu_i^{(k)})^2 \geq \frac{1}{2}\norm{{\vdeltastar}^{(t,k)}}_{\vwtil^{(t,k)}}^2 + \frac{\norm{\vw^{(t,k)}}_1 \cdot \rho^2}{24\cdot n}.
\end{align*}
Finally by \cref{lemma:all-bounds-w}, $\norm{{\vdeltastar}^{(t,k)}}_{\vwtil^{(t,k)}}^2 \leq 6\cdot \norm{\vw^{(t,k)}}_1$. Thus,
\begin{align*}
\frac{1}{2}\norm{{\vdeltastar}^{(t,k+1)}}_{\vwtil^{(t,k+1)}}^2 \geq \left(1+ \frac{\rho^2}{72 \cdot n}\right) \cdot \frac{1}{2}\norm{{\vdeltastar}^{(t,k)}}_{\vwtil^{(t,k)}}^2.
\end{align*}
\end{proof}

We finally bound the number of changes to $\vwhat$. This is the main factor in the running time of the inverse maintenance procedure.

\begin{lemma}
\label{lemma:num-changes}

Let $\widehat{T} := T+K$ be the number of iterations of Algorithm \ref{alg:mwu-inf}, and $k\in[\widehat{T}]$.
For $t\in\widehat{T}$ and $\eta \in \{0,\ldots,\ceil{\log_2(\widehat{T})}\}$, let $c_{t,\eta}$ be the number of entries of $\vw$ that change by a factor in the interval of $(2^{-\eta-1},2^{-\eta}]$. Then 
\[
\sum_{t=1}^{\widehat{T}} c_{t,\eta} \leq \widehat{T} 2^{3(\eta+1)}.
\]
\end{lemma}
\begin{proof}
Note that in a regular step, the relative change (i.e., $(\vw_i^{(t+1,k)}-\vw_i^{(t,k)})/\vw_i^{(t,k)}$) to each entry is at most $\abs{\vs_i \cdot \vdelta_i^{(t,k)}}/\rho < 1$.
Moreover by the upper bound of $\norm{\vs \odot \vdelta^{(t,k)}}_{3}$ in the regular steps and the construction of $\vu^{(k)}$ in the width-reduction steps, we have that
\begin{align}
\norm{\frac{\vw^{(\new)}-\vw^{(t,k)}}{\vw^{(t,k)}}}_3 \leq
1,
\end{align}
where $\vw^{(\new)}$ is either $\vw^{(t+1,k)}$ or $\vw^{(t,k+1)}$ depending on the type of the step. Therefore the number of changes of factor in $(2^{-\eta-1},2^{-\eta}]$ in one step is at most $2^{3\eta+3}$ and
the number of such changes over the coruse of the algorithm is $\widehat{T}\cdot 2^{3\eta+3}$. 
\end{proof}

\begin{proof}[Proof of \cref{thm:main-mixed-2-infty}]
We show that Algorithm \ref{alg:mwu-inf} achieves the desired result if 
\begin{align}
\label{eq:two-infty-bounded-obj}
0.5\leq \min_{\vdelta: \ma^\top \vdelta = \vb} \norm{\vdelta}_{\vr}^2 + \norm{\vs \odot \vdelta}_{\infty} < 1.
\end{align}
We require this assumption to be able to use the results we developed in this section, e.g., \cref{lemma:all-bounds-w}. Note that if we scale all of $r$ and $s$ by a number $\alpha$, the minimum value is also scaled by $\alpha$. Therefore we only need to ``guess'' the correct scaling factor as a power of two. This means that we try to minimize the objective function with different scaling factors and then we take the minimum over the vectors return for these different scaling factors. Note that this only affects the running time of the algorithm. Later in the proof, when we discuss the running time, we take the number of scaling factors we need to try into consideration.

We first need to bound the number of iterations. The number of regular iterations is bounded by $\ceil{2\rho\cdot \log n}$ by construction. Let $K$ be the number of width-reduction steps of the algorithm. Then since $\vw^{(0,0)}=\vec{1}$, by \cref{lemma:first-pot-increase}
for all steps $(t,k)$ of the algorithm,
\begin{align}
\label{eq:main-infty-proof-1-norm-bound}
\norm{\vw^{(t,k)}}_1 \leq \exp\left(\frac{\sqrt{6}+1}{\rho} (T+K) \right) n \leq \exp\left(\frac{\sqrt{6}+1}{\rho} \cdot K \right) \cdot \left(3\sqrt{6}+3 \right) n^2
\end{align}
Now let
$\vdeltahat^{(t,k)} := \argmin_{\vdelta: \ma^\top \vdelta = \vb} \norm{\vdelta}_{\vwtil^{(t,k)}}^2$. Since $\vw^{(t,k)}=\vec{1}$ and $\vwtil^{(t,k)}_i = \norm{\vw^{(t,k)}}_1 \cdot \vr_i + \left( \vw^{(t,k)}_i + \frac{1}{n} \norm{\vw^{(t,k)}}_1 \right) \cdot \vs_i^2$, defining $\vu \in\R^n$ as $ \vu = \vec{1} \cdot \min_{i\in[n]}\{n \cdot \vr_i +2 \cdot \vs_i^2\}$, we have
\[
\vu \leq \vwtil^{(0,0)}.
\]
Therefore by \cref{lemma:QuadraticSolIncrease},
\[
\norm{\vdeltahat^{(0,0)}}_{\vwtil^{(t,k)}}^2 \geq \min_{i\in[n]}\{\vr_i +\vs_i^2\} \cdot  \min_{\vdelta:\ma^\top \vdelta = \vb} \norm{\vdelta}_{2}^2
\]
Since for a linear system, product of the pseudoinverse and the vector gives the solution with minimum $2$-norm and $\ma$ has full column rank,
\[
\argmin_{\vdelta:\ma^\top \vdelta = \vb} \norm{\vdelta}_2^2 = \ma (\ma^\top \ma)^{-1} \ma^\top \vb.
\]
Since by \cref{lemma:width-reduct-pot-increase} for each width reduction step,
\[
\norm{{\vdeltahat}^{(t,k+1)}}_{\vwtil^{(t,k+1)}}^2 \geq \left(1+ \frac{\rho^2}{72 \cdot n}\right) \cdot \norm{{\vdeltahat}^{(t,k)}}_{\vwtil^{(t,k)}}^2,
\]
we have
\[
\norm{{\vdeltahat}^{(T,K)}}_{\vwtil^{(T,K)}}^2 \geq \exp(\frac{K\cdot \rho^2}{144 \cdot n}) \cdot \norm{\pi_{\ma} \vb}_2^2 \cdot  \min_{i\in[n]}\{\vr_i +\vs_i^2\}
\]
Moreover by \eqref{eq:main-infty-proof-1-norm-bound} and \cref{lemma:all-bounds-w},
\[
\norm{\vdeltahat^{(T,K)}}_{\vwtil^{(T,K)}}^2 \leq 6 \cdot \norm{\vw^{(t,k)}}_1 \leq \exp\left(\frac{\sqrt{6}+1}{\rho} \cdot K \right) \cdot \left(18\sqrt{6}+18 \right) n^2.
\]
Therefore
\[
\frac{K\cdot \rho^2}{144 \cdot n} + \log\left( \norm{\pi_{\ma} \vb}_2^2 \cdot  \min_{i\in[n]}\{\vr_i +\vs_i^2\} \right)\leq \frac{\sqrt{6}+1}{\rho} \cdot K + \log\left(\left(18\sqrt{6}+18 \right) n^2 \right).
\]
Therefore
\begin{align}
\label{eq:infty-norm-k-bound}
K \leq \frac{144 \cdot n \rho }{\rho^3 - 144 \cdot (\sqrt{6}+1) n} \log\left(\frac{\left(18\sqrt{6}+18 \right) n^2}{\norm{\pi_{\ma} \vb}_2^2 \cdot  \min_{i\in[n]}\{\vr_i +\vs_i^2\}} \right).
\end{align}
Since $144 \cdot (\sqrt{6}+1) < 500$ and $\rho \geq 8 \cdot n^{1/3}$, $\rho^3 - 144 \cdot (\sqrt{6}+1) n$ is positive, and 
\begin{align}
\label{eq:infty-norm-k-coefficient-bound}
\frac{144 \cdot n \rho }{\rho^3 - 144 \cdot (\sqrt{6}+1) n} \leq \frac{144 \cdot n \rho }{12 \cdot n} = 96 n^{1/3}.
\end{align}
Therefore $K$ and the number of iterations of the algorithm are
\begin{align}
\label{eq:infty-norm-k-final-bound}
K=\Otil\left( n^{1/3}  \log\left(\frac{1}{\norm{\pi_{\ma} \vb}_2 \cdot  \min_{i\in[n]}\{\vr_i +\vs_i^2\}} \right)\right),
\end{align}
which by \cref{remark:polykappa-projection} (since if $\norm{\pi_{\ma} \vb}_2$ is too small, we can return the vector of all zeros as the solution) is 
\begin{align}
\label{eq:infty-norm-k-final-final-bound}
K=\Otil\left( n^{1/3}  \log\left(\frac{\kappa}{  \min_{i\in[n]}\{\vr_i +\vs_i^2\}} \right)\right).
\end{align}
Therefore by \eqref{eq:infty-norm-k-bound}, \eqref{eq:infty-norm-k-coefficient-bound}, and \cref{lemma:first-pot-increase}, for the output of the algorithm $\vyhat:=\frac{1}{T} \sum_{t=1}^T \vy^{(t)}$
we have
\[
\norm{\vyhat}_{\vr}^2 + \norm{\vs \odot \vyhat}_{\infty} \leq  7\cdot (1+6) + 13 \leq 62.
\]
Since by \eqref{eq:two-infty-bounded-obj}, the optimal objective value is at least a half, this implies that we achieve a constant factor approximation. 
Note that for all ${\vdeltastar}^{(t,k)}$, and $\vxstar$,
\[
\pi_{\ma} \vxstar = \ma(\ma^\top \ma)^{-1} \ma^\top \vxstar = \ma(\ma^\top \ma)^{-1} \vb = \ma(\ma^\top \ma)^{-1} \ma^\top {\vdeltastar}^{(t,k)} = \pi_{\ma} {\vdeltastar}^{(t,k)}
\]
Therefore since for all $(t,k)$, $\norm{\pi_{\ma} ({\vdelta}^{(t,k)} - {\vdeltastar}^{(t,k)})} \leq \veps \norm{\pi_{\ma} {\vdeltastar}^{(t,k)}}$, for all $t\in[T]$, $\norm{\pi_{\ma} (\vy^{(t)} - {\vxstar})} \leq \veps \norm{\pi_{\ma} {\vxstar}}$.
Thus
by triangle inequality,
\[
\norm{\pi_{\ma} (\vyhat - \vxstar)} \leq \veps \norm{\pi_{\ma} \vxstar}.
\]
Finally, we need to bound the running time. The number of different scaling factors we need to try to guarantee \eqref{eq:two-infty-bounded-obj}. Note that for any $\vdelta\in\R^n$,
\[
\vr_{\min}\cdot\norm{\vdelta}_2^2 \leq \norm{\vdelta}_{\vr}^2 + \norm{\vs \odot \vdelta}_{\infty} \leq 2\cdot \max\{\vr_{\max} , \vs_{\max}\} \cdot \max\{\norm{\vdelta}_2^2,\norm{\vdelta}_2\},
\]
where $\vr_{\min} = \min_{i\in[n]} \vr_i$, $\vr_{\max} = \max_{i\in[n]} \vr_i$, and $\vs_{\max} \max_{i\in[n]} \vs_i$.
Therefore 
\[
\vr_{\min}\cdot\norm{\pi_{\ma} \vxstar}_2^2 \leq \min_{\ma \vdelta = \vb}\norm{\vdelta}_{\vr}^2 + \norm{\vs \odot \vdelta}_{\infty} \leq 2\cdot \max\{\vr_{\max} , \vs_{\max}\} \cdot \max\{\norm{\pi_{\ma} \vxstar}_2^2,\norm{\pi_{\ma} \vxstar}_2\}.
\]
Therefore the number of scaling factors we need to try to have the guarantee of \eqref{eq:two-infty-bounded-obj} is at most
\[
\log(2 (\vr_{\max}+\vs_{\max})\max\{1, 1/ \norm{\pi_{\ma} \vxstar}_2\} / \vr_{\min})= O(\log(\kappa (\vr_{\max}+\vs_{\max})/(\vr_{\min}\cdot \epsilon))),
\]
where the equality follows from \cref{remark:polykappa-projection}.
We now bound the running time of Algorithm \ref{alg:mwu-inf} in the dense case. We first bound the running time of inverse maintenance. Note that the inverse is either updated through Step \ref{alg:mwu-inf-step-change-p-whole} or Step \ref{alg:mwu-inf-step-change-p-not-whole} of the algorithm. The former is triggered when the $1$-norm of the weights is changed by a factor of two, which only occurs $O(\log n)$ times by \eqref{eq:main-infty-proof-1-norm-bound}, \eqref{eq:infty-norm-k-final-bound}, and because $\vw^{(0,0)}=\vec{1}$. Therefore the cost of such updates is bounded by $\Otil(n^{\omega} \log(\kappa/\epsilon))$. Now consider updates through Step \ref{alg:mwu-inf-step-change-p-not-whole} of the algorithm. For an index $i\in[n]$ suppose the entry $i$ of $\vwhat$ has changed in iterations $s$ and $e$ and has been fixed between these two iterations. Moreover, suppose $1+q_t$ be the relative change of entry $i$ of $\vw$ at step $t$. Since an entry of $\vwhat$ changes only when the corresponding entry of $\vw$ has changed by more than a factor of two, we have
\[
\exp(\sum_{t=e}^{s-1} q_t)\geq \prod_{t=e}^{s-1}(1+q_t)\geq 2.
\]
Now if for all $\eta \in \{0,\ldots,\ceil{\log_2(\widehat{T})}\}$, where $\widehat{T}:=T+K$ is the number of iterations of the algorithm, the number of $q_t$'s for $t\in\{e,\ldots,s-1\}$ is less than $\frac{2^{\eta}}{\ceil{\log_2(\widehat{T})}\cdot \log(2)}$, then $\exp(\sum_{t=e}^{s-1} q_t) < 2$. Therefore for at least one of the $\eta$'s, the number of such $q_t$'s is at least $\frac{2^{\eta}}{\ceil{\log_2(\widehat{T})}\cdot \log(2)}$. Therefore by \cref{lemma:num-changes}, the sum of the rank of the updates caused by changes between $(2^{-\eta-1},2^{\eta}]$ through Step \ref{alg:mwu-inf-step-change-p-not-whole} of the algorithm is at most
\begin{align}
\label{eq:eta-changes}
\widehat{T} 2^{2\eta+3} \cdot \ceil{\log_2(\widehat{T})}\cdot \log(2) = \Otil \left( \widehat{T} 2^{2\eta}\right).
\end{align}
By concavity of $(\cdot)^{\omega-2}$ and since we only add entries that have changed due to accumulations of changes in $(2^{-\eta-1},2^{\eta}]$ once every $2^\eta$ iterations, the cost of such updates is
\begin{align*}
\Otil\left(\frac{\widehat{T}}{2^{\eta}} n^2 (\frac{\widehat{T} 2^{2\eta}}{\widehat{T} / 2^{\eta}})^{\omega-2} \right) = \Otil\left(\widehat{T}\cdot n^2 \cdot 2^{\eta(3(\omega-2)-1)} \cdot \log(\kappa/\epsilon)\right).
\end{align*}
Since $3(\omega-2)-1>0$ for the current value of $\omega$, this is increasing in $\eta$, and therefore the total cost for updates through Step \ref{alg:mwu-inf-step-change-p-not-whole} of the algorithm is
\[
\Otil\left(\widehat{T}^{3(\omega-2)} \cdot n^2 \cdot \log(\widehat{T})\cdot \log(\kappa/\epsilon) \right),
\]
which by \eqref{eq:infty-norm-k-final-final-bound} and definition of $T$ is
\[
\Otil \left(n^{\omega} \log\left(\frac{\kappa}{  \min_{i\in[n]}\{\vr_i +\vs_i^2\}} \right)\cdot \log(\kappa/\epsilon)\right).
\]
By \cref{lemma:constrained-richardson}, the overall cost of solving the constrained weighted linear regression problems is
\[
\Otil\left(\widehat{T} \cdot n^2 \log(\kappa)\log^2(1/\epsilon)\right)=\Otil\left(n^{7/3} \log\left(\frac{\kappa}{  \min_{i\in[n]}\{\vr_i +\vs_i^2\}} \right)\log(\kappa)\log^2(1/\epsilon)\right).
\]
We now consider the sparse case. First, note that the only randomization comes from the construction and reconstruction of the sparse inverse. Taking union bound and upper bounding the number of reconstructions by the total number of iterations of the algorithm gives the probability bound. We now bound the running time. First note that the number of reconstructions of the sparse inverse triggered by Step \ref{alg:mwu-inf-step-change-p-whole} of the algorithm because the $1$-norm of $\vw$ has changed by a factor of two is only $O(\log n)$ as discussed above. Moreover Step \ref{alg:mwu-inf-step-change-p-whole} is triggered once every $(n/m)^{1/3}$ iterations. Therefore the total cost of Step \ref{alg:mwu-inf-step-change-p-whole} is
\[
\Otil \left(\left(\nnz(\ma)\cdot m \cdot n + n^{\omega} m^{2-\omega} \right) \log^2(\kappa/\epsilon) \cdot m^{1/3} \log\left(\frac{\kappa}{  \min_{i\in[n]}\{\vr_i +\vs_i^2\}} \right)\right),
\]
which is
\[
\Otil \left(\left(\nnz(\ma)\cdot m^{4/3} \cdot n + n^{\omega} m^{7/3-\omega} \right) \log^2(\kappa/\epsilon) \cdot \log\left( \alpha_1\kappa \right)\right).
\]
Now note that the only $\eta$ that can cause an index to be added to the set $S$ through Step \ref{alg:mwu-inf-step-change-p-not-whole} are the ones with $\frac{2^{\eta}}{\ceil{\log_2 (16\rho \cdot \log n)}\cdot \log(2)}\leq (n/m)^{1/3}$. Otherwise, the changes are too small to accumulate enough in $(n/m)^{1/3}$ iterations before a total reconstruction of the sparse inverse through Step \ref{alg:mwu-inf-step-change-p-whole} is triggered. Now consider reconstructions of the inverse triggered by Step \ref{alg:mwu-inf-step-change-p-not-whole}. For one $\eta$, by our above bounds on the number of changes \eqref{eq:eta-changes} is $\widehat{T} 2^{2\eta+3}$. Therefore the cost of such reconstruction is
\[
\Otil\left(\left(\nnz(\ma)\cdot m \cdot n + n^{\omega} m^{2-\omega} \right) \log^2(\kappa/\epsilon) \cdot \frac{\widehat{T} 2^{2\eta}}{n/m} \right).
\]
Since this is increasing in $\eta$, taking the large possible $\eta$ and replacing $\widehat{T}$ by its value, this is 
\[
\Otil\left(\left(\nnz(\ma)\cdot m \cdot n + n^{\omega} m^{2-\omega} \right) \log^2(\kappa/\epsilon) \cdot \frac{n^{1/3} \log(\alpha_1 \kappa) (n/m)^{2/3}}{n/m} \right),
\]
which is 
\[
\Otil \left(\left(\nnz(\ma)\cdot m^{4/3} \cdot n + n^{\omega} m^{7/3-\omega} \right) \log^2(\kappa/\epsilon) \cdot \log\left( \alpha_1\kappa \right)\right).
\]
We now bound the cost of updates to the inverse through the Woodbury identity. In this case, by \cref{thm:sparse-ds}, and since $(\cdot)^{\omega-2}$ is a concave function, for any $\eta$, the cost is
\[
\Otil\left(\left(\nnz(\ma)\cdot m^{2} \cdot \widehat{T} 2^{2\eta} + n^{2} \cdot  \frac{\widehat{T}}{2^{\eta}} \cdot(\frac{\widehat{T} 2^{2\eta}}{\widehat{T} / 2^{\eta}})^{\omega-2} \right) \log^2(\kappa/\epsilon) \right).
\]
Since we only need to consider $\eta$ such that $\frac{2^{\eta}}{\ceil{\log_2 (16\rho \cdot \log n)}\cdot \log(2)}\leq (n/m)^{1/3}$, and this is increasing in $\eta$, the total cost of these updates is
\[
\Otil\left(\left(\nnz(\ma)\cdot m^{4/3} \cdot n + n^{\omega} m^{7/3-\omega} \right) \log^{2}(\kappa/\epsilon) \cdot \log(\alpha_1 \kappa) \right).
\]
Finally, by \cref{thm:sparse-ds,lemma:constrained-richardson}, the cost of solving constrained weighted regression problems is
\[
\Otil \left(\left(\nnz(\ma)\cdot m^2 + n^{2} \right) \log^2(\kappa/\epsilon) \cdot\widehat{T} \right),
\]
which since $m< n$ is
\[
\Otil\left(\left(\nnz(\ma)\cdot m^{4/3} \cdot n + n^{\omega} m^{7/3-\omega} \right) \log^{2}(\kappa/\epsilon) \cdot \log(\alpha_1 \kappa) \log^2(1/\epsilon)\right).
\]
Combining these with the number of scaling factors we need to try to guarantee \eqref{eq:two-infty-bounded-obj} gives the running time.
\end{proof}

\subsection{Mixed \texorpdfstring{$(2,p)$}{(2,p)}-Norm Minimization}
\label{sec:two-p-norm}

In this section, we consider the bit complexity of solving the mixed $(2,p)$-norm minimization \eqref{eq:res-prob-with-grad-constraint} directly. Similar to the mixed $(2,\infty)$-norm minimization, we utilize the multiplicative weights update algorithm, width reduction, and inverse maintenance techniques. The main theorem of this section is the following, which can also be improved beyond the (current) matrix multiplication time for sparse matrices by the data structure of \cref{thm:sparse-ds}.

\begin{theorem}
Let $p>1$, $1>\veps>0$, $z\in \R$, $\ma\in\R^{n\times d}$, $\vg\in\R^{n}$, and $\vt\in\R^{n}_{\geq 0}$ such that $n^{-1/p}\leq \vt_i \leq 1$, for all $i\in[n]$. Moreover, suppose
$\norm{\ma^\top \ma}_{\fro}, \norm{(\ma^\top \ma)^{-1}}_{\fro} \leq \kappa$, and
\[
\norm{(\mi-\ma(\ma^{\top} \ma)^{-1} \ma^\top)\vg}_2 \geq \frac{\veps}{\kappa}.
\]
Moreover, suppose the optimal value of the following problem is at most one.
\begin{align}
\min_{\vdelta \in \R^n} ~~ & ~~ \gamma_p(t,\vdelta) \\ \nonumber \text{s.t.} ~~ & ~~
    \vg^\top \vdelta = z, \\ \nonumber
    ~~ & ~~ \ma \vdelta = 0.
\end{align}
Then there exists an algorithm that computes a constant factor approximation to this problem in time $\Otil_p((n^{\omega} + n^{7/3}) \cdot \log(\kappa/\veps))$.
\end{theorem}

\RestyleAlgo{algoruled}
\IncMargin{0.15cm}
\begin{algorithm}[!ht]

\textbf{Input:} $\ma\in \mathbb{R}^{n\times d}$, $\vt\in \mathbb{R}^n$, $\vg\in \mathbb{R}^n$, $z\in \mathbb{R}$, $p\in (1,\infty)$\\

Set $\mabar = \begin{bmatrix} \ma | \vg \end{bmatrix}$ and $\vbbar = \begin{bmatrix}
\vec{0}_d \\ z
\end{bmatrix}$\\
Set $\rho = \tilde{\Theta}_p(n^{\frac{(p^2-4p+2)}{p(3p-2)}})$,
$\beta = \tilde{\Theta}_p(n^{\frac{p-2}{3p-2}})$,
$\alpha = \tilde{\Theta}_p \left(n^{-\frac{(p^2-5p+2)}{p(3p-2)}} \left(\log(n \norm{\mabar}_2^2 /\norm{\vbbar}_2^2)\right)^{-\frac{p}{(3p-2)}} \right)$, and
 $\tau = \tilde{\Theta}_p \left(n^{\frac{(p-1)(p-2)}{(3p-2)}} \left(\log(n \norm{\mabar}_2^2 /\norm{\vbbar}_2^2)\right)^{\frac{p(p-1)}{(3p-2)}}\right)$ \DontPrintSemicolon \tcp*[h]{\textcolor{blue}{$\rho$ is width parameter, $\beta$ is threshold for $\vr$, $\alpha$ is step size, $\tau$ is threshold for $p$-norm. The constants are picked so the relations in \cref{lemma:p-norm-smooth-bound,lemma:p-norm-quad-pot-grow-fast} are satisfied.}}\\
Set $T = \alpha^{-1} n^{1/p}$,
$i=k=0$, $\vw^{(i,k)} = \vec{0}$ and
$\vx = \vec{0}$\\
$\vrhat_j^{(i,k)} \leftarrow (n^{1/p} \vt_j)^{p-2}, \forall j\in [n]$ \label{alg-step:mixed-2-p-mwu-init-r}\\
Set $\ds$ to be the inverse maintenance data structure \DontPrintSemicolon \tcp*{\textcolor{blue}{either the dense data structure $\dds$ or the sparse data structure $\sds$ with parameter $m$}}
$\ds.\initialize(\mabar,\vrhat^{(i,k)},\frac{\epsilon}{10^6 (\kappa\cdot n)^{30}})$\\
\While{$i<T$}{
{\bf (1) Find the significant buckets and update the preconditioner.}\\
$\vr^{(i,k)} \leftarrow (n^{1/p} \vt)^{p-2} + (\vw^{(i,k)})^{p-2}$ \label{alg-step:mixed-2-p-mwu-update-r}\\
For all $j\in [n]$ find the least non-negative integer $\eta_j$ such that $\frac{1}{2^{\eta_j}} \leq \frac{\vr_j^{(i,k)} - \vr_j^{(\old)}}{\vrhat_j}$\\
$\vr^{(\old)} \leftarrow \vr^{(i,k)}$\\
For all $j \in [n]$, $c_{j,\eta_j} \leftarrow c_{j,\eta_j} + 1$\\
\If{$\ds$ is sparse and $\floor{n^{(p-2)/(3p-2)}/ m^{1/3}}$ divides $i$}{
$S \leftarrow [n]$\\
} \Else {
$S \leftarrow \displaystyle \bigcup_{\eta: i+1 \mod{2^\eta}\equiv 0} \{j: c_{j,\eta} \geq 2^\eta\}$
}
$\vrhat^{(i,k)}_{j} \leftarrow \vr_j^{(i,k)}, \forall j \in S$\\
$c_{j,\eta} \leftarrow 0$ for all $(j,\eta)$ such that $j\in S$.\\
$\ds.\update(S, \vrhat_S^{(i,k)})$

{\bf (2) Solve the weighted linear regression by Richardson's iteration and preconditioning (\cref{lemma:constrained-richardson}).}
\\
Let $\vdeltabar^{(i,k)}\in\R^n$ such that $\norm{\pi_{\mabar}(\vdeltabar^{(i,k)} - {\vdeltastar}^{(i,k)})}_{2} \leq \frac{\epsilon}{T} \norm{\pi_{\mabar}{\vdeltastar}^{(i,k)}}_{2}$ and $\norm{\vdeltabar^{(i,k)} - {\vdeltastar}^{(i,k)}}_2 \leq \frac{\epsilon}{2\kappa^6} \cdot \norm{{\vdeltastar}^{(i,k)}}_{\vr^{(i,k)}}$ where
${\vdeltastar}^{(i,k)} = \argmin_{\Delta:\mabar^\top \vdelta = \vbbar} \norm{\vdelta}_{\vr^{(i,k)}}^2$\\
{\bf (3) Update the weights.}\\
\If(\tcp*[h]{\textcolor{blue}{regular step}}){$\norm{ \vdeltabar }_p^p \leq \tau$}{
$\vw^{(i+1,k)}_j \leftarrow \vw^{(i,k)}_j + \alpha |\vdeltabar_j|, \forall j \in [n]$\\
$\vx \leftarrow \vx + \alpha \vdeltabar$\\
Set $i=i+1$
} \Else(\tcp*[h]{\textcolor{blue}{width-reduction step}}){
For all $j\in [n]$ with $|\vdeltabar_j| \geq \rho$ and $\vr_j \leq \beta$, set $\vw^{(i,k+1)}_j = 4^{1/(p-2)} \max \{n^{1/p} \vt_j, \vw^{(i,k)}_j\}$.\\
For rest of $j\in[n]$, set $\vw^{(i,k+1)}_j = \vw^{(i,k)}_j$.\\
Set $k=k+1$
}
}
\Return $n^{-1/p} \vx$
\caption{Algorithm for the adjusted mixed $(2,p)$-norm problem.}
\label{alg:res-problem}
\end{algorithm}

We start by adjusting the vector $\vt$ and the number $z$, so that all of the entries of $\vt$ are within a polynomial (in $n$) bound, the corresponding problem has an optimal value less than or equal to one, and an approximate solution to the \emph{adjusted} problem gives an approximate solution to the original mixed $(2,p)$-norm problem. 

\begin{lemma}
\label{lemma:t-adjustment}
Let $p\geq 2$, $\vt,\vg\in\R^n$, $\vt\geq 0$, and $j\in\Z$ such that the following is feasible for some $\vdelta\in\R^n$.
\begin{align}
\label{eq:smoothed-p-norm-feasible}
\gamma_p(\vt, \vdelta) & \leq \frac{p}{p-1} 2^{j+p}, \\
\vg^\top \vdelta & = 2^j, \nonumber \\
\ma^\top \vdelta & = 0 \nonumber.
\end{align}
Moreover for all $i\in[n]$, let $\widehat{z} = \left(\frac{2}{p}\right)^{1/2} \left( \frac{p-1}{p} \right)^{1/p} 2^{j(1-1/p)-2}$
\[
\vthat_i = \min\biggl\{\max\biggl\{\left( \frac{p-1}{p} \right)^{1/p} \frac{1}{2^{1+j/p}} \vt_i, n^{-1/p}\biggr\}, 1\biggr\}.
\]
Also let
\[
\vdeltastar = \argmin_{\vdelta: \ma^\top\vdelta=0, \vg^\top \vdelta = \widehat{z}} \gamma_p(\vthat, \vdelta) \leq 1.
\]
Then $\gamma_p(\vthat, \vdeltastar) \leq 1$, and for $\vthat\in\R^n$ such that $\gamma_p(\vthat, \vdeltahat) \leq \beta$, 
\[
\gamma_p(\vt, \vdeltatil) \leq \left( \frac{p}{2} \right)^{p/2} \cdot \frac{p \cdot 2^{p+j}}{p-1} \cdot (\beta + 1),
\]
where $\vdeltatil = \left( \frac{p}{2} \right)^{1/2} \cdot \left(\frac{p}{p-1}\right)^{1/p} \cdot 2^{1+j/p} \cdot (\beta + 1)$.
\end{lemma}

Note that the construction of $\vthat$ in the above lemma guarantees that $n^{-1/p}\leq \vthat \leq 1$
Equipped with the above, we focus on the following problem for the rest of the section.

\begin{definition}[Adjusted mixed $(2,p)$-norm problem]
Let $p\geq 2$, $\mabar\in\R^{n\times d}$, $\vbbar \in \R^d$ and $\vt \in \R^n$ with $n^{-1/p}\leq \vt\leq 1$ such that the optimal value of the following problem is at most one.
\begin{align*}
\min & ~~  \gamma_p(\vt,\vdelta), \\
\st 
& ~~ \mabar^\top \vdelta = \vbbar.
\end{align*}
Then we call this problem an adjusted mixed $(2,p)$-norm problem.
\end{definition}

Note that in our case, $\mabar$ is the matrix $\ma$ concatenated with the gradient vector $\vg$, and $\vbbar$ is the vector zero concatenated by an adjusted version of the value $\vg^\top \vdelta$ for the optimal solution. Then our goal is to find a solution with $\gamma_p(\vt,\vdelta) < \beta$ for the above problem for some constant $\beta>1$.

Similar to the mixed $(2,\infty)$-norm minimization, we solve a series of weighted linear regression problems of the form explained in \cref{subsec:weighted-lin-reg}. Our main contributions are two folds. We show that the algorithm of \cite{AdilKPS19} outputs an approximate and almost feasible solution under fixed-point arithmetic with appropriate bit complexity. Moreover, we show that by using our inverse maintenance technique for the sparse solver, the running time improves beyond the current matrix multiplication time for poly-conditioned sparse matrices. 

\begin{lemma}[\cite{AdilKPS19}]
\label{lemma:p-norm-quad-potential-bound}
Let $p\geq 2$, $\vt,\vw,\vr \in \R_{\geq 0}^n$ with $\vt_j \geq n^{-1/p}$, $\vr_j = (n^{1/p} \vt_j)^{p-2} + \vw_j^{p-2}$, for all $j\in[n]$. Let $\ma\in\R^{n\times d}$, $\vb\in\R^d$, and 
\[
\vdeltahat = \argmin_{\vdelta: \ma^\top \vdelta = \vb} \norm{\vdelta}_{\vr}^2.
\]
Moreover, suppose $\min_{\vdelta:\ma^\top \vdelta \vb} \gamma_{p}(\vt, \vdelta) \leq 1$. 
Then
\begin{enumerate}
    \item 
$
\norm{\vdeltahat}_2^2 \leq \norm{\vdeltahat}_{\vr}^2 \leq n^{(p-2)/p} + \norm{\vw}_p^{p-2}
$
\item $
\abs{\vdeltahat}^\top \abs{\nabla \gamma (n^{1/p} \vt, \vw)} \leq p \gamma (n^{1/p} \vt, \vw)^{(p-1)/p} + p \cdot n^{(p-2)/2p} \gamma (n^{1/p} \vt, \vw)^{1/2}.
$
\end{enumerate}
\end{lemma}

Note that if we replace $\vdeltahat$ with $\vdeltabar$ that is close to $\vdeltahat$ according to \cref{lemma:constrained-richardson}, then the bounds hold by multiplying an appropriate constant with the right-hand side.

A proof similar to \cite{AdilKPS19} implies the following about the growth of the potential function $\gamma_p(n^{1/p} \vt, \vw^{(i,k)})$. Note that the main difference between this and the result of \cite{AdilKPS19} is that our solution to the weighted linear regression problem has some error.

\begin{lemma}[\cite{AdilKPS19}]
\label{lemma:p-norm-smooth-bound}
Let $p\geq 2$ and $i,k$ be nonnegative integers. Let $\varepsilon$ be the error of solving the weighted linear regression problems.
Given $\alpha^{p-1} \tau \leq n^{(p-1)/p}$ and $k \leq n^{2/p} \beta^{-2/(p-2)} \rho^2$,
\[
\gamma_p(n^{1/p} \vt, \vw^{(i,k)}) \leq (1+\varepsilon)^p\left(p^2 2^p \alpha i + n^{1/p} \right)^p \exp\left( \frac{\zeta k}{n^{2/p} \beta^{-2/(p-2)} \rho^2}\right),
\]
where $\zeta := \frac{p}{2} 4^{p/(p-2)} ((p^2 2^p +1)^{p-2} \exp(\frac{p-2}{p}) + 1)$ is just a function of $p$.
\end{lemma}

\noindent
A direct application of \cref{lemma:p-norm-quad-potential-bound} and noting that $\norm{\vw^{(i,k)}}_p \leq \gamma_p(n^{1/p} \vt, \vw^{(i,k)})^{1/p}$ implies the following. 
\begin{lemma}[\cite{AdilKPS19}]
\label{lemma:p-norm-quad-smooth-bound}
Let 
\[
{\vdeltastar}^{(i,k)} = \argmin_{\mabar^\top \vdelta = \vbbar} \norm{\vdelta}_{\vr^{(i,k)}}^2.
\]
Then $\norm{{\vdeltastar}^{(0,0)}}_{\vr^{(0,0)}}^2 \geq \frac{\norm{\vbbar}^2}{\norm{\mabar}_2^2}$, and 
\[
\norm{{\vdeltastar}^{(i,k)}}_{\vr^{(i,k)}}^2 \leq n^{(p-2)/p} + \gamma_p(n^{1/p} \vt, \vw^{(i,k)})^{(p-2)/p}.
\]
\end{lemma}

Since 
\cref{lemma:p-norm-quad-smooth-bound,lemma:p-norm-smooth-bound} imply a bound on the growth of the function $\norm{{\vdeltastar}^{(i,k)}}_{\vr^{(i,k)}}^2$, if we show that in width reduction steps, it grows larger, then we have a bound on the maximum number of width reduction steps. The following shows that this function grows large in the width reduction step.

\begin{lemma}[\cite{AdilKPS19}]
\label{lemma:p-norm-quad-pot-grow-fast}
Consider a width reduction step in Algorithm \ref{alg:res-problem}, i.e., $\norm{\vdeltabar^{(i,k)}}_p > \tau$. Let $q\geq 1$ $\gamma_p(n^{1/p} \vt, \vw^{(i,k)}) \leq q n$, $\tau^{2/p} \geq  2 q \cdot n^{(p-2)/p} \beta^{-1}$, and $\tau \geq 10 q \cdot \rho^{p-2} n^{(p-2)/p}$. Moreover let 
\[
{\vdeltastar}^{(i,k)} = \argmin_{\mabar^\top \vdelta = \vbbar} \norm{\vdelta}_{\vr^{(i,k)}}^2.
\]
Then 
\[
\norm{{\vdeltastar}^{(i,k+1)}}_{\vr^{(i,k+1)}}^2 \geq \norm{{\vdeltastar}^{(i,k)}}_{\vr^{(i,k)}}^2 \cdot  (1+ q \frac{\tau^{2/p}}{n^{(p-2)/p}}).
\]
Moreover, for regular steps, $\norm{{\vdeltastar}^{(i+1,k)}}_{\vr^{(i+1,k)}}^2 \geq \norm{{\vdeltastar}^{(i,k)}}_{\vr^{(i,k)}}^2$.
\end{lemma}

Note that in the above lemma $q$ is a function of only $p$ and comes from \cref{lemma:p-norm-smooth-bound}. Now directly combining \cref{lemma:p-norm-smooth-bound,lemma:p-norm-quad-smooth-bound,lemma:p-norm-quad-pot-grow-fast} gives the following bound for the number of iterations of Algorithm \ref{alg:res-problem}.
\[
O_p \left(n^{(p-2)/(3p-2)} \log^{p/(3p-2)} \left(n \norm{\mabar}_2^2 / \norm{\vbbar}_2^2 \right) \right).
\]

The last piece is to bound the number and distribution of changes in the vector $\vrhat$. Then we can use our data structure results to give the desired running time bounds for both the sparse and dense cases. Note that even though the following result of \cite{AdilKPS19} is with respect to exact solutions for the weighted linear regression problems, since we have the guarantee of $\norm{\vdeltabar - \vdeltahat}_2 \leq \epsilon \norm{ \vdeltahat}_{\vr}$, for $\vdeltahat = \argmin_{\mabar^\top \vdelta = \vbbar} \norm{\vdelta}_{\vr}$, from \cref{lemma:constrained-richardson}, we can guarantee that the error is small enough so that no constant factor change happens due to the error of the regression solution over the course of the algorithm.

\begin{theorem}[\cite{AdilKPS19}]
\label{thm:res-k}
Let $\ell_{e,\eta}$ be the number of indices $j$ that are added to $S$ at iteration $e := i+k$
(where $i$ and $k$ are the numbers of regular and width-reduction steps, respectively)
due to changes between $2^{-\eta}$ and $2^{-\eta+1}$ in Algorithm \ref{alg:res-problem}. Let $T+K=\tilde{\Theta}_p(n^{\frac{p-2}{3p-2}} \log^{p/(3p-2)} \left(n \norm{\mabar}_2^2 / \norm{\vbbar}_2^2 \right))$ be the number of iterations (consisting of $T$ regular steps and $K$ width reduction steps). Then
\begin{align}
\sum_{e=1}^{T+K} \ell_{e,\eta} = \begin{cases}
0 & ~  ~ \text{ if } 2^\eta > T+K \\
\Otil_p \left(n^{\frac{p+2}{3p-2}} \log^{p/(3p-2)} \left(n \norm{\mabar}_2^2 / \norm{\vbbar}_2^2 \right) 2^{2\eta}\right) & ~ ~ \text{ otherwise.}
\end{cases}
\end{align}
\end{theorem}

Now note that for $q$ iterations, only $\eta$ with $2^{\eta+1} < q$ can cause an index to be added to the set $S$. 

\begin{theorem}[continues=thm:main-p-norm]
Let $\ma\in \mathbb{R}^{n\times d}$ be a matrix with condition number bounded by $\kappa$, and $\vb\in\R^{d}$ be a vector with the bit complexity bounded by $\log(\kappa)$. Let $\vxstar=\argmin_{\ma^\top \vx=\vb} \norm{\vx}_p^p$. Let $m\leq n^{(\omega-2)/(\omega-1)}$ be the number of blocks in the block Krylov matrix used by the sparse linear system solver. For $2 \geq p$, there is an algorithm that finds $\vxhat$ such that $\norm{\pi_\ma(\vxhat - \vxstar)}_2 \leq \epsilon \norm{\pi_{\ma}\vxstar}_2$ and 
\[
\norm{\vxhat}_p^p \leq (1+\epsilon) \norm{\vxstar}_p^p
\]
in time 
\[
\Otil_p \left( \left( n^{\omega} + n^{7/3}\log(1/\epsilon) \right)\log^2(1/\epsilon) \log^{1.5}(\kappa / \epsilon) \right).
\]
Moreover, for sparse matrices, there is an algorithm that returns an output with the same guarantees with probability at least $1- n^{-10}$ in time
\begin{align*}
\Otil_p\left(\left(n^{\omega} m^{7/3-\omega} + \nnz(\ma)\cdot m^{4/3} \cdot n \cdot \log(1/\epsilon) + n^{7/3} \cdot \log(1/\epsilon) \right) \log^{2.5}(\kappa/\epsilon) \log^2(1/\epsilon)\right).
\end{align*}
\end{theorem}

\begin{proof}
First, by \cref{lemma:p-norm-initial-solution}, the solution to the linear regression problem is polynomially close to the solution of the $p$-norm problem. Therefore by \cref{lemma:res-improvement}, we only need to solve $O_p(\log(n/\epsilon))$ instances of the residual problem to constant approximation. To do so by \cref{lemma:binary-search,lemma:binary-search-range}, we only need to solve $O_p(\log(n/\epsilon)\log(n/\epsilon))$ instances of the smoothed $p$-norm minimization problems to constant factor approximation. Then \cref{lemma:t-adjustment} implies that to approximately solve each such instance, we only need to solve an adjusted smoothed $p$-norm minimization problem to constant factor approximation. 

Now note that by \cref{lemma:p-norm-smooth-bound,lemma:p-norm-quad-smooth-bound,lemma:p-norm-quad-pot-grow-fast}, in Algorithm \ref{alg:res-problem}, the number of width-reduction steps is bounded by $\Otil_p(n^{2/p} \beta^{-2/(p-2)}\rho^2)$. Therefore
by construction and \cref{lemma:p-norm-smooth-bound}, Algorithm \ref{alg:res-problem} outputs a vector $\vx$ such that 
\[
\gamma_p(n^{1/p}\vt, \vx) \leq  \gamma_p(n^{1/p}\vt, \vw^{(T,K)}) = O_p(1) \cdot n,
\]
where $T$ and $K$ are the numbers of regular steps and width-reduction steps, respectively. Therefore
\[
\gamma_p(\vt, n^{-1/p}\vx) = n^{-1/p} \gamma_p(n^{1/p}\vt, \vx) = O_p(1).
\]
Thus the output of the algorithm is a constant-factor approximation to the smoothed $p$-norm problem.

We now bound the time complexity of Algorithm \ref{alg:res-problem} for both the dense and the sparse case. We first consider the dense case. By \cref{lemma:p-norm-smooth-bound,lemma:p-norm-quad-smooth-bound,lemma:p-norm-quad-pot-grow-fast}, the number of iterations of the algorithm is $\Otil_p \left(n^{(p-2)/(3p-2)} \log^{p/(3p-2)} \left(n \norm{\mabar}_2^2 / \norm{\vbbar}_2^2 \right) \right)$ which since $p/(3p-2) \leq 0.5$ for $p\geq 2$ is $\Otil_p \left(n^{(p-2)/(3p-2)} \log^{0.5} (\frac{\kappa}{\epsilon})\right)$. In each iteration, we iteratively solve a constrained weighted regression problem by accessing a precondition. 
since $\norm{\vw}_{\infty}^p \leq \norm{\vw}_p^p \leq \gamma_p(n^{1/p}\vt,\vw)$, by \cref{lemma:p-norm-smooth-bound}, $\norm{\vr}_{\infty} \leq \poly(n^p)$.
Therefore by \cref{lemma:constrained-richardson}, each constrained weighted regression problem is solver in $\Otil_p(n^2 \log(1/\epsilon)\log(\kappa/\epsilon))$ time. 
Since $\frac{p-2}{3p-2}<1/3$,
this gives a total running time of 
\[
\Otil_p(n^{7/3} \log^3(1/\epsilon) \log^{1.5}(\kappa / \epsilon)),
\]
for solving the constrained weighted regression problems given the preconditioner.

We now bound the running time of inverse maintenance. Consider the cost of inverse maintenance for updates that come from changes that are between $2^{-\eta}$ and $2^{-\eta+1}$. By \cref{thm:dense-ds}, the cost of an update of rank $r$ is $O(\mtime(n,n,r) \log(\kappa/\epsilon))$.
Therefore by \cref{thm:res-k}, and because we only need to consider $\eta$'s with $(T+K)2^{-\eta} > 2$ (larger $\eta$'s do not cause a constant-factor change over the course of the algorithm), the total cost of inverse maintenance over the course of the algorithm is 
\begin{align*}
\sum_{\eta=0}^{\frac{p-2}{3p-2}\log(n)} \sum_{e=0}^{T+K} \Otil_p(\mtime (n,n,\ell_{e,\eta}) \log(\kappa/ \epsilon)).
\end{align*}
Since 
\[
\mtime (n,n,\ell_{e,\eta}) = \Otil(n^{2} \ell_{e,\eta}^{\omega-2}),
\]
$(\cdot)^{\omega-2}$ function is concave, the number of iterations is $\Otil_p \left(n^{(p-2)/(3p-2)} \log^{p/(3p-2)} \left(n \norm{\mabar}_2^2 / \norm{\vbbar}_2^2 \right) \right)$, and because we only perform inverse maintenance updates that come from the changes between $2^{-\eta}$ and $2^{-\eta+1}$ once every $2^{\eta}$ iterations, by \cref{thm:res-k}, the total cost of inverse maintenance is
\begin{align*}
&
n^2 \log^{1/2}(\kappa) \cdot \log(\kappa/\epsilon) \cdot\sum_{\eta=0}^{\frac{p-2}{3p-2}\log(n)} \Otil_p \left(n^{(p-2)/(3p-2)} 2^{-\eta} \left( n^{4/(3p-2)} 2^{3\eta} \right)^{\omega-2} \right) 
\\ & = 
n^2 \log^{1/2}(\kappa)\cdot \log(\kappa/\epsilon) \cdot \sum_{\eta=0}^{\frac{p-2}{3p-2}\log(n)} \Otil_p \left(n^{\frac{p-2+4(\omega-2)}{3p-2}} 2^{\eta(3(\omega-2)-1)} \right)
\end{align*}
Since $3(\omega-2) - 1>0$ for current value of $\omega$, $2^{\eta(3(\omega-2)-1)}$ is increasing in $\eta$.
Since there are only $O(\log(n))$ many different $\eta$, the total cost of inverse maintenance (above) is 
\[
n^2 \log^{1/2}(\kappa) \cdot \log(\kappa/\epsilon) \cdot \log(n) \cdot \Otil_p \left(n^{\frac{p-2+4(\omega-2)}{3p-2}} n^{\frac{p-2}{3p-2}(3(\omega-2)-1)} \right) = \Otil_p(n^{\omega} \log(\kappa/\epsilon) \cdot \log^{1/2}(\kappa/\epsilon)).
\]
Combining this with the cost of solving the constrained weighted regression problems and considering the number of residual problems and smoothed $p$-norm problems we solve gives the final running time for the dense case.

We now analyze the sparse case. First, note that for the sparse case, the randomness only comes from the probability of failure of the inverse operator in \cref{thm:sparse_inverse}. 
Note that even if this reconstruction happens in every iteration, by the above discussion regarding the number of iterations, the probability of failure is less than 
\[
\Otil_p(n^{-20} n^{(p-2)/(3p-2)} \log^2(1/\epsilon) \log^{0.5}(\kappa/\epsilon) ).
\]
Since $1/\epsilon$ and $\kappa$ are at most exponential in $n$, this implies that the total failure probability is bounded by $n^{-10}$ for large enough $n$.

We trigger the reconstuction of the inverse operator once every $\floor{n^{(p-2)/(3p-2)}/m^{1/3}}$ iterations in Algorithm \ref{alg:res-problem}. Therefore by \cref{thm:sparse-ds}, the total cost for such reconstructions is
\[
\Otil_p \left(\left(\nnz(\ma)\cdot m \cdot n + n^{\omega} m^{2-\omega} \right) \log^2(\kappa/\epsilon) \cdot m^{1/3} \log^{p/(3p-2)} \left(n \norm{\mabar}_2^2 / \norm{\vbbar}_2^2 \right)  \right),
\]
which is
\[
\Otil_p\left(\left(\nnz(\ma)\cdot m^{4/3} \cdot n + n^{\omega} m^{7/3-\omega} \right)\log^{2.5}(\kappa/\epsilon) \right).
\]
The other way Algorithm \ref{alg:res-problem} might trigger reconstruction of the inverse operator is that the sparse data structure (\cref{thm:sparse-ds}) receives an update of rank greater than $n/m$. Note that since we force a reconstruction once every $\floor{n^{(p-2)/(3p-2)}/m^{1/3}}$ iterations, the only $\eta$'s that can trigger this second kind of construction should satisfy $2^\eta < \floor{n^{(p-2)/(3p-2)}/m^{1/3}}$. The cost for such reconstructions is then
\[
\Otil_p \left(\left(\nnz(\ma)\cdot m \cdot n + n^{\omega} m^{2-\omega} \right) \log^2(\kappa/\epsilon) \cdot \frac{n^{(p+2)/(3p-2)} 2^{2\eta}}{n/m} \log^{p/(3p-2)} \left(n \norm{\mabar}_2^2 / \norm{\vbbar}_2^2 \right)  \right).
\]
Since this is increasing in $\eta$, and
\[
\frac{n^{(p+2)/(3p-2)} n^{2(p-2)/(3p-2)}/m^{2/3}}{n/m} = m^{1/3},
\]
the total cost of this kind of reconstruction is also
\[
\Otil_p\left(\left(\nnz(\ma)\cdot m^{4/3} \cdot n + n^{\omega} m^{7/3-\omega} \right) \log^{2.5}(\kappa/\epsilon) \right).
\]
The final part of the inverse maintenance running time for the sparse case is when a Woodbury-type update happens, which only occurs when the rank of the update is less than $n/m$. Note that in this case, by \cref{thm:sparse-ds}, and since $(\cdot)^{\omega-2}$ is a concave function and $p/(3p-2)<0.5$, for any $\eta$, the cost is
\[
\Otil_p\left(\left(\nnz(\ma)\cdot m^{2} \cdot n^{(p+2)/(3p-2)} 2^{2\eta} + n^{2} n^{(p-2)/(3p-2)} 2^{-\eta} \left(n^{4/(3p-2)2^{3\eta}}\right)^{\omega-2} \right) \log^{1/2}(\kappa/\epsilon)\log^2(\kappa/\epsilon) \right).
\]
Since we only need to consider $\eta$ such that $2^\eta < \floor{n^{(p-2)/(3p-2)}/m^{1/3}}$, and this is increasing in $\eta$, the total cost of these updates is
\[
\Otil_p\left(\left(\nnz(\ma)\cdot m^{4/3} \cdot n + n^{\omega} m^{7/3-\omega} \right) \log^{2.5}(\kappa/\epsilon) \right).
\]
Finally, by \cref{thm:sparse-ds,lemma:constrained-richardson}, the cost of solving constrained weighted regression problems is
\[
\Otil_p \left(\left(\nnz(\ma)\cdot m^2 + n^{2} \right) \log^2(\kappa/\epsilon) \cdot \log(1/\epsilon) \cdot  n^{(p-2)/(3p-2)} \log^{p/(3p-2)} \left(n \norm{\mabar}_2^2 / \norm{\vbbar}_2^2 \right) \right).
\]
Now note that since $(p-2)/(3p-2)<1/3$, and $m< n$, this is
\[
\Otil_p \left(\left(\nnz(\ma)\cdot m^{4/3} \cdot n + n^{7/3} \right) \log^2(\kappa/\epsilon) \cdot \log(1/\epsilon) \cdot \log^{1/2} (\kappa/\epsilon) \right).
\]
Combining these running times with the number of residual problems and smoothed $p$-norm problems we have to solve gives the desired result.
\end{proof}

\cref{thm:main-p-norm} implies that for $\nnz(A)=O(n)$, polynomially bounded $\kappa/\epsilon$, and current value of $\omega$, the running time is $\Otil_p(n^{2.363})$.
\section{Open Problems}

In this paper, we discussed the bit-complexity of the modern approaches for solving linear regression, $p$-norm regression, and LPs to high accuracy, settling the actual running times of these algorithms. In the following, we discuss some directions and open problems for improving the running times for these and other problems.

\paragraph{Tall cases.}\cite{lee2013path} has shown that for matrices $\ma\in\R^{n\times d}$ with $d\leq n$, an LP problem can be solved in $\Otil(\sqrt{d})$ iterations instead of $\Otil(\sqrt{n})$ iterations. This led to many exciting works, with the most recent one achieving an algorithm with $\Otil((nd+d^{2.5} )\cdot \log(W/\epsilon))$ arithmetic operations \cite{lee2015efficient,BrandLSS20,BLLSS0W21}, where $W$ is a parameter bounding the absolute value of all the numbers in the problem. Even more recently \cite{jambulapati2021improved} showed the $p$-norm minimization problems can be solved in $\Otil_p(d^{(p-2)/(3p-2)})$ iterations instead of $\Otil_p(n^{(p-2)/(3p-2)})$. However, they do not analyze the number of arithmetic operations for their algorithm. It is very interesting to settle the bit complexity and running time of these algorithms. We believe that our techniques and results would be helpful for these, but due to the more complex nature of the inverse maintenance in these problems, further tools are required as well.

\paragraph{Weighted linear regression in matrix multiplication time.}
The works on shifted numbers \cite{storjohann2005shifted} have shown that a linear system can be solved in time $\Otil(n^{\omega} \cdot \ell)$ instead of $\Otil(n^{\omega} \cdot \log(\kappa/\epsilon))$. As we discussed in \cref{subsec:results}, this approach leads to an algorithm (\cref{thm:inverse_free_ipm}) that is faster than the algorithm rising from the approach of \cite{DBLP:conf/stoc/CohenLS19,DBLP:conf/soda/Brand20} (\cref{thm:robust_ipm}) in some settings. However, if we consider the worst case, the latter approach is faster. The main reason for this is that we have to solve linear systems of the form $\ma^{\top} \mw \ma \vx = \vg$ in each iteration of IPM. The running time of approaches based on shifted numbers has a linear dependence on the bit complexity of the matrix, but the dependence on the bit complexity of vector $\vg$ is $\ell_{\vg}/n$, where $\ell_{\vg}$ is the bit complexity of $\vg$. However, note that the multiplication with $\mw$ is changing the bit complexity of the matrix and the only bound we have for entries of $\mw$ come from the inner and outer radius of the LP. Then an important problem is that whether linear systems of the form $\ma^\top \mw \ma \vx = \vg$ can be solved in time $\Otil(n^{\omega}\cdot (\ell_{\ma}+\frac{\ell_{\mw}}{n}+\frac{\ell_{\vg}}{n}))$. An immediate consequence of such a result is an algorithm for solving LPs faster than the approach of \cite{DBLP:conf/stoc/CohenLS19,DBLP:conf/soda/Brand20} in the worst case. Moreover this might lead to faster algorithms for \emph{exact} LP solvers.

\paragraph{Inverse maintenance with shifted numbers.} A drawback of the shifted numbers approach is that it does not work with current techniques for inverse maintenance since it does not produce the inverse as one single explicit matrix. Developing inverse maintenance techniques for solving dynamically changing linear systems using shifted number would improve the running times in \cref{thm:inverse_free_ipm}.

\paragraph{Inverse maintenance with the sparse solver.} We showed that the $p$-norm minimization problem can be solved faster than matrix multiplication for sparse polyconditioned matrices for the current value of $\omega\approx 2.372$. However, for LPs, we can only show such a result for values of $\omega>2.5$. This is mainly due to the bit complexity of the sparse solver that does not allow solving a large batch (of size close to $n$) of linear systems faster than matrix multiplication time and consequently prevents inverse maintenance for LPs for the current value of $\omega$. An approach to resolve this is to find a representation of the inverse in the sparse inverse solver with bit complexity $\Otil(1)$. This then allows for solving a large batch of linear systems.

\paragraph{Bit complexity of general matrix data structures.} Very recently \cite{van2021unifying} has presented an approach for maintaining general matrix formulas. The general approach is that any matrix formula can be considered as a block of the inverse of some larger matrix. This is similar to the approach we utilized for maintaining $\ma (\ma^\top \mx \ms^{-1} \ma)^{-1} \ma^\top$ for linear programs. Exploring the bit complexity bounds in this general form would be interesting. The main questions here are the dependence of required error and condition number bounds on the input matrices and number of them.
\addcontentsline{toc}{section}{References}
\newpage
\bibliographystyle{alpha}
\bibliography{main}
\newpage
\appendix
\section{Low-Rank Matrices}
\label{sec:low-rank}

In this section, we first address the problem of $p$-norm minimization with low-rank matrices. We show that the matrix can be concatenated with a small multiple of the identity matrix, and this only slightly changes the solution.

\LowRankLemma*
\begin{proof}
First note that for any $\vx\in\R^{n}$ such that $\ma^\top \vx = \vb$, a padded with zero version $\vxbar\in\R^{n+d}$ of $\vx$ satisfies $\mabar^\top \vxbar = \vb$. In addition $\norm{\vx}_p^p = \norm{\vxbar}_p^p$. Therefore \[\min_{\vx:\mabar^\top \vx = \vb} \norm{\vx}_p^p \leq \norm{\vxstar}_p^p.\]
Therefore
\[
\norm{\vxtil}_p^p \leq \norm{\vxhat}_p^p \leq (1+\veps_3) \min_{\vx:\mabar^\top \vx = \vb} \norm{\vx}_p^p \leq (1+\veps_3) \norm{\vxstar}_p^p.
\]
Now let
\[
\vystar = \argmin_{\vy:\ma^\top \vy = \vb} \norm{\vy}_2^2.
\]
Let $\vz\in\R^d$ be a vector with entries equal to the last $d$ entries of $\vxhat$. We have
\[
\mabar^\top \vxhat = \ma^\top \vxtil + \veps_2 \vz.
\]
Therefore since $\norm{\mabar^\top \vxhat - \vb}_2 \leq \veps_3$, by triangle inequality 
\[
\norm{\ma^\top \vxtil - \vb}_2 = \norm{\ma^\top \vxtil + \veps_2 \vz - \vb - \veps_2 \vz}_2 \leq \veps_3 + \norm{\veps_2 \vz}_2
\] 

Moreover note that $\norm{\vxhat}_p^p = \norm{\vxtil}_p^p + \norm{\vz}_p^p$. Therefore $\norm{\vz}_p^p \leq (1+\veps_3) \norm{\vxstar}_p^p$. By Holder's inequality and definition of $\vxstar,\vystar$, we have
\begin{align*}
\norm{\vz}_2^p \leq d^{(p-2)/2} \norm{\vz}_p^p \leq d^{(p-2)/2} \cdot (1+\veps_3) \cdot \norm{\vxstar}_p^p \leq d^{(p-2)/2} \cdot (1+\veps_3) \cdot\norm{\vystar}_p^p \leq d^{(p-2)/2} \cdot (1+\veps_3) \cdot \norm{\vystar}_2^p.
\end{align*}
Now note that $\vystar = (\ma^\top)^{\dagger} \vb$, since the $(\ma^\top)^{\dagger} \vb$ is the solution to $\ma^\top \vy = \vb$ that has the minimum $2$-norm \cite{planitz19793}. Moreover since $0<\veps_3<1$, and $p\geq 2$, $(1+\veps_3)^{1/p}<2$. Therefore
\[
\norm{\vz}_2 \leq 2\cdot d^{(p-2)/2p} \norm{\vystar}_2 \leq 2\cdot d^{(p-2)/2p} \norm{(\ma^\top)^\dagger}_2 \norm{\vb}_2 \leq 2\cdot d^{(p-2)/2p} \frac{\norm{\vb}_2}{\sigma}.
\]
Thus 
\[\norm{\ma^\top \vxtil - \vb}_2 \leq \veps_3 + \norm{\veps_2 \vz}_2 \leq \veps_3 + \veps_1 \cdot \norm{\vb}_2.\]
\end{proof}

\end{document}